\documentclass{aa}

\usepackage{float}
\usepackage{graphicx}
\usepackage{txfonts}
\usepackage[colorlinks = true,
linkcolor = blue,
urlcolor  = blue,
citecolor = blue,
anchorcolor = blue]{hyperref}
\begin{document}

\title{MEGADES: MEGARA galaxy disc evolution survey}
\subtitle{Ionised gas diagnosis}

\author{M. Chamorro-Cazorla\inst{1,2},
A. Gil de Paz\inst{1,2},
A. Castillo-Morales\inst{1,2},
A. Camps-Fari\~{n}a\inst{1,2},
J. Gallego\inst{1,2},
E. Carrasco\inst{3},
J. Iglesias-P\'aramo\inst{4},
R. Cedazo\inst{5},
M. L. Garc\'ia-Vargas\inst{6},
S. Pascual\inst{1,2},
N. Cardiel\inst{1,2},
A. P\'erez-Calpena\inst{6},
P. G\'omez-\'Alvarez\inst{6},
I. Mart\'inez-Delgado\inst{6},
C. Catal\'an-Torrecilla\inst{1,2}
\and
J. Zamorano\inst{1,2}
}

\institute{\inst{1}Departamento de F\'isica de la Tierra y
Astrof\'isica, Universidad Complutense de Madrid, E-28040 Madrid,
Spain\\
\email{mchamorro@ucm.es}\\
\inst{2}Instituto de F\'isica de Part\'iculas y del Cosmos IPARCOS, Facultad de Ciencias F\'isicas, Universidad Complutense de Madrid, E-28040 Madrid,
Spain\\
\inst{3}Instituto Nacional de Astrof\'isica, \'Optica
y Electr\'onica, Luis Enrique Erro No.1, C.P. 72840, Tonantzintla,
Puebla, Mexico\\
\inst{4}Instituto de Astrof\'isica de Andaluc\'ia-CSIC,  Glorieta de
la Astronom\'ia s/n, 18008, Granada, Spain\\ 
\inst{5}Universidad Polit\'ecnica de Madrid, Madrid, Spain\\
\inst{6}FRACTAL S.L.N.E. C/ Tulip\'an 2, p13, 1A. E-28231 Las Rozas de Madrid (Spain)
}

\date{\today}


\abstract
   {We present the ionised gas properties and metallicity gradients of the central area of a sample of 43 galaxies using observations obtained by the MEGADES survey. Exploiting the technical capabilities of MEGARA (\textit{Multi-Espectrógrafo en GTC de Alta Resolución para Astronomía}) our observations combine relatively high spectral (R $\sim$ 6000) and spatial (0.62") resolution to study the properties of the ionised gas, such as using the classic diagnostic BPT diagrams in its $[\ion{N}{II}]$ and $[\ion{S}{II}]$ variants. We explore how the diagrams vary as a function of both radius and velocity dispersion of the H$\alpha$ line. We also propose a new diagnostic diagram to assess the relative contributions of AGN, shocks and H II regions in each spatial region as the ratio between the velocity dispersion of the $[\ion{N}{II}]\lambda6584$ and H$\alpha$ lines. A considerable number of regions, regardless of their galactocentric distance, have emission line spectra associated with shocks. This inference follows both from their line ratios, typically characterised by high $[\ion{N}{II}]\lambda6584$/H$\alpha$ and intermediate $[\ion{O}{III}]\lambda5007$/H$\beta$, and from their position in our diagnostic diagram where they lie between the areas associated with HII regions and with AGN. The better selection of HII-region like emission allows for a robust oxygen abundance determination using the N2 indicator, which we use to measure precise abundance gradients. Most galaxies show negligible metallicity gradients, especially among the low abundance (<8.37 dex) fast rotators. The mean value of the slope of the metallicity gradients for this subset is 0.005\,dex\,R$_{\rm e}^{-1}$, with a dispersion of 0.422\,dex\,R$_{\rm e}^{-1}$. Above 8.37 dex the fast rotators consistently show slightly negative metallicity gradients, with a weak correlation between the slope and the y-intercept. The mean slope of these galaxies is $-$0.681\,dex\,R$_{\rm e}^{-1}$, with a dispersion of 0.933\,dex\,R$_{\rm e}^{-1}$. The overall mean value of the gradients for all galaxies in the MEGADES sample is $-$0.025\,dex\,R$_{\rm e}^{-1}$, with a dispersion of 0.766\,dex\,R$_{\rm e}^{-1}$. We discuss the possible implications of these results for the impact of galactic winds on the abundance gradients of galaxies.}


\authorrunning{M. Chamorro-Cazorla et al.}
\maketitle
%

\section{Introduction}

The emission spectra of galaxies are a treasure trove of information which can be exploited to measure several physical parameters of the interstellar medium (ISM) such as gas density, temperature, ionisation, and abundance, among others. The precise properties of the ISM can be determined with detailed photoionization models such as \texttt{Cloudy} \citep{Ferland_1998}, PyNeb \citep{Luridiana_2015}, MAPPINGS \citep{Sutherland_2018} or HCm \citep{Perez_Montero_2014} which calculate the proper energy balance between atomic species and energy levels of each species. Most of the time, however, it is far more practical or even only possible to use line emission diagnostics based on strong emission lines which are easy to measure with significant signal-to-noise \citep[e.g.][]{Kewley_2002,Marino_2013,Pilyugin_2016,Curti_2017}.

These line emission diagnostics have become one of the main tools employed in extragalactic astronomy in order to understand the physics of galaxies with a multitude of applications of particular diagnostics which can provide information on virtually any relevant physical parameter of ISM of the galaxies \citep[see reviews by][]{Kewley_2019, Maiolino_2019}. One of the most used diagnostics is the BPT diagram \citep{Baldwin_1981} based on the H$\alpha$, H$\beta$, [\ion{N}{II}] and [\ion{O}{III}] emission lines ([\ion{N}{II}] can be substituted by [\ion{S}{II}] and [\ion{O}{I}] lines). In this diagram, the combination of lines informs the user of the hardness of the ionising spectra acting on the ISM as well as how prevalent shocks are as an ionisation source. The BPT diagram allows for the classification of galaxies or regions inside them regarding the origin of the ionisation and, consequently, the emission that is observed which is tied to physical processes such as SFR-based ionisation by young massive stars, AGN feedback or ionisation by old stellar populations. However, it has long been recognised that the original [\ion{N}{II}]-based BPT diagram has intrinsic limitations. In particular, objects with composite ionisation or “transition” properties often fall in areas of overlap between star formation and AGN regions, making their classification ambiguous. This is largely due to the sensitivity of the [\ion{N}{II}]/H$\alpha$ ratio to both metallicity and ionisation conditions, which can blur the separation between mechanisms. A more robust approach is therefore to make use of the full set of BPT diagrams, including those based on [\ion{S}{II}] and [\ion{O}{I}], which respond differently to metallicity and hardness of the ionising spectrum and thus help to disentangle these cases. Recent work has shown that the combined use of all three diagrams significantly reduces the fraction of ambiguous sources and improves the classification of transition objects \citep[e.g.][]{Perez-Diaz_2021,Oliveira_2024}. Despite these improvements, some limitations of BPT diagnostics remain, particularly regarding the identification of weak AGN. In such cases, contamination from other sources of hard ionising spectra, such as hot low-mass stars (HOLMES; \citealt{Binette_1994}), can mimic AGN-like emission. This has motivated the development of alternative approaches, among which the WHAN diagram \citep{Cid_Fernandes_2011} is one of the most widely used. Its distinguishing feature is the inclusion of the equivalent width in H$\alpha$ (EW$_\mathrm{H\alpha}$), which provides an additional discriminant to separate genuine AGN activity from ionisation by evolved stellar populations. In particular, AGN are expected to show high values of EW$_\mathrm{H\alpha}$, which enhances their detectability in this diagnostic.

The ability to detect and measure ionisation via shocks is of key importance to address the effects that stellar and AGN feedback have on the host galaxies. Other than in the flux ratios of the diverse emission lines in the spectra, the presence of shocks is also apparent in the velocity dispersion of the emission lines which is much wider in the presence of strong shocks. This can be due to additional kinematic components arising from expanding shells or to higher pressure of the ISM, both arising from mechanical energy input from sources such as supernovae or AGN \citep[e.g.,][]{Ho_2008, Heckman_1990,Veilleux_2001,Ho_2014,Camps-Farina_2018,Camps-Farina_2020}.
As a result, the velocity dispersion has long been used as a diagnostic for the presence of strong shocks in emission spectra in galaxies, typically those that produce galactic winds such as powerful starbursts \citep[e.g.,][]{Rich_2010, Rich_2011} or those produced by AGN \citep[e.g.,][]{D'Agostino_2019}. In \cite{D'Agostino_2019} it is shown that velocity dispersion can act as a differentiating parameter in identifying AGN, complementing the position of the objects in the BPT diagram such that objects with similar positions can be distinguished. Whereas the presence of very wide velocity dispersion measurements is indicative of strong shocks and therefore of the presence of an AGN, its absence is not indicative of the absence of an AGN, which can present themselves with relatively narrow lines. This is especially the case for weaker AGN classified as LINERS \citep[e.g.][]{Cazzoli_2022}.

The advent of integral field spectroscopy studies has allowed for a transition from studying galaxies based on their integrated or central spectra to being able to measure the spectra of each individual spatial region in the galaxy. Surveys such as SAMI \citep{Croom_2012}, CALIFA \citep{Sanchez_2012} or MaNGA \citep{Bundy_2015} thus allow us to extend the aforementioned analysis of the spectra of galaxies into spatially resolved scales by providing 2D maps of the flux and kinematics of the emission lines as well as gradients and 2D maps of diagnostics and several physical parameters \citep[e.g.,][]{Sanchez_2022}. One of the most fruitful of these is the study of the metallicity gradients using IFU data \citep{Vila-Costas_1992,Sanchez_2014,Belfiore_2017}, which are important as tracers of various physical processes such as mergers \citep{DiMatteo_2009,Rich_2012}, gas accretion \citep{Cresci_2010, Sharda_2021} or radial mixing \citep{Sellwood_2002, Schonrich_2009}.

Spatially mapping diagnostics such as the BPT and WHAN is especially useful as one of the main applications of these diagnostics is determining whether a galaxy hosts an AGN. As such, finding the signature ratios for the presence of an AGN located at the centre and surrounding areas of the galaxy provides a more robust detection. Additionally, spatially mapping the BPT ratios allows for the detection of conical outflows of shocked gas ionised by the AGN \citep[e.g.,][]{Lopez-Coba_2019,Mingozzi_2019,Wylezalek_2020}. The physical processes by which the ISM is ionised have been extensively studied but the detailed interplay between the ionising radiation and the multi-phase ISM, as well as the extent to which each ionising source (such as OB stars, AGN, HOLMES, etc.) contributes to the overall energy balance is still not completely understood. High resolution studies using a diverse array of diagnostics will be key in understanding this topic.

In this work we study the emission line diagnostics applied to the MEGADES \citep{Chamorro_Cazorla_2023} sample observed with the MEGARA instrument (\citealt{armando_2018SPIE};  \citealt{carrasco_2018SPIE}). The unprecedented combination of high spatial and spectral resolution in the IFU spectra allows us to probe the ionisation mechanisms in greater detail than has been possible until now. We also combine the BPT and WHAN diagrams with the velocity dispersion of the emission lines to create a novel method for ionised gas diagnostics which allows for a robust determination on the origin of the excitation in the emitted light we observe. Our main objective is to present and validate a new diagnostic method that combines classical line-ratio diagrams with the velocity dispersion of the emission lines, enabling a more robust determination of the ionisation mechanisms at play, which is only feasible thanks to the unprecedented spatial and spectral resolution achieved with MEGARA. As a practical application, we also derive metallicity gradients in the MEGADES galaxies to demonstrate the usefulness of the diagnostic in a context of astrophysical relevance, but this is not intended as a comprehensive chemical abundance study.

The paper is structured as follows: In Sec. \ref{sec: data} we present the sample and data employed as well as how it was reduced and in Sec. \ref{sec: analysis} we explain how the emission line fluxes, velocity dispersions and effective radii were measured. The results and proposed line diagnostics are presented in Sec. \ref{section:Results} before briefly discussing their implications in Sec. \ref{section:Discussion}.

Throughout this paper, we assume a standard $\Lambda$CDM universe whose cosmological parameters are H$_{0}$ = 70\,km\,s$^{-1}$\,Mpc$^{-1}$, $\Omega_{\Lambda}$ = 0.7 and $\mathrm{\Omega_{m}}$ = 0.3.

\section{Data}
\label{sec: data}

All data employed to conduct the analyses presented in this paper belong to MEGADES Survey Data Release I \citep[DR1;][]{Chamorro_Cazorla_2023}. MEGADES includes data from the central regions of 43 different galaxies observed with the MEGARA instrument configured in its LCB (Large Compact Bundle) mode. This observing mode provides integral field spectroscopy (IFS) observations using an integral field unit (IFU) composed of 567 fibres that form a hexagonal tessellation whose individual size is 0.62 arcsec in diameter. These fibres allow the measurement of spectra from spatially connected regions of the sky covering a field of view of 12.5 x 11.3 arcsec$^{2}$. This corresponds to the innermost regions of the galaxies, covering radii from 0.19 up to 3.4\,kpc depending on distance. In total, each MEGARA LCB observation delivers 567 individual spectra, one per fibre, sampling these central regions. Figures illustrating the exact field of view covered by MEGARA for each galaxy are provided in the sample presentation paper \citep{Chamorro_Cazorla_2023}, to which we refer the reader for a visual representation of the covered area. In addition, the LCB mode has 56 extra fibres spread over 8 different regions in the outermost parts of the multi-object mode (MOS) unit of the instrument, located at distances between 1.75 and 2.5\,arcmin from the centre of the IFU, which allow us to take measurements of the sky background simultaneously with the observations. MEGARA is equipped with 18 volume-phase holographic (VPH) gratings to provide observations with spectral coverage ranging from 3650\,$\AA$ to 9700\,$\AA$ with three different spectral resolutions: low (LR; R$\sim$6000), medium (MR; R$\sim$12000) and high (HR; R$\sim$20000).

\begin{figure*}[h]
\center
\includegraphics[trim={5mm 0 0 2.05cm}, clip, width=1\linewidth]{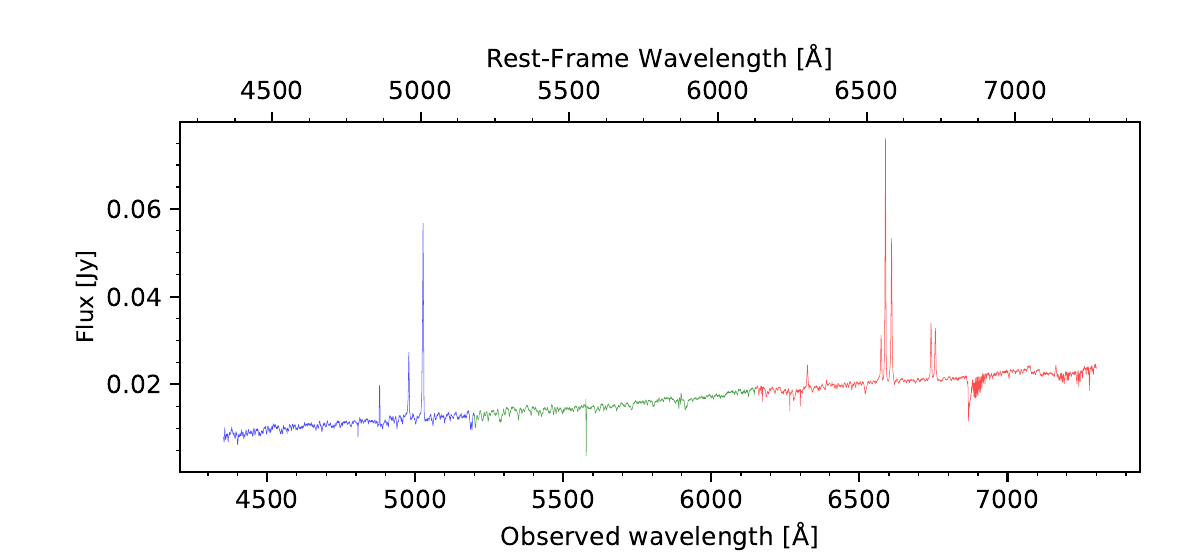}
\caption{NGC~3982 concatenated spectra observed with the LR-B, LR-V and LR-R VPHs in blue, green and red, respectively. Each spectrum is obtained by combining the 567 individual spectra measured simultaneously with the MEGARA LCB IFU.}
\label{fig:MEGARA_spectrum}
\end{figure*}

The observations included in MEGADES DR1 were made using three low-resolution VPHs: VPH480-LR (LR-B), VPH570-LR (LR-V) and VPH675-LR (LR-R). Relevant information on the characteristics of these gratings is included in Table~\ref{table:VPH_characteristics}. The combination of the observations of the three gratings allows us to cover a very wide spectral range from \text{$\sim$ 4350\,$\AA$} to 7288\,\AA. This means that our observations provide us access to the information supplied by several emission lines, from H$\beta$ to the sulphur doublet $[\ion{S}{II}]\lambda\lambda6717,6731$, including all the information regarding the stellar emission of the galaxies. MEGADES includes galaxies from two previous samples, on the one hand galaxies that are included in the S4G sample and on the other hand galaxies that are part of the CALIFA sample. For the galaxies belonging to S4G we have observations obtained with the 3 low-resolution gratings mentioned above, while for the galaxies belonging to CALIFA, we have observations made with LR-V and LR-R (5166 -- 7288 \AA). To illustrate the spectral coverage and the typical appearance of our observations, Figure~\ref{fig:MEGARA_spectrum} shows the integrated spectrum of NGC~3982 obtained by combining the 567 fibres of the MEGARA LCB observation. The figure displays the observations made with the three low-resolution VPHs (LR-B, LR-V and LR-R). The main emission lines used in this work are clearly visible in the spectrum, serving as the basis for the BPT and WHAN diagnostic analysis in Section~\ref{section:Results}.

No reddening correction was applied to the emission-line fluxes, as our analysis relies on relative line ratios involving transitions close in wavelength (e.g. those used in the BPT and WHAN diagrams), which are essentially insensitive to extinction effects. This is a standard practice in emission-line diagnostics (\citealt{Veilleux_1987}; \citealt{Dopita_2016}), and has also been emphasised in recent high-redshift studies \citep{Tripodi_2024}.

\begin{table}[h]
\caption{MEGARA VPHs characteristics.}              
\label{table:VPH_characteristics}      
\centering             
\resizebox{0.49\textwidth}{!}{
\begin{tabular}{c c c c c}          
\hline\hline                     
\noalign{\smallskip}
MEGARA VPH & Spectral Coverage  & Line Res. & $\mathrm{\Delta\lambda_{FWHM}}$ & R \\    
& [$\AA$] & [$\AA$ px$^{-1}$] & [$\AA$] \\
\hline                                   
\noalign{\smallskip}
LR-B & 4350.61 -- 5250.83 & 0.23 & 0.792 & 6061 \\ 
LR-V & 5165.57 -- 6176.18 & 0.27 & 0.937 & 6078 \\
LR-R & 6158.34 -- 7287.67 & 0.31 & 1.106 & 6100 \\
\hline                                             
\end{tabular}}
\end{table}

The MEGADES DR1\footnote{\href{https://www.megades.es}{https://www.megades.es} \\ Username: "public", password: "6BRLukU55E".} provides both the reduced observations and some analyses performed on the stellar component and the interstellar gas component of the galaxies in the sample. The process followed for the reduction of the observations is the standard MEGARA IFU data reduction procedure using the MEGARA data reduction pipeline (DRP) v0.12.0 \citep{pascual_2022}, as described in the MEGARA Reduction Cookbook \citep{africa_castillo_morales_2020_3932063}\footnote{\href{https://doi.org/10.5281/zenodo.1974953}{DOI: 10.5281/zenodo.1974953}} and in the sample presentation paper \citep{Chamorro_Cazorla_2023}, where the DR1 was first introduced and whose preparation included the data reduction carried out by our team. In this paper we make use of those publicly released reduced data products.

\section{Analysis}
\label{sec: analysis}

\subsection{Emission lines}
\label{subsec: Emission lines}

To perform the analysis of the ionised gas present in the interstellar medium of the galaxies in the MEGADES sample we used the already reduced and processed data presented in Section~\ref{sec: data}. More specifically, among all the data products generated for the MEGADES DR1, we analysed the files containing the independent fit of each emission line (named \texttt{[OBJECT]\_[VPH]\_[SPECTRAL\_LINE]}.fits). These files contain the information of each emission line fitting obtained by applying Gauss-Hermite models. These emission-line fits are obtained by using the megaratool \texttt{analyze\_rss} created for this purpose and for other MEGARA emission-line data. More detailed information on the analysis conducted on these spectral features can be found in \cite{Chamorro_Cazorla_2023}. In particular, the fluxes of the Balmer lines (H$\beta$ and H$\alpha$) used throughout this work are corrected for the underlying stellar absorption. This correction was performed by fitting and subtracting the stellar continuum using a full spectral fitting approach, as described in detail in \cite{Chamorro_Cazorla_2023}. Therefore, the resulting emission line fluxes are free from contamination by the stellar population and suitable for diagnostic analysis.

The broad spectral coverage of the MEGADES observations provides us with information on the ionised gas based on different strong emission lines: H$\beta$ and $[\ion{O}{III}]\lambda5007$ lines, H$\alpha$, $[\ion{N}{II}]\lambda6584$ and the two [\ion{S}{II}] lines, $[\ion{S}{II}]\lambda6717$ and $[\ion{S}{II}]\lambda6731$. We detail the definition of the windows set for the emission-line fittings in Table~\ref{table:MEGADES_II_emission_lines_def}.

\begin{table}[h]
	\caption{Line fitting windows definitions in rest frame.}              
	\label{table:MEGADES_II_emission_lines_def}      
	\centering                    
    \resizebox{0.49\textwidth}{!}{
	\begin{tabular}{c c c c}          
		\hline \hline
		\noalign{\smallskip}
		Ion & $\lambda_{0}$ & Line window & Continuum windows\\
		& [\AA] & [\AA] & [\AA] \\
		\hline
		\noalign{\smallskip}
		H$\beta$ & 4861.333 & 4848 -- 4877 &  4828 -- 4848  \& 4877 -- 4892 \\
		$[\ion{O}{III}]\lambda5007$ & 5006.843 & 4997 -- 5017 & 4977 - 4997 \& 5017 -- 5037 \\
		NaI D & 5889.950 & 5883 -- 5905 & 5850 -- 5870 \& 5910 -- 5930 \\
		H$\alpha$ & 6562.819 & 6555 -- 6573 & 6513 -- 6533 \& 6598 -- 6618 \\
		$[\ion{N}{II}]\lambda6584$ & 6583.460 & 6575 -- 6594 & 6513 -- 6533 \& 6598 -- 6618 \\
		$[\ion{S}{II}]\lambda6717$ & 6716.440 & 6707 -- 6724 & 6680 -- 6700 \& 6751 -- 6771 \\
		$[\ion{S}{II}]\lambda6731$ & 6730.810 & 6724 -- 6741 & 6680 -- 6700 \& 6751 -- 6771 \\
		\hline                                             
	\end{tabular}}
\end{table}

\begin{figure}[h]
	\centering
	\includegraphics[trim={0cm 0mm 0cm 0mm},clip, width=1\linewidth]{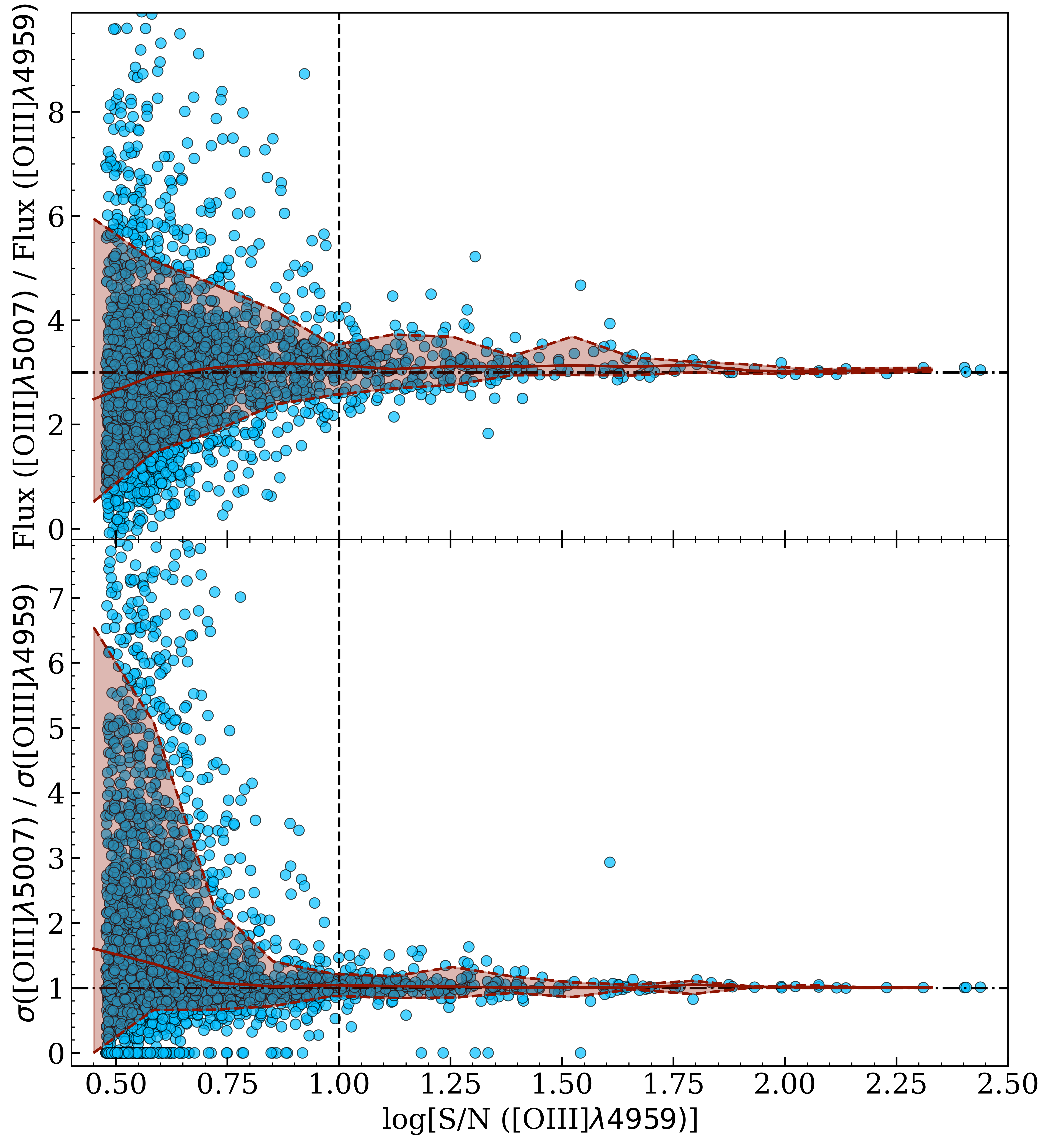}
	\caption{Flux (top panel) and velocity dispersion (bottom panel) ratios of $[\ion{O}{III}]\lambda5007$ to $[\ion{O}{III}]\lambda4959$ as a function of the S/N of $[\ion{O}{III}]\lambda4959$ measured on the peak of the line using all data included in the MEGADES survey. In both cases, the shaded region between the dark red dashed lines marks the position of the 10th and 90th percentiles. The dark red solid line indicates the median value of the flux or velocity dispersion ratio at each S/N value.}
	\label{fig:OIII_sigma_ratios_vs_sn}
\end{figure}

Before starting to analyse the ionised gas emission lines, we wanted to verify to what extent we can be confident in our results. In order to do so, we have checked the limit of the signal-to-noise ratio of our data at which our measurements become unreliable. The method used for this verification is based on comparing the measurements of the flux and the line-of-sight velocity dispersion ($\sigma$) of two different emission lines produced by the same ion, the same excitation mechanism and starting from the same upper level. For this purpose, we apply this technique to the emission line doublet $[\ion{O}{III}]\lambda4959$ and $[\ion{O}{III}]\lambda5007$. Figure~\ref{fig:OIII_sigma_ratios_vs_sn} shows the relation between the flux (top panel) and the velocity dispersion (bottom panel) measured on both lines as a function of the signal-to-noise ratio (S/N; per pixel) at the peak of the weakest line ($[\ion{O}{III}]\lambda4959$) for measurements performed on all data included in the MEGADES survey.

As expected, Figure~\ref{fig:OIII_sigma_ratios_vs_sn} shows that the lower the S/N the greater the scatter on the ratio obtained when measuring the flux and velocity dispersion of each of these two lines. In order to get similar values for $\sigma([\ion{O}{III}]\lambda5007)$ and $\sigma([\ion{O}{III}]\lambda4959)$ we have to reach at least a signal-to-noise ratio at the peak of the line of S/N > 10. This value will be the threshold that we will use throughout this work whenever we make use of ionised-gas data.

Although Galactic extinction correction was not applied to our spectra, the impact of such correction on our derived line ratios is negligible. Our calculations indicate that the relative extinction affecting the $A_{{\rm H}\alpha}/A_{{\rm [\ion{N}{II}]}\lambda 6584}$ ratio is 1.004, while the $A_{{\rm H}\beta}/A_{[\ion{O}{III}]\lambda5007}$ ratio is affected by a relative extinction of 1.039. The average extinction in the case of our sample of galaxies in the g-band is approximately 0.15\,mag, which implies changes in the line ratios of $-$0.0001 dex for H$\alpha$/$[\ion{N}{II}]\lambda6584$ and $-$0.0013 dex for H$\beta$/$[\ion{O}{III}]\lambda5007$. Based on these results, we can conclude that the relevance of Galactic extinction in the context of nearby-line ratios is not significant.

\subsection{Effective radii estimation}
\label{subsec: effective_radii}

\begin{figure*}[ht!]
	\centering
	\includegraphics[trim={0cm 0mm 0cm 0mm},clip, width=0.33\linewidth]{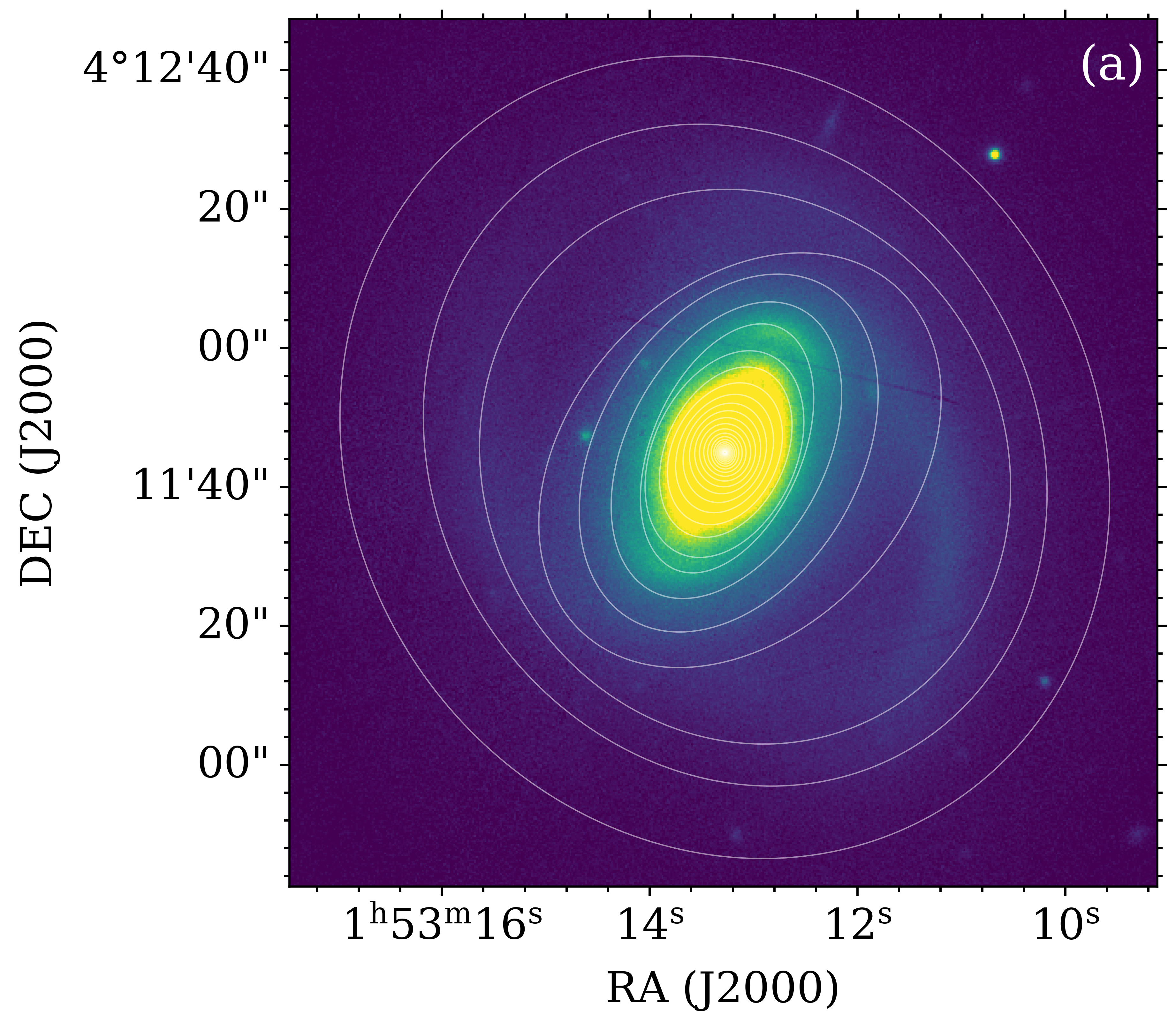}
    \includegraphics[trim={0cm 0mm 0cm 0mm},clip, width=0.3\linewidth]{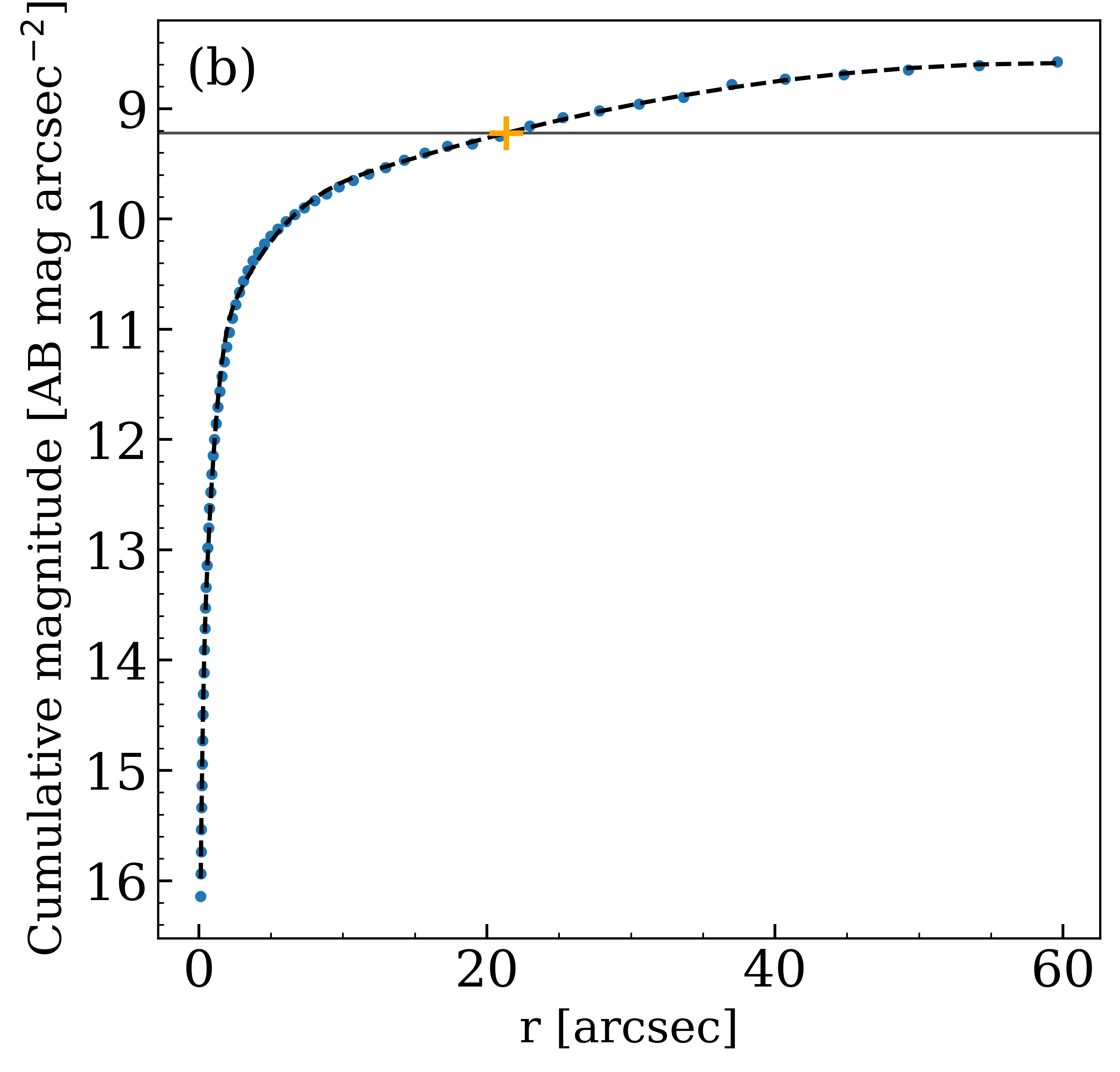}
	\includegraphics[trim={0cm 0mm 0cm 0mm},clip, width=0.325\linewidth]{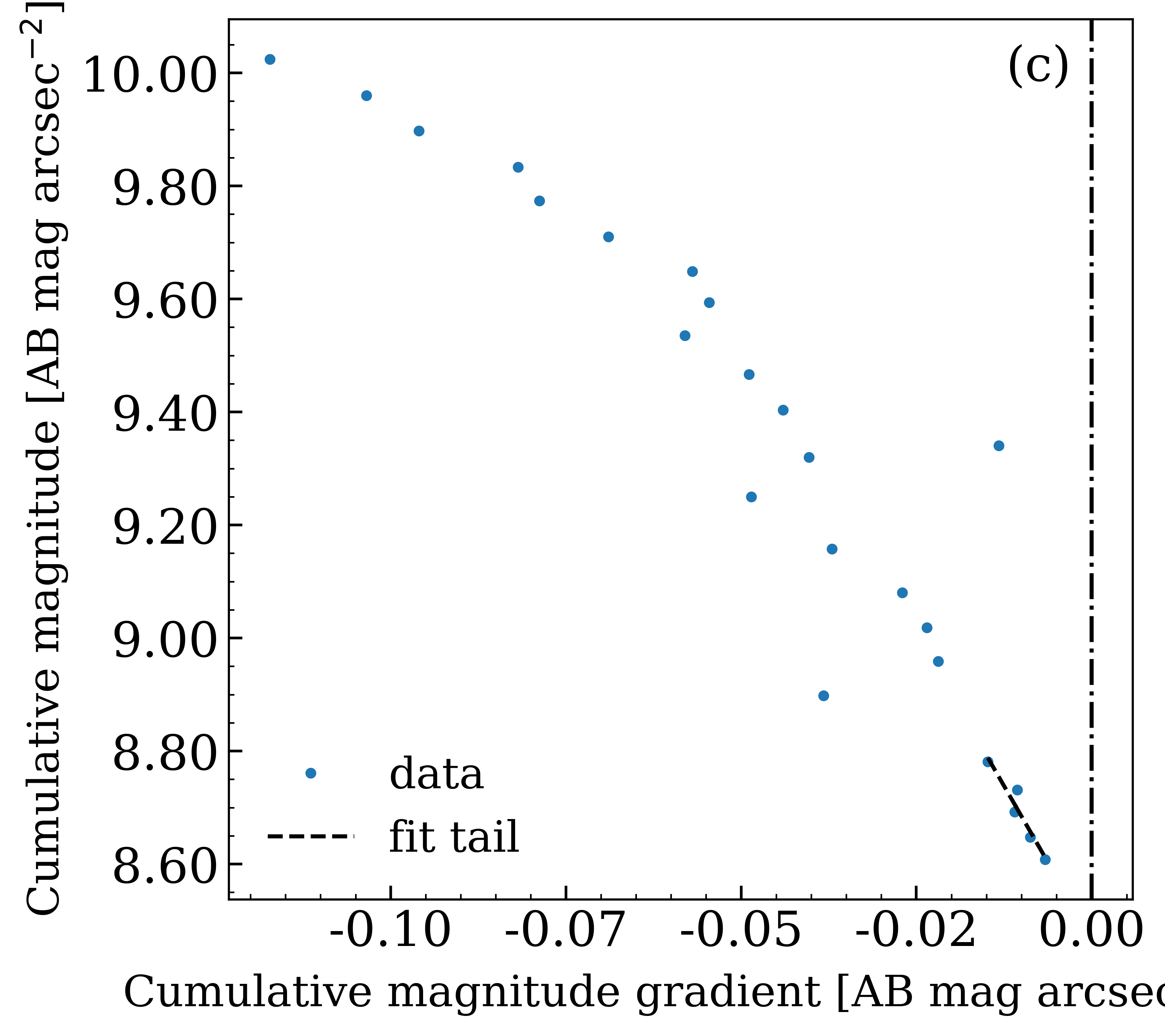}
    \caption{Panel (a): PanSTARRS g-band image of the galaxy NGC~0718 with the isophotes calculated using Photutils overplotted in white. Panel (b): Cumulative magnitude derived from the previously measured isophotes as a function of the semi-major axis. Blue dots correspond to the integrated magnitude within each isophote, and the black dashed line is a smoothed interpolation of the distribution. The solid blue horizontal line indicates the asymptotic magnitude of the galaxy, and the orange cross marks the semi-major axis at which the cumulative magnitude reaches a value equal to the asymptotic magnitude plus 2.5log(2), corresponding to the effective radius. Panel (c): Cumulative magnitude as a function of the cumulative magnitude gradient. Each blue dot corresponds to a given isophote. The black dashed line shows a linear fit to the data points at the lowest gradients, and the vertical black dashed-dotted line marks the zero gradient. The intercept of the linear fit with the vertical axis is used to estimate the asymptotic magnitude of the galaxy.}
	\label{fig:A1}
\end{figure*}

In order to establish comparisons in our estimates of metallicity gradients between the various galaxies in our sample, we have proceeded to estimate the effective radius. This measure allows us to better contextualise the metallicity variations in relation to the physical dimensions of the galaxies. For the estimation of the effective radii of the galaxies in the sample, available in Table~\ref{table:gradients_summary}, we have applied the method based on the asymptotic magnitude estimation described in \cite{Cairos_2001} and applied in works such as \cite{Gil_de_Paz_2007}. Therefore, we use the brightness profiles of each of the galaxies calculated from their photometric decomposition elliptical isophotes (Panel (a) in Figure~\ref{fig:A1}) to estimate the cumulative flux at each radius (Panel (b) in Figure~\ref{fig:A1}) and how it changes as a function of the cumulative flux gradient (Panel (c) Figure~\ref{fig:A1}). If we perform a linear fit of the cumulative flux gradient as a function of its gradient we can use the y-intercept of this fit as the asymptotic magnitude of the galaxy. Once we have this information, to calculate the effective radius of the galaxy we simply locate the semimajor axis at which the cumulative flux is equal to the asymptotic magnitude plus 2.5log(2).

The images we use to perform the photometric decomposition of the MEGADES galaxies come from the PanSTARRS survey \citep{Chambers_2016}. Specifically, we have used the $g$-band cutout images of 250\,\arcsec $\times$ 250\,\arcsec (1000 $\times$ 1000 pixels) in size available at the Pan-STARRS1 Database \citep{Flewelling_2020}. It is worth remembering that the field of view of the MEGARA IFU is 12.5 x 11.3 arcsec$^{2}$ so we can only cover the most central regions of the PanSTARRS image (50 $\times$ 45 pixels).

The photometric analysis has been carried out using the \texttt{Photutils} \citep{larry_bradley_2023} software package included in \texttt{Astropy}, which is specifically designed to perform this kind of task. The \texttt{isophote} package within \texttt{Photutils} fits different elliptical isophotes to the input data and allows us to obtain a model of the data from these fits based on the iterative method created by \cite{Jedrzejewski_1987}.

\section{Results}
\label{section:Results}

\subsection{Diagnostic diagrams}

\begin{figure*}[ht!]
	\centering
	\includegraphics[trim={0cm 0mm 0cm 0mm},clip, width=0.45\linewidth]{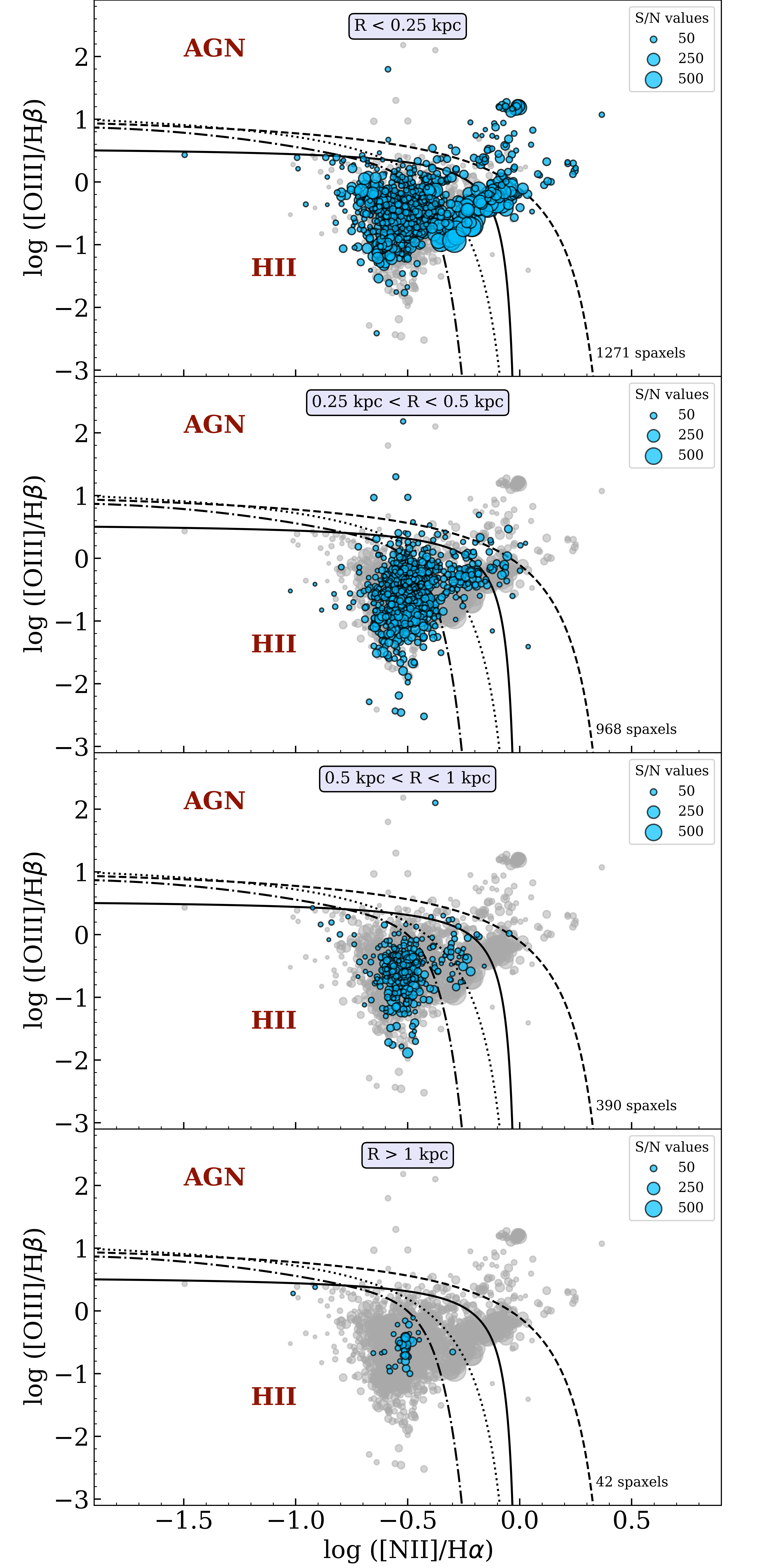}
	\includegraphics[trim={0cm 0mm 0cm 0mm},clip, width=0.45\linewidth]{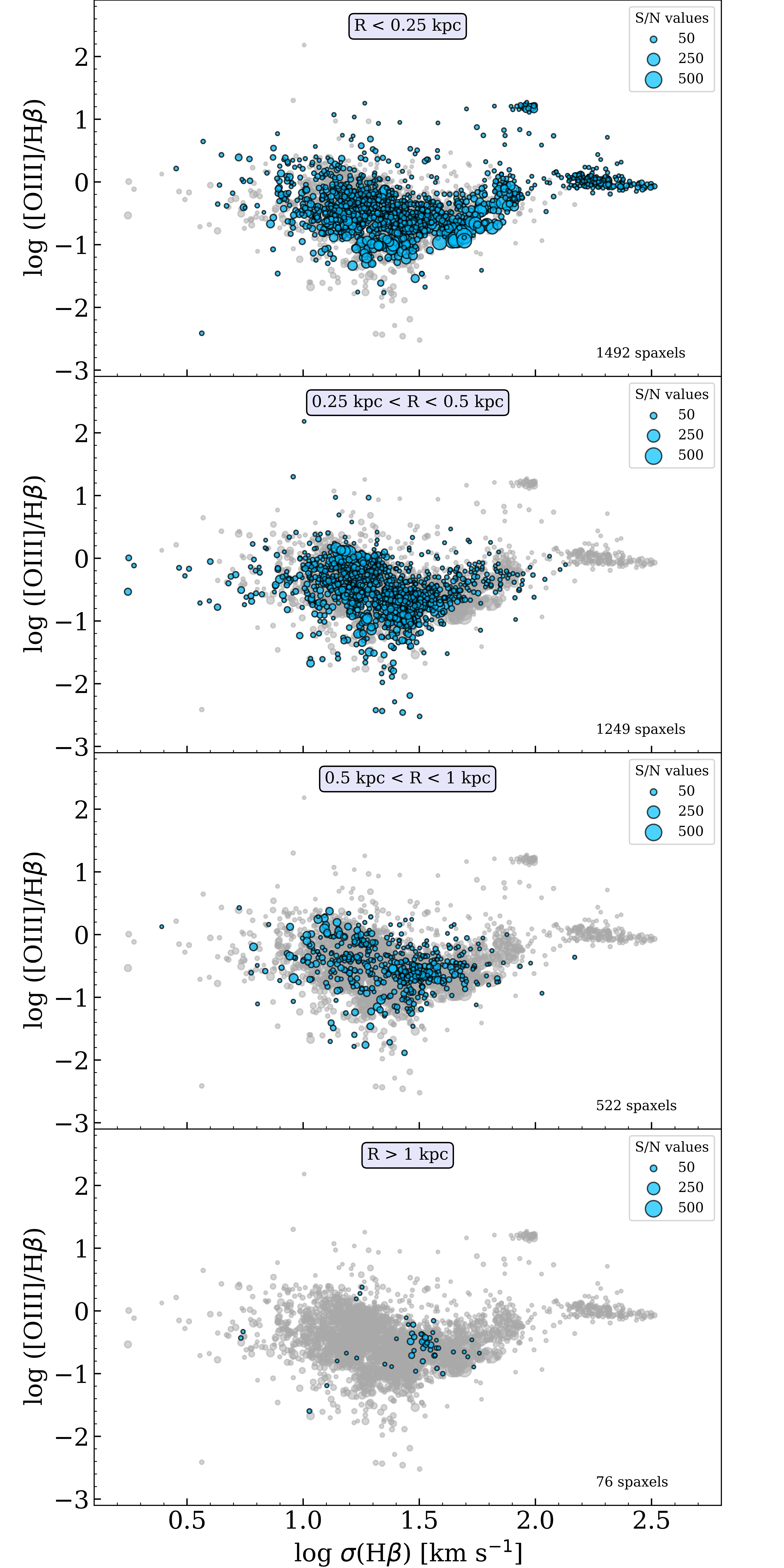}
	
	\caption{\textit{Left panels}: BPT $[\ion{O}{III}]\lambda5007$/H$\beta$ versus $[\ion{N}{II}]\lambda6584$/H$\alpha$ diagram of all spaxels in the MEGADES sample. From top to bottom this diagram is shown in different panels with galactocentric distance cuts ranging from < 0.25\,kpc, between 0.25 and 0.5\,kpc, between 0.5 and 1\,kpc, and \text{> 1\,kpc,} respectively. The size of each point depends on the signal-to-noise ratio (S/N) measured at the peak of the $[\ion{N}{II}]\lambda6584$ line. The dashed, dotted, dash-dotted, and solid lines represent the \cite{Kewley_2001}, \cite{Kauffmann_2003}, \cite{Stasinska_2006} and \cite{Espinosa_2020} demarcation curves, respectively. \textit{Right panels}: relationship between $[\ion{O}{III}]\lambda5007$/H$\beta$ and the velocity dispersion measured on the H$\beta$ line for different galactocentric distances as shown in the left panels. The size of each point depends on the S/N measured at the peak of the H$\beta$ line. Grey dots in the background of the figures correspond to all points in the sample.}
	\label{fig:dd_OIII}
\end{figure*}

Since diagnostic diagrams were first proposed by~\cite{Baldwin_1981}, comparing the flux ratios $[\ion{O}{III}]\lambda5007$/H$\beta$ versus $[\ion{N}{II}]\lambda6584$/H$\alpha$, BPT diagrams (named after the authors that first used them: Baldwin, Phillips and Terlevich) have been very useful to discern the excitation mechanisms responsible for the ionisation and line emission in the interstellar medium. In the left panel of Figure~\ref{fig:dd_OIII} we show the results obtained from measuring the aforementioned flux ratios in all the spaxels of the sample (when they meet the quality criteria explained in Section~\ref{subsec: Emission lines}). Different panels in this figure show the evolution of the distribution of the points when plotting these diagrams using spaxels at different galactocentric distances. These cuts at different galactocentric distances (R < 0.25\,kpc, 0.25\,kpc < R < 0.5\,kpc, 0.5\,kpc < R < 1\,kpc and R > 1\,kpc) show that most of the spaxels with emission lines excited by the influence of the central AGN disappear beyond 0.25\,kpc. This may be expected since as we move away from the galactic centre the influence of the AGN becomes gradually weaker at a rate that depends on the brightness of the AGN itself and the seeing of the observations. At all distances we have regions excited by photoionisation from massive stars and, as we get further away from the innermost region, it becomes the only existing excitation mechanism. Note that, in principle, the ionisation by the AGN should only reach the Broad- (BLR) and Narrow-Line Region (NLR), both much more compact that the MEGARA spaxel size for any of our objects. The fact that we find spaxels with AGN-like line ratios beyond the innermost bin in galactocentric distance might be due to contamination coming from extended Point-Spread Function (PSF) associated to the nuclear (point-like) AGN emission.

In the right panels of Figure~\ref{fig:dd_OIII} we show the velocity dispersion values measured on the H$\beta$ lines ($\sigma_{{\rm H}\beta}$) and display them against $[\ion{O}{III}]\lambda5007$/H$\beta$ in order to compare these results with those of the BPT diagrams to their left, for the same cuts in  galactocentric distances. As expected, the spaxels with lines showing the greatest velocity dispersion ($\sigma_{{\rm H}\beta}$ > 100\,km\,s$^{-1}$) appear in the regions with R < 0.25 kpc, the innermost regions, although we find values corresponding to different ionisation mechanisms in this region. This is also related to the influence of the BLR of the AGN. Here we should note that no multi-component decomposition of the Balmer lines was attempted in these innermost regions of AGN-host galaxies, so the H$\beta$ line widths reported are likely intermediate between those of the BLR and NLR, even if these fits were performed adopting a rather flexible Gauss-Hermite functional form. In this diagram we also report on the presence of a U-shape in the $[\ion{O}{III}]\lambda5007$/H$\beta$ ratio that is particularly noticeable at the innermost regions \text{(R < 0.5 kpc)} mainly because the contribution of spaxels with high-$\sigma$ but also high-$[\ion{O}{III}]\lambda5007$/H$\beta$ ratios. This deviates from the general trend of having spaxels with progressively lower $[\ion{O}{III}]\lambda5007$/H$\beta$ ratios as  $\sigma$ increases, until we reach the turning point at log($\sigma_{\mathrm{H}\beta}$(km\,s$^{-1}$)) $\sim$ 1.5. The most likely explanation for that trend is that as $\sigma$ increases in regions away from the influence of the AGN the contribution of shocks associated with sites of star formation (stellar winds and supernovae) increases, which also leads to a larger contribution of these mechanisms to the line emission and these mechanisms are known to lead to lower  ionisation. This naturally boosts lower-ionisation species (and emission lines), such as $[\ion{N}{II}]\lambda6584$ (see below), but dims higher-ionisation ones, such as $[\ion{O}{III}]\lambda5007$.

\begin{figure*}[ht!]
	\centering
	\includegraphics[trim={0cm 0mm 0cm 0mm},clip, width=0.45\linewidth]{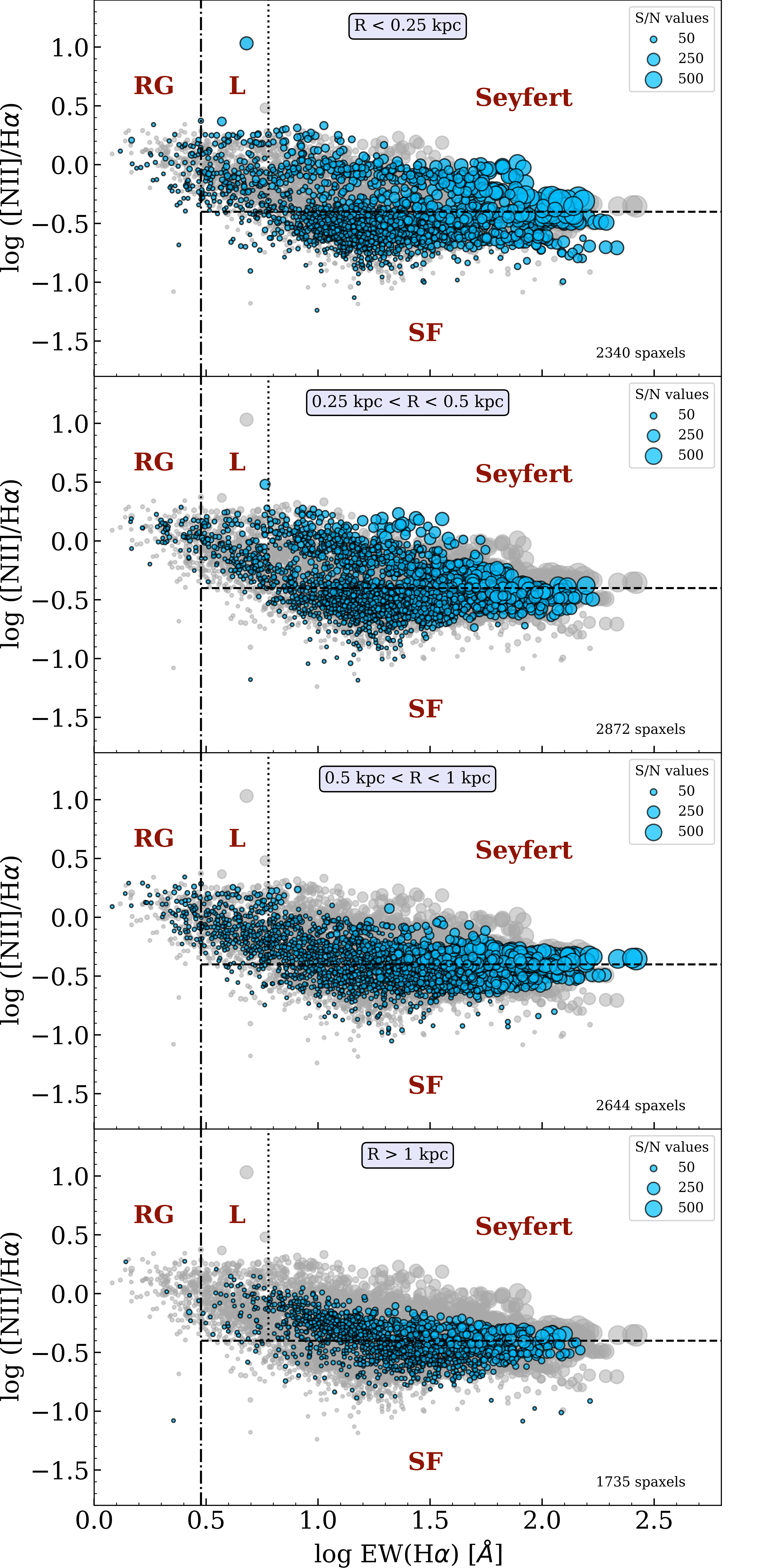}
	\includegraphics[trim={0cm 0mm 0cm 0mm},clip, width=0.45\linewidth]{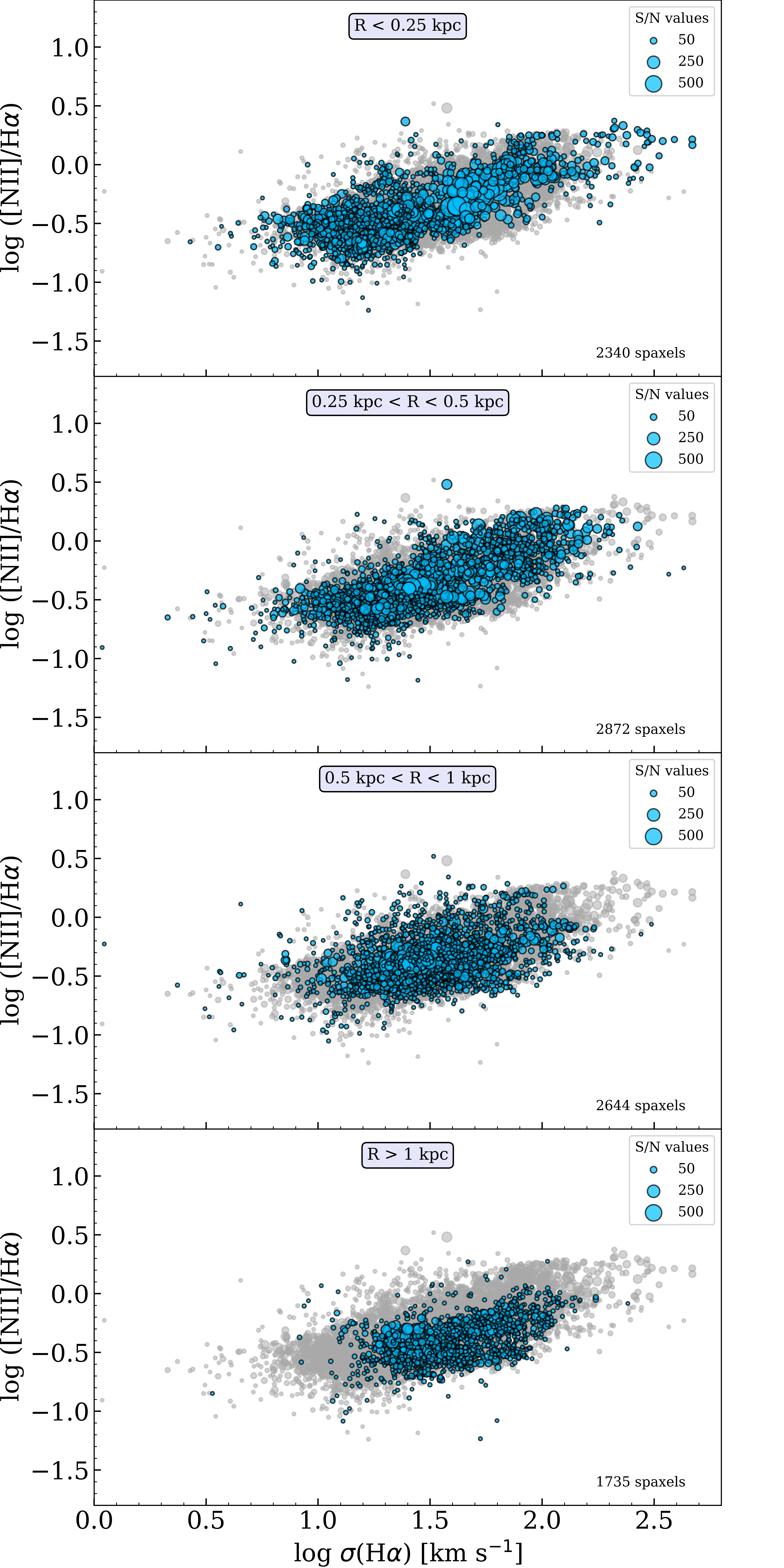}
	
	\caption{\textit{Left panels}: Diagram showing the $[\ion{N}{II}]\lambda6584$/H$\alpha$ versus the equivalent width of the H$\alpha$ line for different galactocentric distance cuts. The distance cuts are respectively from top to bottom; $<$ 0.25\,kpc, between 0.25 and 0.5\,kpc, between 0.5 and\,1 kpc, and $>$ 1 kpc. The dashed line marks the boundary proposed by \cite{Stasinska_2006} to distinguish between star-forming regions (labelled as SF in the plot) and AGN. The dotted line indicates the boundary established by \cite{Kewley_2006} to separate, within the AGN area, between Seyfert and LINERs (labelled as L). The dash-dotted line separates the region populated by lines produced in "retired galaxies" (labelled as RG) proposed by~\cite{Cid_Fernandes_2011}. \textit{Right panels}: diagram $[\ion{N}{II}]\lambda6584$/H$\alpha$ versus velocity dispersion of the H$\alpha$ line for different galactocentric distance cuts, as in the case of the left panel. In both panels the size of each point depends on the signal-to-noise ratio measured at the peak of the $[\ion{N}{II}]\lambda6584$ line. Grey dots in the background of the figures correspond to all points in the sample.}
	\label{fig:dd_NII}
\end{figure*}

The difference in the number of spaxels between the two panels of Figure~\ref{fig:dd_OIII} arises because the left-hand panels combine lines observed with different gratings (LR-B for H$\beta$ and $[\ion{O}{III}]\lambda5007$, LR-R for H$\alpha$ and $[\ion{N}{II}]\lambda6584$). In this case only spaxels with information available in both gratings can be used. In addition, the S/N criterion is applied to different lines on each axis, which introduces further differences in the selected sample. Finally, the effective overlap of the IFU fields depends on how well the telescope pointings of the LR-B and LR-R observations are aligned. The pointings are not always perfectly accurate and in some cases the offset can be significant, as discussed in \citet{Chamorro_Cazorla_2023}. By contrast, the right-hand panels, based solely on LR-B data, is not affected by these limitations and therefore includes more spaxels. In Figure~\ref{fig:dd_NII}, both axes rely on LR-R lines, so the same spaxels enter in both panels.

Another method used to establish the origin of emission lines is the so-called WHAN diagram~\citep{Cid_Fernandes_2010}. This diagram represent $[\ion{N}{II}]\lambda6584$/H$\alpha$ versus the equivalent width of H$\alpha$ (EW$_{{\rm H}\alpha}$) and enable us to distinguish regions excited by star formation from those excited by AGN, differentiating Seyfert from LINER (or LINER-like; see below) zones. The innovative aspect of this classification is the inclusion of a region that has classically been considered to be low-luminosity AGN (and, in particular, LINERs) but should be better designed as presenting LINER-like excitation conditions. According to these authors, this new mechanism occurs in "retired galaxies" (RG), which are galaxies that have stopped forming stars and whose source of ionisation comes from evolved hot low-mass stars (HOLMES,~\citep{Binette_1994}; see also the discussions by~\cite{Singh_2013, Papaderos_2013, Belfiore_2016} on a similar type of objects, the LIERs, Low-Ionization Emission-line Regions). It should be noted that studies carried out by~\cite{Byler_2019} show that the EW$_{{\rm H}\alpha}$ produced by a population whose ionisation radiation is dominated by post-AGB stars typically varies between 0.1\,$\AA$ and 2.5\,$\AA$, specially at ages over log$_{10}$(t/yr) $\gtrsim$ 9.

An important feature of the WHAN diagram is its use of a simple constant threshold in log($[\ion{N}{II}]\lambda6584$/H$\alpha$) to separate star-forming and AGN-like regions. This threshold, typically set at log($[\ion{N}{II}]\lambda6584$/H$\alpha$) = $-$0.4 originates from the work of~\cite{Stasinska_2006}, who showed that this ratio alone can effectively classify galaxies into normal star-forming (log($[\ion{N}{II}]\lambda6584$/H$\alpha$) $\leq$ $-$0.4), hybrid ($-$0.4 < log($[\ion{N}{II}]\lambda6584$/H$\alpha$) $\leq$ $-$0.2), and AGN-dominated (log($[\ion{N}{II}]\lambda6584$/H$\alpha$) > $-$0.2) categories. Thus, the WHAN diagram inherits this empirical demarcation, making it particularly useful when other diagnostic lines are unavailable or suffer from low S/N.

We show the WHAN diagram using the observations of the MEGADES sample in Figure~\ref{fig:dd_NII} (left panels). We find that the distribution of points in our sample is similar to the one obtained by~\citep{Cid_Fernandes_2011} using SDSS data (see their Figure 6). According to this diagram, part of the emitting regions we detect (those with EW$_{{\rm H}\alpha}$ < 3\,$\AA$) come from ionisation by HOLMES. This mechanism is present at practically all galactocentric distances. Although the WHAN and BPT diagrams share a similar criterion in terms of log($[\ion{N}{II}]\lambda6584$/H$\alpha$), the inclusion of EW(H$\alpha$) in WHAN leads to differences in the classification of individual spaxels. In particular, we find that $\sim$ 95~\% of the regions identified as AGN in the BPT diagram using the~\cite{Stasinska_2006} criterion are also classified as AGN in the WHAN diagram, highlighting both the consistency and the added diagnostic value of this diagram.

In the right panel of Figure~\ref{fig:dd_NII} we present the measurements obtained for $[\ion{N}{II}]\lambda6584$/H$\alpha$ ratio versus the H$\alpha$ velocity dispersion. In this case we observe a clear correlation with larger values of $[\ion{N}{II}]\lambda6584$/H$\alpha$ ratio showing larger H$\alpha$ velocity dispersion. According to the scenario proposed above to explain the U-shape in the $[\ion{O}{III}]\lambda5007$/H$\beta$ versus $\sigma$ diagram, this trend could be explain as due to the enhanced contribution of shock-excitation to the emission from low-ionisation species such as $[\ion{N}{II}]\lambda6584$ at progressively larger $\sigma$ values. 

\begin{figure*}[ht!]
	\centering
	\includegraphics[trim={0cm 0mm 0cm 0mm},clip, width=0.45\linewidth]{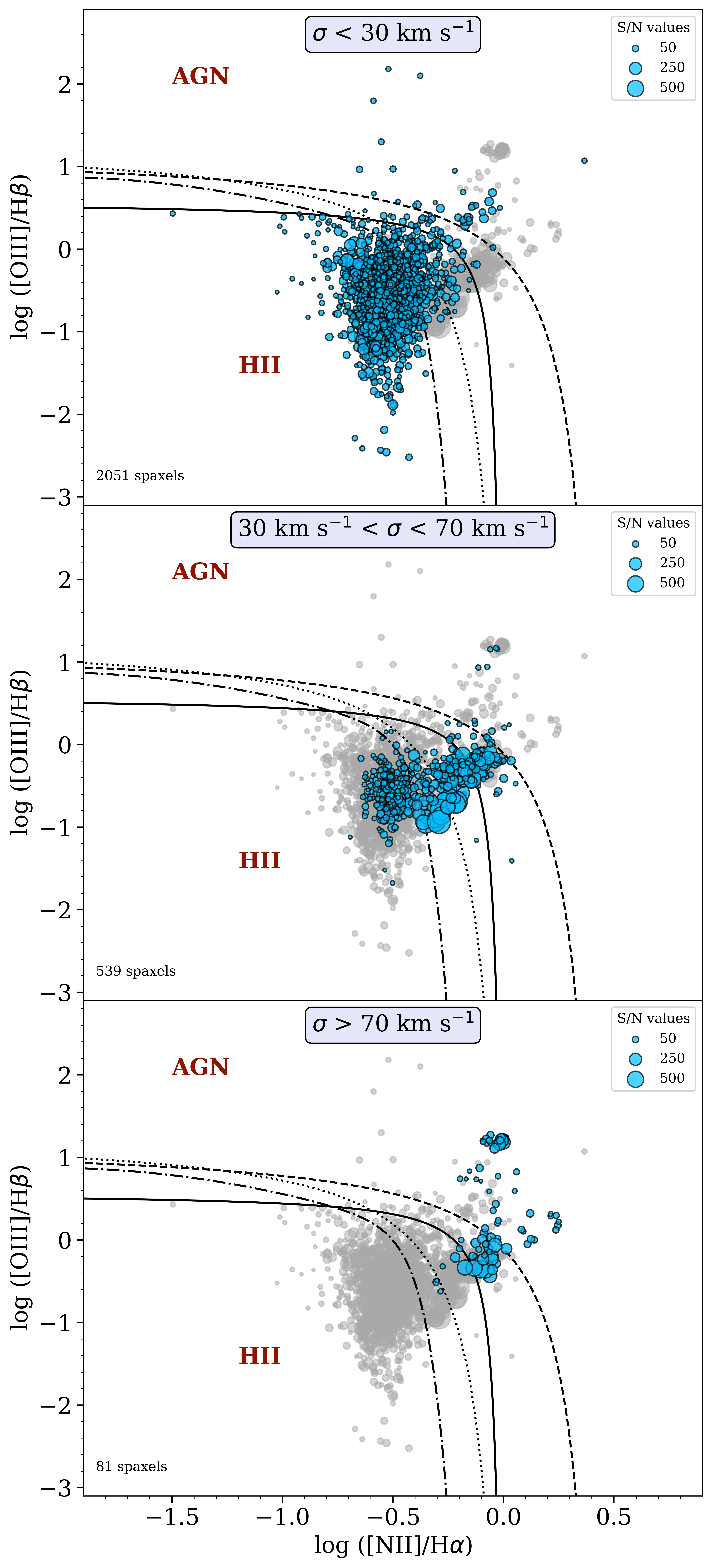}
	\includegraphics[trim={0cm 0mm 0cm 0mm},clip, width=0.45\linewidth]{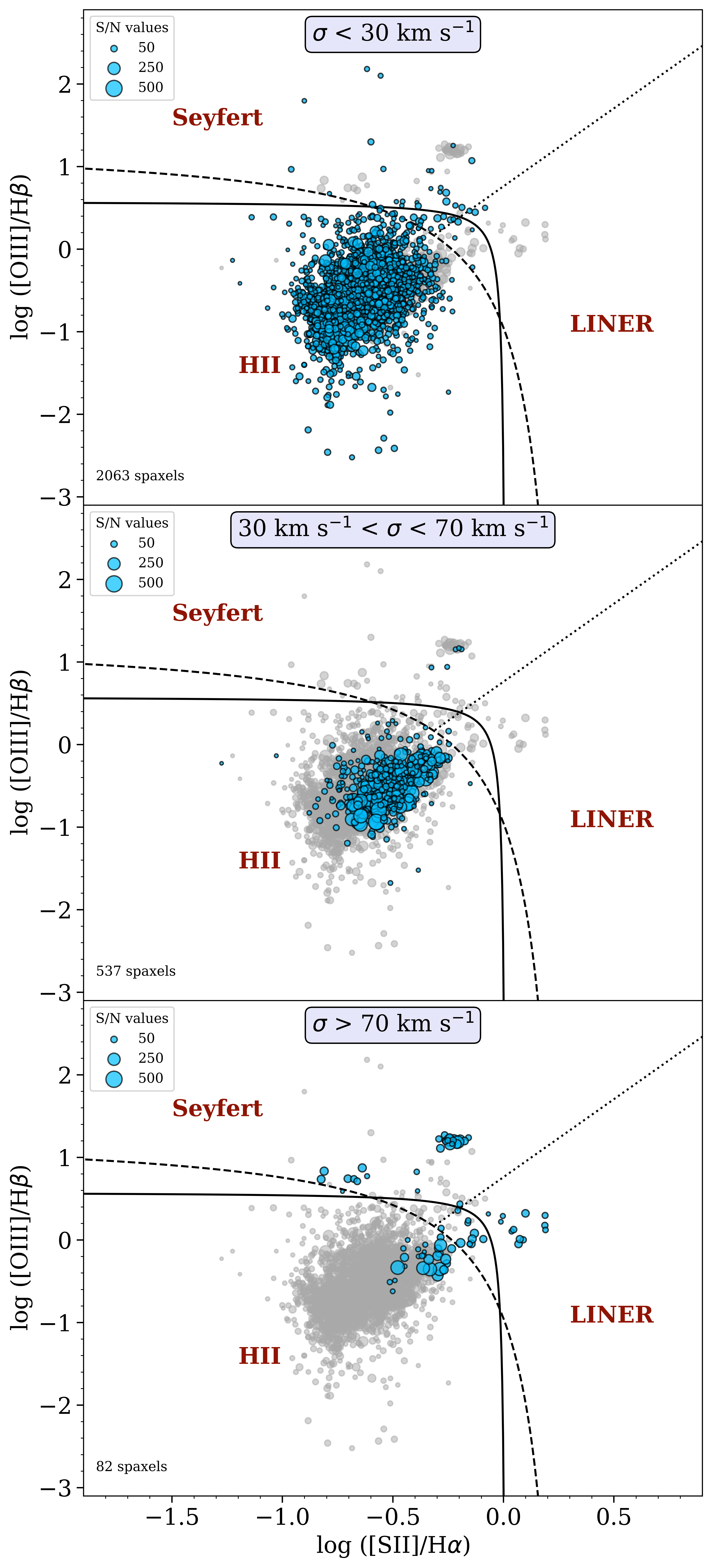}
	\caption{BPT diagrams $[\ion{O}{III}]\lambda5007$/H$\beta$ versus $[\ion{N}{II}]\lambda6584$/H$\alpha$ (left panel) and $[\ion{O}{III}]\lambda5007$/H$\beta$ versus ($[\ion{S}{II}]\lambda6717$+$[\ion{S}{II}]\lambda6731$)/H$\alpha$ (right panel). Both panels show the diagrams for three different cuts as a function of the velocity dispersion measured on the H$\alpha$ line. From top to bottom the cutoffs are: $<$ 30\,km\,s$^{-1}$, between 30 and 70\,km\,s$^{-1}$ and  $>$ 70\,km\,s$^{-1}$, respectively. In the left panel, the dashed, dotted, dash-dotted, and solid lines represent the \cite{Kewley_2001}, \cite{Kauffmann_2003}, \cite{Stasinska_2006} and \cite{Espinosa_2020} demarcation curves, respectively. In the right panel, the dashed, the dotted and the solid lines represent the \cite{Kewley_2001}, \cite{Kewley_2006} and \cite{Espinosa_2020} demarcation curves, respectively. The size of the points in the left and right panels depends on the S/N measured at the peak of the $[\ion{N}{II}]\lambda6584$ line and $[\ion{S}{II}]\lambda6717$, respectively. Grey dots in the background of the figures correspond to all points in the sample.}
	\label{fig:BPT_sigma_cuts}
\end{figure*}

The fact that the measurement of the ionised-gas velocity dispersion allowed us to identify spaxels affected by (broad) AGN emission, have led us to evaluate the BPT diagrams of our emission-line regions in the MEGADES galaxies as a function of this parameter. Thus, in Figure~\ref{fig:BPT_sigma_cuts} we show again the classical BPT diagrams of $[\ion{O}{III}]\lambda5007$/H$\beta$ versus $[\ion{N}{II}]\lambda6584$/H$\alpha$ (left panels) and $[\ion{O}{III}]\lambda5007$/H$\beta$ versus ($[\ion{S}{II}]\lambda6717$+$[\ion{S}{II}]\lambda6731$)/H$\alpha$ (right panels) but now split in different ranges of $\sigma_{{\rm H}\alpha}$, namely $\sigma_{{\rm H}\alpha}$ < 30\,km\,s$^{-1}$ (top panels), 30\,km\,s$^{-1}$ < $\sigma_{{\rm H}\alpha}$ < 70\,km\,s$^{-1}$ (middle panels) and \text{$\sigma_{{\rm H}\alpha}$ > 70\,km\,s$^{-1}$} (bottom panels).

Both diagrams show that for lower velocity dispersions \text{($\sigma_{{\rm H}\alpha}$ < 30\,km\,s$^{-1}$)} the spaxels are mostly located where the excitation is due to photoionisation by massive stars. For intermediate $\sigma_{{\rm H}\alpha}$ values (30\,km\,s$^{-1}$ < $\sigma_{{\rm H}\alpha}$ < 70\,km\,s$^{-1}$) we still have a majority of photoionisation-dominated line ratios but they occupy regions closer to the AGN dividing line. Finally, for the highest velocity dispersion values (\text{$\sigma_{{\rm H}\alpha}$ > 70\,km\,s$^{-1}$}) we find that all spaxels are located in the AGN excitation region for the $[\ion{N}{II}]\lambda6584$/H$\alpha$ diagram and in the case of the $[\ion{S}{II}]\lambda6717$+$[\ion{S}{II}]\lambda6731$ diagram most of them are also located in this region, although this depends on the specific demarcation considered.

The identification of relations between the velocity dispersion of the ionised gas emission lines and the mechanisms driving their excitation is a predictable finding, supported by previous research, as evidenced by the work of \cite{D'Agostino_2019}. They establish a relation between the velocity dispersion and the way in which the regions of a galaxy are distributed in the BPT diagram (emission line ratio information, ELR) in order to define a criterion that enables the determination of the predominant excitation mechanism in each area of the galaxy. However, this criterion is based on a data-dependent method, as the authors themselves mention in the paper, and since they only have the analysis of one galaxy (NGC 1068) we consider it unsuitable to apply it to our sample data. The fact that we include data from 43 different galaxies, each of them with different metallicities and with different spatial resolutions due to their different distances, means that our data may introduce a large scatter in the BPT diagram of the sample, especially along the boundary that divides the excitation mechanisms.

The potential to identify the mechanisms responsible for the excitation of the different components of interstellar gas by measuring their velocity dispersion highlights the need for instruments with high spectral (and spatial) resolution in order to have sufficient resolving power to be able to measure the lines with the lowest velocity dispersion. This study has only been possible thanks to the unique combination of high spatial and spectral resolution of the observations collected with MEGARA.

\subsection{Dynamical selection of excitation mechanisms}

Due to the fact that the behaviour of the different spaxel-by-spaxel line ratios as a function of line width can be interpreted in terms of the relative contribution of AGN, shocks and H~II regions photoionisation within each spaxel, we now attempt to provide a diagnosis based exclusively on dynamical properties. In order to do that, we first assume that in many cases more than one mechanism might be at play but that different lines have a different degree of sensitivity to those mechanisms. Thus, while the $[\ion{N}{II}]$ lines are very sensitive to low-velocity shocks, narrow Balmer lines are sensitive to all mechanisms (including the  Narrow-Line Region of AGN at intermediate line widths) and very broad Balmer lines are almost exclusively produced in the BLR of AGN. Therefore, there where different mechanisms at play, the comparison between the line widths of these lines (besides the impact on line ratios described in the previous section) could provide further information, and additional diagnostics, on the nature of the dominant excitation mechanism. 

\begin{figure*}[ht!]
	\centering
	\includegraphics[trim={0cm 0mm 0cm 0mm},clip, width=0.45\linewidth]{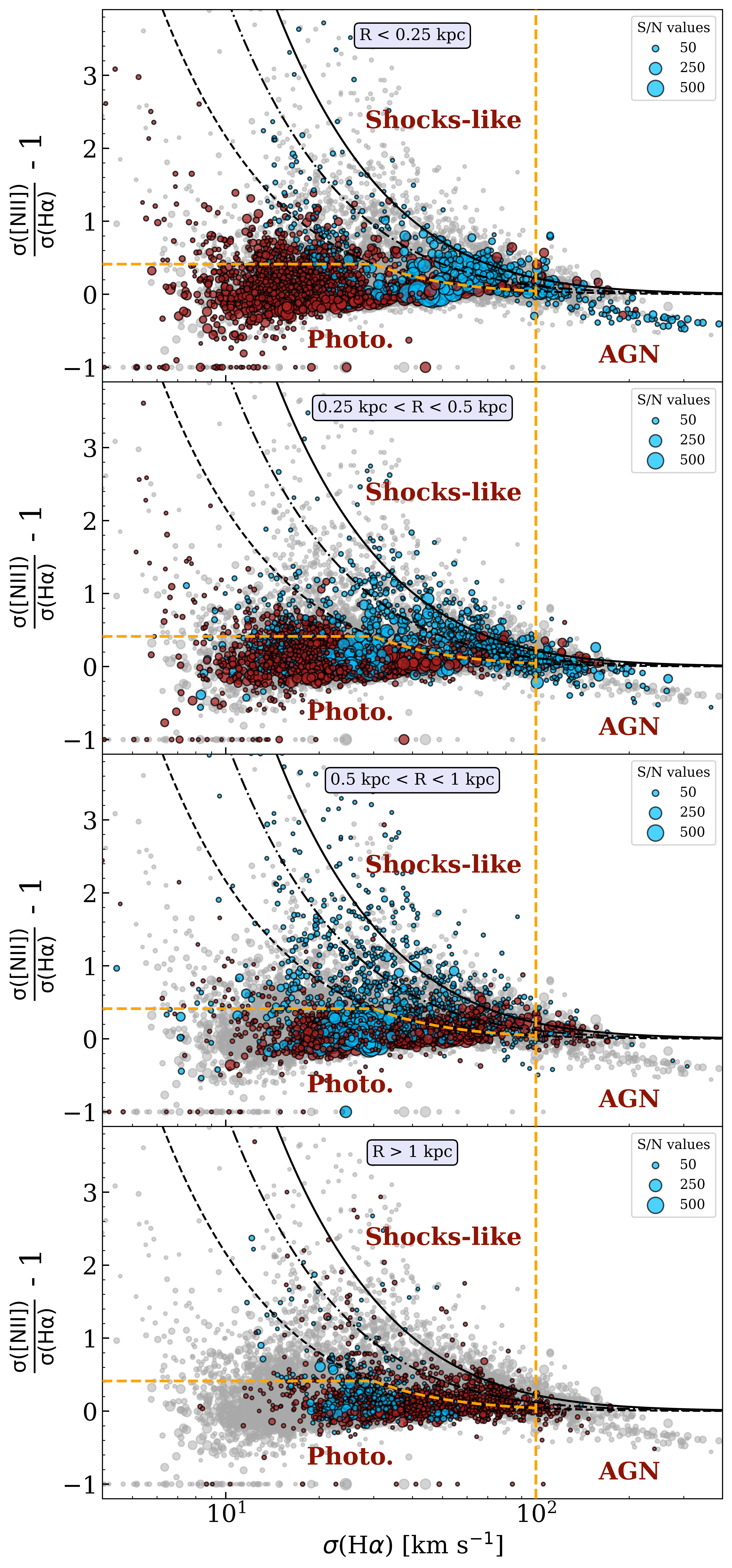}
	\includegraphics[trim={0cm 0mm 0cm 0mm},clip, width=0.435\linewidth]{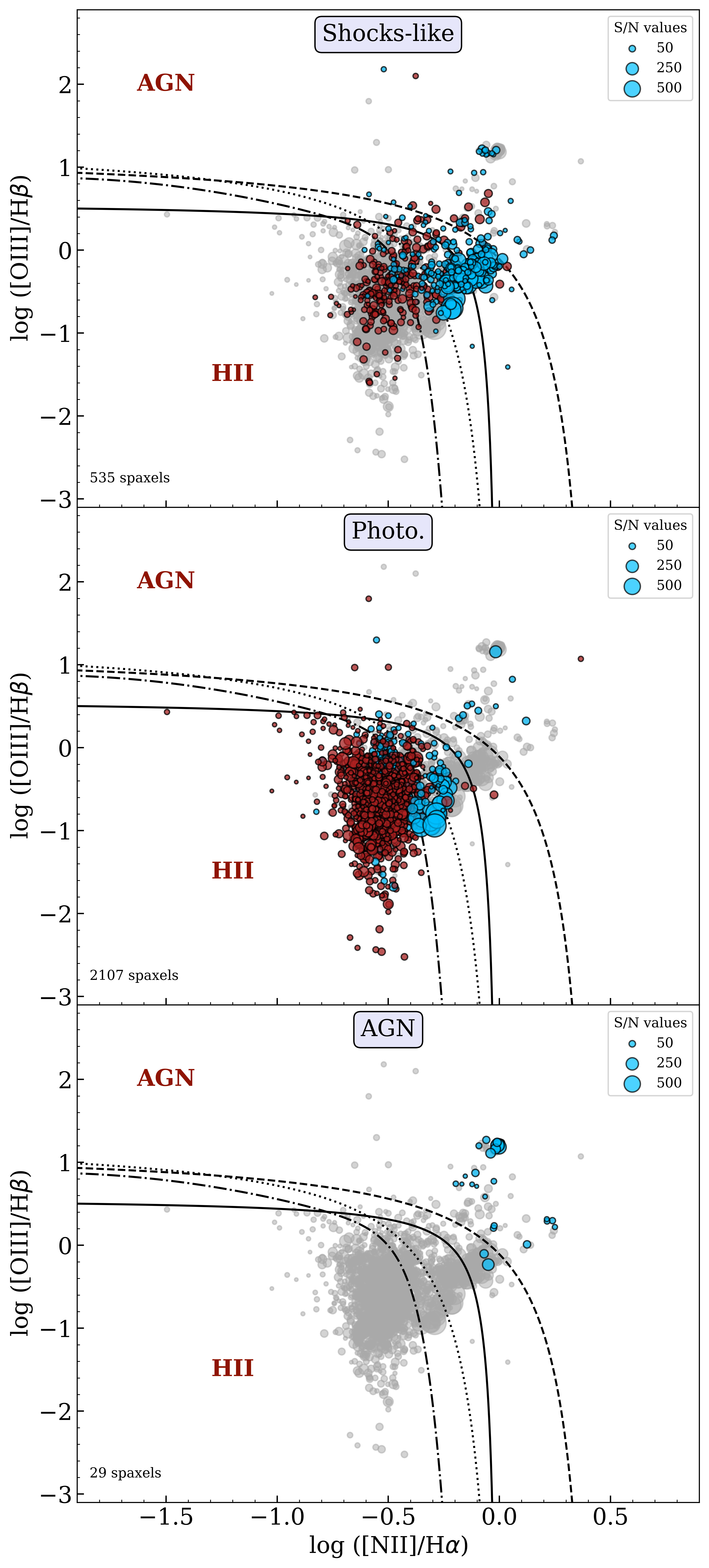}
	
	\caption{\textit{Left panels}: relationship between the measured ratio of velocity dispersions of $[\ion{N}{II}]\lambda6584$ and H$\alpha$ lines versus the velocity dispersion on the H$\alpha$ line for different galactocentric distance cuts. The dashed, dash-dotted and solid black curves represent lines with a constant broadening with respect to the H$\alpha$ line of 30, 50 and 70\,km\,s$^{-1}$, respectively. The dashed orange lines indicate the boundaries between the regions defined by the dynamic selection of excitation mechanisms explained in this section. \textit{Right panels}: BPT diagram $[\ion{O}{III}]\lambda5007$/H$\beta$ versus $[\ion{N}{II}]\lambda6584$/H$\alpha$ corresponding to the points falling within each of the regions defined in the figure on the left. From top to bottom we have the BPT diagram for the lines excited by shocks-like mechanism, photoionisation and AGNs, respectively. In the right panel, the dashed, dotted, dash-dotted, and solid lines represent the \cite{Kewley_2001}, \cite{Kauffmann_2003}, \cite{Stasinska_2006} and \cite{Espinosa_2020} demarcation curves respectively. In both panels blue dots denote lines measured in galaxies classified as Seyfert or LINER (see Table~\ref{table:gradients_summary}) while the red dots represents the rest of the galaxies. The size of the points depends on the signal-to-noise ratio measured at the peak of the $[\ion{N}{II}]\lambda6584$ line. Grey dots in the background of the figures correspond to all points in the sample.}
	\label{fig:BPT_dynamic_cuts}
\end{figure*}

In the left panel of Figure~\ref{fig:BPT_dynamic_cuts} we have plotted the velocity dispersion measured in the $[\ion{N}{II}]\lambda6584$ and H$\alpha$ lines against the velocity dispersion in the H$\alpha$ line. By doing so, we try to establish a purely dynamical criterion to distinguish the excitation mechanisms that have led to the appearance of these emission lines. 

The boundaries separating the regions populated by lines excited by (or dominated by) different excitation mechanisms have been selected on the basis of the following criteria. First, the line separating the AGN region from the shock and photoionisation regions has been adopted according to the full width at half maximum (FWHM) lower limit of 200\,km\,s$^{-1}$ given in~\cite{Vaona_2012} for lines belonging to the NLR. This, translated into velocity dispersion, implies a $\sigma \sim$ 85\,km\,s$^{-1}$. To add some conservative margin to this limit, we have taken $\sigma_{{\rm H}\alpha}$ = 100\,km\,s$^{-1}$ as the proposed division between the zone of excitation produced by AGN and the rest of the mechanisms.

For $\sigma_{{\rm H}\alpha}$ < 100\,km\,s$^{-1}$, we have followed two different criteria. First, for low $\sigma_{{\rm H}\alpha}$ values, we have made a cut at constant $\sigma_{[\ion{N}{II}]\lambda6584}$/$\sigma_{{\rm H}\alpha} -$ 1. This constant value represents regions that would show a  $[\ion{N}{II}]\lambda6584$ line broader than H$\alpha$ by a constant fraction of the line width of H$\alpha$. Then, for $\sigma_{{\rm H}\alpha}$ > 30\,km\,s$^{-1}$ we apply a constant (quadratic) broadening in velocity of $\Delta$=30\,km\,s$^{-1}$. In Figure~\ref{fig:BPT_dynamic_cuts} we show the predictions in $\sigma_{[\ion{N}{II}]\lambda6584}$/$\sigma_{{\rm H}\alpha} -$ 1 for such 30\,km\,s$^{-1}$ broadening but also for 50 and 70\,km\,s$^{-1}$. These lines correspond to the following equation:

\begin{equation}
	\frac{\sigma_{[\ion{N}{II}]\lambda6584}}{\sigma_{{\rm H}\alpha}} = \sqrt{1 + \left ( \frac{\Delta}{\sigma_{{\rm H}\alpha}} \right ) ^2}
\end{equation}

In order to verify that the cutoffs proposed do a good job in isolating the different excitation mechanisms, we split the different spaxels by galactocentric distance and also include information on the nature of the objects in our sample regarding the nuclear activity as published in the literature. Thus, we plot in blue those spaxels (independently of their galactocentric distance) coming from galaxies that have been identified in the literature as LINER or Seyfert (see Table~\ref{table:gradients_summary}; taken from the NASA/IPAC Extragalactic Database), while in red we identify spaxels from galaxies spectroscopically classified as H~II galaxies, starburst, or star-forming galaxies in general. Given that our observations target the central $\sim$\,0.5-3\,kpc of our galaxies, when nuclear activity is present, it should play a major role on the ionizing properties of our spectra. As such, the Seyfert/LINER classification from the literature already provides a reasonable proxy for assessing whether AGN-related processes dominate the excitation mechanisms in the corresponding spaxels. In summary, the proposed dynamical criteria adopts (from high to low velocity dispersion): (1) a minimum $\sigma_{{\rm H}\alpha}$ of 100\,km\,s$^{-1}$ for AGN-dominated (NLR or BLR) spaxels, (2) a (limiting) broadening of $\Delta$ = 30\,km\,s$^{-1}$ for spaxels with H$\alpha$ line widths below 100\,km\,s$^{-1}$ but above 30\,km\,s$^{-1}$, to separate between the spaxels ionised by shocks and UV photons, and (3) a linear offset in $\sigma_{[\ion{N}{II}]\lambda6584}$ $-$ $\sigma_{{\rm H}\alpha}$ that is proportional to $\sigma_{{\rm H}\alpha}$ in the case of the spaxels with the narrowest lines, i.e$.$ $\sigma_{{\rm H}\alpha}$ < 30\,km\,s$^{-1}$.

Applying the criteria described above, we obtain a distribution of points in the BPT diagram as shown in the right-hand panels of Figure~\ref{fig:BPT_dynamic_cuts}. We can see that the dynamic distinction we have made separates the different zones quite well from each other. As for the left-hand panels, we have represented the measurements of the lines coming from galaxies classified as Seyfert or LINER in the literature with blue dots. These points are distributed within the BPT diagram in the regions of either AGN-excited lines or in the intermediate zones between that region and the emission region of the HII regions. Regarding the presence of blue points in the middle panel, we note that the non-nuclear regions of AGN-host galaxies are expected to have line ratios (and velocity dispersion values) compatible with being ionised by massive stars. 

\subsection{Ionised-gas metallicity gradients}
\label{gas-metallicities}

The metallicity of ionised-gas can be estimated from several strong--line indicators including the $[\ion{N}{II}]\lambda6584$/H$\alpha$ ratio (also called N2 indicator; see~\citealt{Marino_2013} and references therein). These authors established a relation between the Oxygen abundance derived from electron temperature-sensitive lines (namely $[\ion{O}{III}]\lambda4363$ and $[\ion{N}{II}]\lambda5755$) and the value of $[\ion{N}{II}]\lambda6584$/H$\alpha$ in the interstellar medium of galaxies using measurements mainly from the literature but also from the CALIFA survey. Using this relation we can easily convert our N2 measurements into measurements of the Oxygen abundance in the ionised-gas and, therefore, of the overall gas metallicy. However, this relation is only valid when the excitation mechanism that has given rise to the line emission is photoionisation, since the data used for its calibration come from photoionised HII regions in very nearby galaxies, so before calculating abundances it is necessary to discriminate the origin of the lines we are studying. It is also important to note that this calibration carries an intrinsic scatter of about 0.16 dex \citep{Marino_2013}, mostly driven by galaxy-to-galaxy variations in the reference sample. Within a given galaxy, spaxels from the same observation are correlated, so this scatter acts essentially as an upper bound on the uncertainty and in practice it affects the absolute zero-point of the gradients more strongly than their relative slopes.

\begin{figure}[ht!]
	\centering
	\includegraphics[trim={0cm 0mm 0cm 0mm},clip, width=1\linewidth]{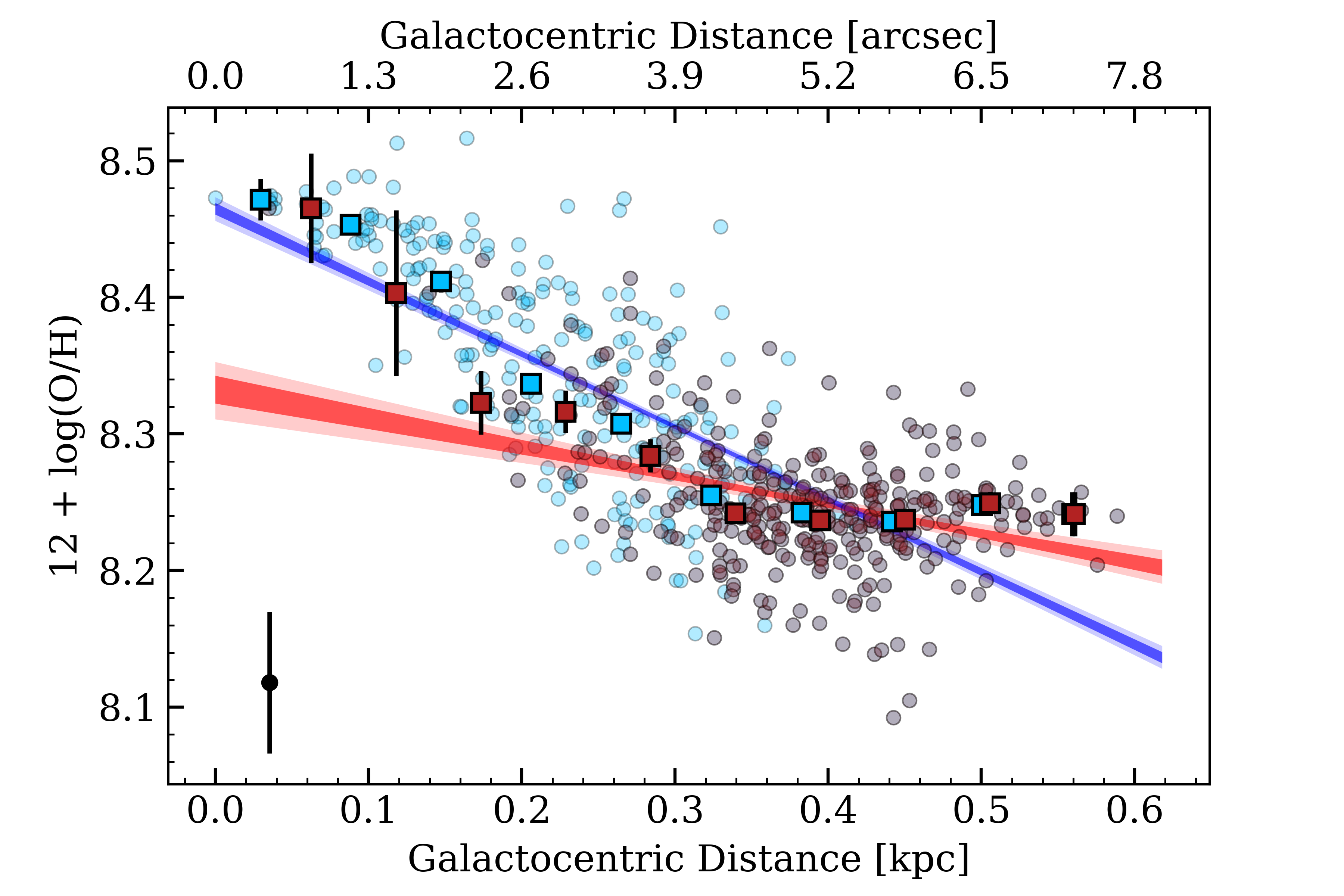}
	\caption{Oxygen abundance (i.e$.$ $\rm 12+log(O/H)$)  estimations based in the calibration by~\cite{Marino_2013} as a function of galactocentric distance for the galaxy NGC~3982. Gradients have been derived using the median values estimated for different galactocentric distance intervals. Within each galactocentric distance interval, the median values of the data are represented by blue (red) squares. The blue dots represent metallicity measurements obtained from all the spaxels while the red dots indicate those spaxels where the emission lines are originated in star-forming regions (see text). The blue and red bands indicate the results of our Bayesian linear best fits to the data with the same colour coding as for the individual measurements. The dark and light shaded bands correspond to the confidence intervals at 1-$\sigma$ and 2-$\sigma$ levels, respectively. The black dot at the bottom left of the plot represents the median error of all (blue) points.}
	\label{fig:gas_metallicity_gradient}
\end{figure}

Based on the studies we have carried out in the previous section (see Figure~\ref{fig:BPT_dynamic_cuts}), we are in the situation of being able to identify the main mechanism that has produced the line emission of the spaxels in our galaxies and therefore derive proper metal abundance in the ionised-gas phase. We can also compare the results obtained when measuring metallicities using all the $[\ion{N}{II}]\lambda6584$ line ratios available in the sample  with those estimated  using spaxels whose line emission is produced by photoionisation exclusively. During this section we will focus on the radial variation of the metal abundance, including the determination of metallicity gradients, since azimuthal variations are beyond the scope of this paper. Thus, in Figure~\ref{fig:gas_metallicity_gradient} we first show the results of the best-fitting Oxygen abundance gradients obtained for the galaxy NGC~3982 from our N2 measurements calculated in the two ways, including and excluding line ratios that could be arising from spaxels ionised by mechanisms other than photoionisation. These gradients have been calculated by dividing the range of galactocentric distances covered in each observation into 10 equal intervals and calculating the median value of the points belonging to each interval. Red dots represent metallicity values measured using photoionised-only spaxels while the blue dots represent all the N2 measurements (i.e$.$ those with S/N > 10 in the peak of the $[\ion{N}{II}]\lambda6584$ line), independently of their line ratios or velocity dispersion. The red and blue bands represent the Bayesian linear fitting we have performed on each dataset (see \cite{Chamorro_Cazorla_2022} for technical details). The best fits for all the galaxies in the sample are shown in Appendix~\ref{section: gas_metallicity_gradients_plots} and the results can be found in Table~\ref{table:gradients_summary}. Hereafter, every time we discuss about photoionised-only spaxels we will be referring those spaxels selected to have their line emission due to UV photons, according to the dynamical criteria defined in Figure~\ref{fig:BPT_dynamic_cuts}. 

\begin{figure}[ht!]
	\centering
	\includegraphics[trim={0cm 0mm 0cm 0mm},clip, width=1\linewidth]{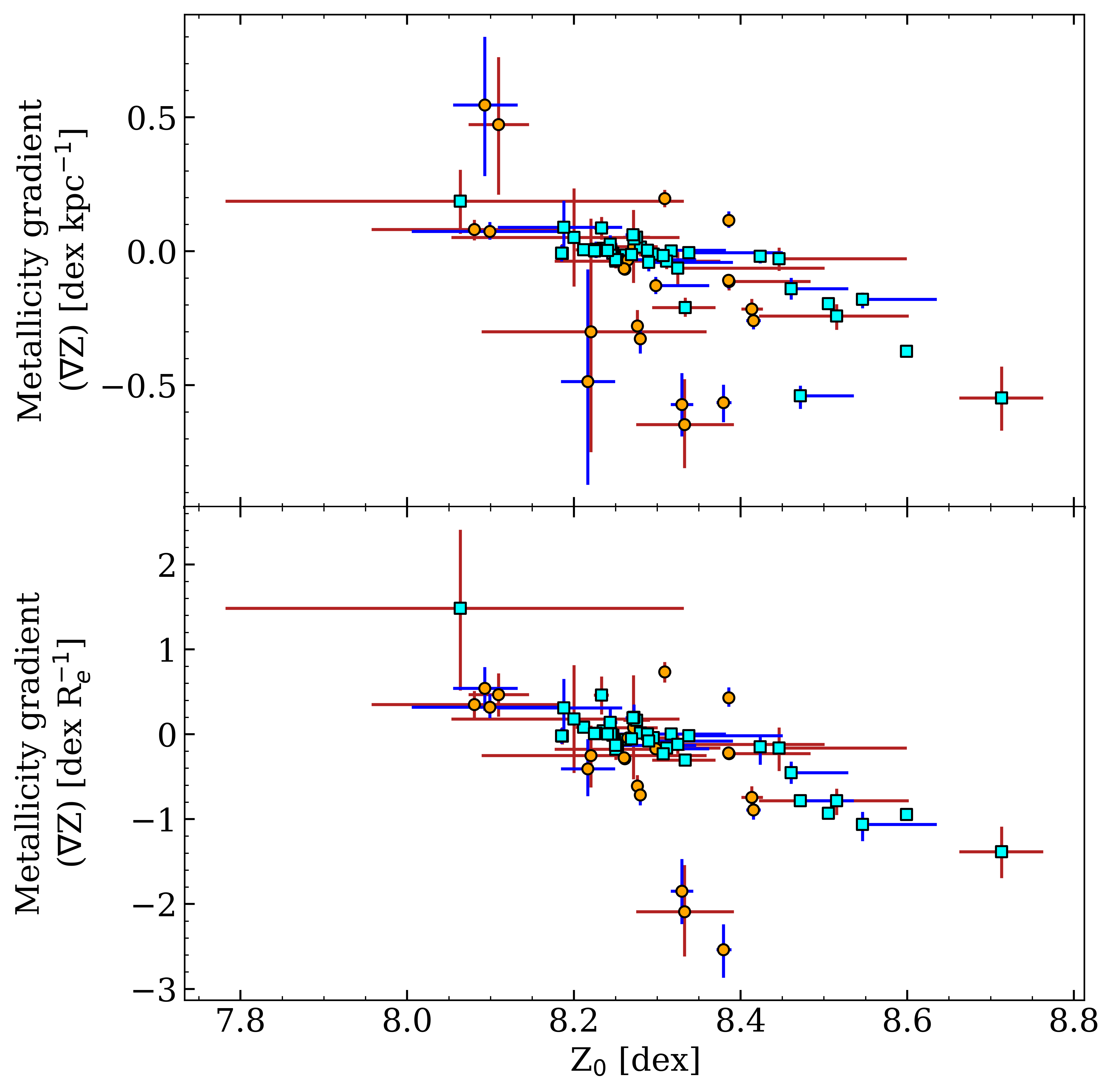}
	\caption{Metallicity gradients of the ionised gas measured for all galaxies in our sample. The orange circles indicate galaxies that were classified as slow rotators (dynamical classes 1 and 2) while the cyan squares indicate fast rotators (classes 3 through 5). The top panel presents the gradients computed as a function of galactocentric distance (in kpc), while the bottom panel shows the gradients computed as a function of the effective radius. The x-axis indicate the y-intercept of the linear fit to the metallicity gradient. Each galaxy is represented twice as we include the results of the best fits obtained using both photo-ionised spaxels (points with red error bars) and all spaxels (points with blue error bars).}
	\label{fig:gas_photo_and_gas_completo_values_by_classes}
\end{figure}

The resulting best-fitting ionised-gas metallicity gradients and central extrapolated metallicities are shown in Figure~\ref{fig:gas_photo_and_gas_completo_values_by_classes}. This figure also includes information on the kinematic class of the objects analysed. These kinematic classes have been determined based on the work by \cite{van_de_Sande_2017} and, although these results are already available, the full analyses of our application of these criteria to the MEGADES sample data will be published in a forthcoming publication (Chamorro-Cazorla et al. in prep.). To understand this kinematic classification in a simple way we can summarise that objects classified in classes 1 and 2 can be considered as slow rotators while objects between classes 3 and 5 are considered fast rotators. Thus, orange circles represent galaxies classified as slow rotators (classes 1 and 2) while the cyan squares represent fast rotators (classes 3 through 5). Note that in this figure every galaxy appears twice since we plot together the results of the fits obtained using both photoionised-only (points with red error bars) and all the spaxels (points with blue error bars).

Figure~\ref{fig:gas_photo_and_gas_completo_values_by_classes} shows that the central abundances of the galaxies in the sample vary between 8 and 8.7\,dex in the 12+log(O/H) scale (for comparison, the Oxygen abundance in the solar photosphere is 8.69$\pm$0.05;~\citealt{Asplund_2009}). On the other hand, we see that most of the galaxies show very low or almost null metallicity gradients, especially if we look at the fast rotators with central metallicities below 8.37\,dex, with a median value for these gradients of 0.005\,dex\,R$_{\rm e}^{-1}$ and a dispersion of 0.422\,dex\,R$_{\rm e}^{-1}$. Above that value, fast rotators show slightly negative metallicity gradients with some degree of correlation between slope and y-intercept with a median value for the slope of $-$0.681\,dex\,R$_{\rm e}^{-1}$ and a dispersion of 0.933\,dex\,R$_{\rm e}^{-1}$. The median value estimated for all the galaxies included in MEGADES is $-$0.025\,dex\,R$_{\rm e}^{-1}$ with a dispersion of 0.766\,dex\,R$_{\rm e}^{-1}$. To place our results in context, previous work measuring the oxygen abundance in nearby galaxies shows results like those of \cite{Sanchez-Menguiano_2016}, with median values of the slopes measured in galaxies belonging to the CALIFA sample of $-$0.07\,dex\,R$_{\rm e}^{-1}$ $\pm$ 0.05\,dex\,R$_{\rm e}^{-1}$ using the same calibration proposed by \cite{Marino_2013} used in this work. It is worth noting here that the data included in the CALIFA sample have observations covering a field of view of 1\,arcmin$^{2}$ while ours stay in much more inner regions of galaxies, as well as having observations of 939 galaxies. We note here that although we show only the marginalised errors in both quantities, these two quantities are correlated for each individual fit in the same direction \citep{Chamorro_Cazorla_2022}. However, the wide range in properties (0.7\,dex in the case of the central metallicity), much larger than the individual errors, indicate that the correlation found is not only driven by the covariance between the two parameters of each individual fit. Besides, this relatively tight correlation is only found in fast rotators, with slow rotators showing a much larger dispersion in metallicity gradient for a given central metallicity. For the majority of slow rotators the gradients are more pronounced (more negative) than those of the fast-rotators of the same central metallicity, although their most notorious difference is their significantly larger dispersion across this plot. 

From this figure, we also find that the error bars for the fits  obtained from photoionised-only spaxels are larger than for those obtained using all data, mainly because the former have fewer points to fit and, in many cases, their range of distances for the calculation is significantly reduced. Note that, in general, the radial gradients derived correspond to regions that are at galactocentric distances below 2\,kpc (see the individual fits in  Appendix~\ref{section: gas_metallicity_gradients_plots}). 

\begin{figure}[ht!]
	\centering
	\includegraphics[trim={0cm 0cm 0cm 0mm},clip, width=1\linewidth]{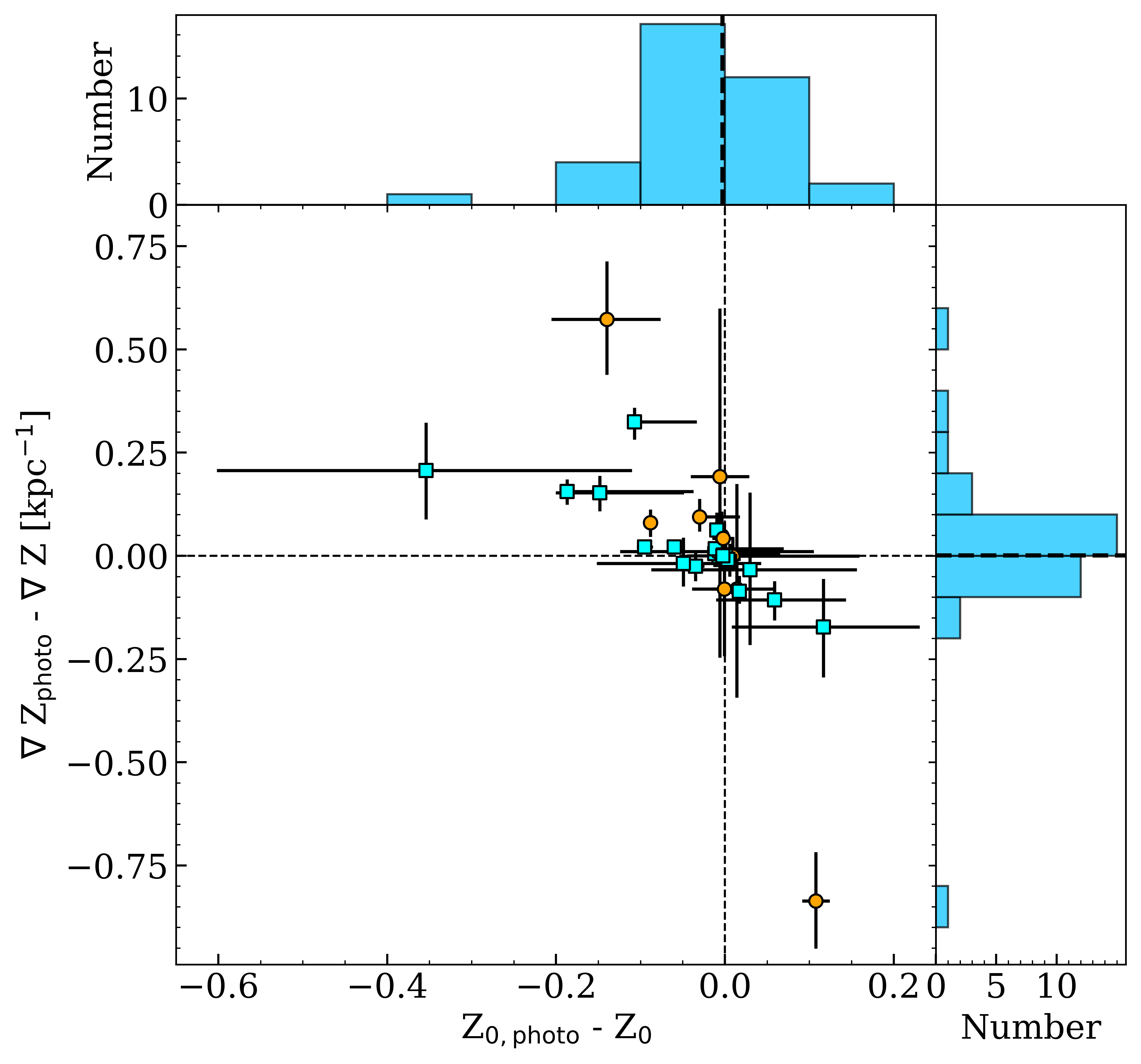}
	\caption{Comparison between ionised-gas metallicity gradients computed when photoionised-only spaxels and when all spaxels are used. The difference between y-intercepts is shown on the x-axis while the difference between slopes is included on the y-axis. Orange circles represent the results of the fits to our slow rotators while the cyan squares are for our fast rotators. The dashed lines in the central plot mark the zero position on each axis. The thick dashed lines in the frequency histograms mark the median value of the data on each axis.}
	\label{fig:gas_photo_and_gas_completo_diff_by_classes}
\end{figure}

In order to evaluate the difference in the best-fitting parameters obtained with the different datasets (photoionised-only and all spaxels) Figure~\ref{fig:gas_photo_and_gas_completo_diff_by_classes} shows the comparison between them colour coded by their kinematical state. Again, as for Figure~\ref{fig:gas_photo_and_gas_completo_values_by_classes}, slow rotators are shown as orange circles and fast rotators are shown as cyan rectangles. The corresponding frequency distributions for the entire sample are shown along with their medians (marked as dashed lines). We find a long tail of fast rotators towards more positive metallicity gradients and lower central metallicities when computed using photoionised-only spaxels. This tail can create the impression of a correlation between slope and intercept for the fast rotators (cyan squares in Figure~\ref{fig:gas_photo_and_gas_completo_diff_by_classes}). This trend could be mostly driven by the natural degeneracy between the two parameters, since steeper negative slopes are typically associated with higher intercepts, which makes the points distribute along this axis. The effect is accentuated by two outliers with particularly large uncertainties, and once these are excluded the correlation becomes much less evident. Whereas in Figure~\ref{fig:gas_photo_and_gas_completo_values_by_classes}, where the broader dynamic range of the data dominates over such degeneracies, the restricted range in Figure~\ref{fig:gas_photo_and_gas_completo_diff_by_classes} enhances this impression. Confirming whether a genuine correlation exists would require additional data to improve the statistics.

At the same time, when considering the full distribution, the median values of these differences remain very close to zero ($-$0.003 in the case of abundance and $-$0.001 in the case of gradient). Thus, the median values of these differences are practically null in both cases.

\begin{table*}[h!]
	\caption{Results for the ionised-gas metallicity gradients for the MEGADES sample, both as a function of the effective radius and galactocentric distance.}              
	\label{table:gradients_summary}      
	\centering                                      
	\resizebox{0.83\textwidth}{!}{
		\begin{tabular}{l l l l l l l l l c}   
			\hline\hline                     
			\noalign{\smallskip}
			Name & \multicolumn{2}{c}{[M/H]$_{\mathrm{gas, \hspace{0.7mm} photo}}$} & \multicolumn{2}{c}{[M/H]$_{\mathrm{gas, \hspace{0.7mm} all}}$} &  \multicolumn{2}{c}{[M/H]$_{\mathrm{gas, \hspace{0.7mm} photo}}$} & \multicolumn{2}{c}{[M/H]$_{\mathrm{gas, \hspace{0.7mm} all}}$} & R$_{\mathrm{e}}$ \\
			& slope  & y-intercept & slope & y-intercept & slope & y-intercept & slope & y-intercept & \\    
			& [dex\,R$_{\rm e}^{-1}$] & [dex] & [dex\,R$_{\rm e}^{-1}$] & [dex] & [dex\,kpc$^{-1}$] & [dex] & [dex\,kpc$^{-1}$] & [dex] & [arcsec]\\
			\hline                                   
			\noalign{\medskip}
            IC1683 & $-$0.062$^{+0.022}_{-0.021}$ & $-$1.876$^{+0.025}_{-0.025}$ & $-$0.055$^{+0.020}_{-0.021}$ & $-$1.846$^{+0.022}_{-0.022}$ & $-$0.014$^{+0.005}_{-0.005}$ & $-$0.428$^{+0.006}_{-0.006}$ & $-$0.013$^{+0.005}_{-0.005}$ & $-$0.421$^{+0.005}_{-0.005}$ & 13.25 \\
            \noalign{\medskip}
            NGC0023 & \phantom{$-$}0.003$^{+0.098}_{-0.000}$ & $-$1.145$^{+0.017}_{-0.017}$ & $-$0.016$^{+0.078}_{-0.000}$ & $-$1.079$^{+0.000}_{-0.345}$ & \phantom{$-$}0.001$^{+0.032}_{-0.000}$ & $-$0.373$^{+0.005}_{-0.006}$ & $-$0.005$^{+0.025}_{-0.000}$ & $-$0.352$^{+0.000}_{-0.113}$ & 9.89 \\
            \noalign{\medskip}
            NGC0600 & $-$0.287$^{+0.066}_{-0.069}$ & $-$1.845$^{+0.033}_{-0.032}$ & $-$0.277$^{+0.068}_{-0.065}$ & $-$1.848$^{+0.031}_{-0.032}$ & $-$0.067$^{+0.015}_{-0.016}$ & $-$0.429$^{+0.008}_{-0.007}$ & $-$0.064$^{+0.016}_{-0.015}$ & $-$0.430$^{+0.007}_{-0.007}$ & 33.86 \\
            \noalign{\medskip}
            NGC0716 & \phantom{$-$}0.041$^{+0.009}_{-0.009}$ & $-$1.748$^{+0.021}_{-0.022}$ & $-$0.040$^{+0.006}_{-0.007}$ & $-$1.518$^{+0.014}_{-0.014}$ & \phantom{$-$}0.011$^{+0.002}_{-0.002}$ & $-$0.455$^{+0.006}_{-0.006}$ & $-$0.010$^{+0.002}_{-0.002}$ & $-$0.395$^{+0.004}_{-0.004}$ & 12.40 \\
            \noalign{\medskip}
            NGC0718 & \multicolumn{1}{c}{$-$} & \multicolumn{1}{c}{$-$} & \multicolumn{1}{c}{$-$} & \multicolumn{1}{c}{$-$} & \multicolumn{1}{c}{$-$} & \multicolumn{1}{c}{$-$} & \multicolumn{1}{c}{$-$} & \multicolumn{1}{c}{$-$} & 21.34 \\
            \noalign{\medskip}
            NGC1042 & \phantom{$-$}0.076$^{+0.605}_{-0.619}$ & $-$1.881$^{+0.123}_{-0.130}$ & $-$2.536$^{+0.331}_{-0.298}$ & $-$1.395$^{+0.039}_{-0.043}$ & \phantom{$-$}0.017$^{+0.135}_{-0.138}$ & $-$0.418$^{+0.027}_{-0.029}$ & $-$0.564$^{+0.074}_{-0.066}$ & $-$0.310$^{+0.009}_{-0.010}$ & 47.81 \\
            \noalign{\medskip}
            NGC1087 & $-$0.130$^{+0.106}_{-0.092}$ & $-$1.813$^{+0.015}_{-0.015}$ & $-$0.131$^{+0.108}_{-0.092}$ & $-$1.823$^{+0.000}_{-0.400}$ & $-$0.031$^{+0.026}_{-0.022}$ & $-$0.438$^{+0.004}_{-0.004}$ & $-$0.032$^{+0.026}_{-0.022}$ & $-$0.440$^{+0.000}_{-0.097}$ & 39.45 \\
            \noalign{\medskip}
            NGC2500 & \phantom{$-$}0.468$^{+0.259}_{-0.249}$ & $-$0.575$^{+0.035}_{-0.036}$ & \phantom{$-$}0.541$^{+0.263}_{-0.251}$ & $-$0.591$^{+0.038}_{-0.039}$ & \phantom{$-$}0.473$^{+0.262}_{-0.252}$ & $-$0.581$^{+0.036}_{-0.037}$ & \phantom{$-$}0.546$^{+0.265}_{-0.254}$ & $-$0.597$^{+0.038}_{-0.040}$ & 28.29 \\
            \noalign{\medskip}
            NGC2537 & $-$0.251$^{+0.376}_{-0.353}$ & $-$0.393$^{+0.110}_{-0.116}$ & $-$0.406$^{+0.322}_{-0.350}$ & $-$0.396$^{+0.027}_{-0.028}$ & $-$0.300$^{+0.449}_{-0.423}$ & $-$0.470$^{+0.131}_{-0.139}$ & $-$0.486$^{+0.386}_{-0.418}$ & $-$0.473$^{+0.032}_{-0.033}$ & 26.97 \\
            \noalign{\medskip}
            NGC2543 & $-$0.161$^{+0.129}_{-0.157}$ & $-$1.642$^{+0.000}_{-0.279}$ & \phantom{$-$}0.199$^{+0.157}_{-0.151}$ & $-$1.812$^{+0.029}_{-0.029}$ & $-$0.037$^{+0.030}_{-0.036}$ & $-$0.379$^{+0.000}_{-0.064}$ & \phantom{$-$}0.046$^{+0.036}_{-0.035}$ & $-$0.418$^{+0.007}_{-0.007}$ & 25.66 \\
            \noalign{\medskip}
            NGC2552 & \multicolumn{1}{c}{$-$} & \multicolumn{1}{c}{$-$} & \multicolumn{1}{c}{$-$} & \multicolumn{1}{c}{$-$} & \multicolumn{1}{c}{$-$} & \multicolumn{1}{c}{$-$} & \multicolumn{1}{c}{$-$} & \multicolumn{1}{c}{$-$} & 40.81 \\
            \noalign{\medskip}
            NGC2967 & \phantom{$-$}0.350$^{+0.173}_{-0.158}$ & $-$2.638$^{+0.533}_{-0.493}$ & \phantom{$-$}0.320$^{+0.130}_{-0.152}$ & $-$2.557$^{+0.406}_{-0.391}$ & \phantom{$-$}0.081$^{+0.040}_{-0.037}$ & $-$0.610$^{+0.123}_{-0.114}$ & \phantom{$-$}0.074$^{+0.030}_{-0.035}$ & $-$0.591$^{+0.094}_{-0.090}$ & 25.61 \\
            \noalign{\medskip}
            NGC3104 & \multicolumn{1}{c}{$-$} & \multicolumn{1}{c}{$-$} & \multicolumn{1}{c}{$-$} & \multicolumn{1}{c}{$-$} & \multicolumn{1}{c}{$-$} & \multicolumn{1}{c}{$-$} & \multicolumn{1}{c}{$-$} & \multicolumn{1}{c}{$-$} & $-$ \\
            \noalign{\medskip}
            NGC3485 & $-$2.091$^{+0.526}_{-0.549}$ & $-$1.155$^{+0.188}_{-0.191}$ & $-$1.849$^{+0.385}_{-0.379}$ & $-$1.166$^{+0.044}_{-0.044}$ & $-$0.647$^{+0.163}_{-0.170}$ & $-$0.357$^{+0.058}_{-0.059}$ & $-$0.572$^{+0.119}_{-0.117}$ & $-$0.360$^{+0.013}_{-0.014}$ & 32.67 \\
            \noalign{\medskip}
            NGC3507$^{\dag}$ & $-$3.278$^{+0.272}_{-0.285}$ & $-$0.057$^{+0.039}_{-0.039}$ & $-$1.293$^{+0.134}_{-0.129}$ & $-$0.313$^{+0.018}_{-0.019}$ & $-$1.379$^{+0.114}_{-0.120}$ & $-$0.024$^{+0.016}_{-0.016}$ & $-$0.544$^{+0.056}_{-0.054}$ & $-$0.132$^{+0.008}_{-0.008}$ & 34.97 \\
            \noalign{\medskip}
            NGC3780 & \multicolumn{1}{c}{$-$} & \multicolumn{1}{c}{$-$} & \multicolumn{1}{c}{$-$} & \multicolumn{1}{c}{$-$} & \multicolumn{1}{c}{$-$} & \multicolumn{1}{c}{$-$} & \multicolumn{1}{c}{$-$} & \multicolumn{1}{c}{$-$} & 36.24 \\
            \noalign{\medskip}
            NGC3893 & $-$0.010$^{+0.058}_{-0.057}$ & $-$0.643$^{+0.000}_{-0.077}$ & \phantom{$-$}0.011$^{+0.048}_{-0.042}$ & $-$0.672$^{+0.005}_{-0.005}$ & $-$0.007$^{+0.040}_{-0.040}$ & $-$0.443$^{+0.000}_{-0.053}$ & \phantom{$-$}0.008$^{+0.033}_{-0.029}$ & $-$0.463$^{+0.004}_{-0.003}$ & 21.65 \\
            \noalign{\medskip}
            NGC3982$^{\dag}$ & $-$0.305$^{+0.050}_{-0.054}$ & $-$0.517$^{+0.058}_{-0.052}$ & $-$0.782$^{+0.070}_{-0.055}$ & $-$0.317$^{+0.000}_{-0.093}$ & $-$0.210$^{+0.034}_{-0.037}$ & $-$0.356$^{+0.040}_{-0.036}$ & $-$0.540$^{+0.049}_{-0.038}$ & $-$0.218$^{+0.000}_{-0.064}$ & 18.83 \\
            \noalign{\medskip}
            NGC3998$^{\dag}$ & \multicolumn{1}{c}{$-$} & \multicolumn{1}{c}{$-$} & \multicolumn{1}{c}{$-$} & \multicolumn{1}{c}{$-$} & \multicolumn{1}{c}{$-$} & \multicolumn{1}{c}{$-$} & \multicolumn{1}{c}{$-$} & \multicolumn{1}{c}{$-$} & 15.17 \\
            \noalign{\medskip}
            NGC4037 & $-$0.610$^{+0.133}_{-0.129}$ & $-$0.907$^{+0.016}_{-0.017}$ & $-$0.716$^{+0.121}_{-0.123}$ & $-$0.899$^{+0.016}_{-0.016}$ & $-$0.278$^{+0.061}_{-0.059}$ & $-$0.414$^{+0.007}_{-0.008}$ & $-$0.327$^{+0.055}_{-0.056}$ & $-$0.410$^{+0.007}_{-0.007}$ & 34.23 \\
            \noalign{\medskip}
            NGC4041 & $-$0.043$^{+0.045}_{-0.059}$ & $-$0.568$^{+0.000}_{-0.073}$ & $-$0.171$^{+0.043}_{-0.043}$ & $-$0.524$^{+0.000}_{-0.086}$ & $-$0.032$^{+0.034}_{-0.044}$ & $-$0.425$^{+0.000}_{-0.054}$ & $-$0.128$^{+0.032}_{-0.032}$ & $-$0.392$^{+0.000}_{-0.064}$ & 15.91 \\
            \noalign{\medskip}
            NGC4189 & \phantom{$-$}0.462$^{+0.230}_{-0.219}$ & $-$2.432$^{+0.049}_{-0.045}$ & \phantom{$-$}0.140$^{+0.179}_{-0.180}$ & $-$2.377$^{+0.044}_{-0.045}$ & \phantom{$-$}0.087$^{+0.043}_{-0.041}$ & $-$0.457$^{+0.009}_{-0.009}$ & \phantom{$-$}0.026$^{+0.034}_{-0.034}$ & $-$0.446$^{+0.008}_{-0.008}$ & 36.98 \\
            \noalign{\medskip}
            NGC4278$^{\dag}$ & \multicolumn{1}{c}{$-$} & \multicolumn{1}{c}{$-$} & \multicolumn{1}{c}{$-$} & \multicolumn{1}{c}{$-$} & \multicolumn{1}{c}{$-$} & \multicolumn{1}{c}{$-$} & \multicolumn{1}{c}{$-$} & \multicolumn{1}{c}{$-$} & 17.55 \\
            \noalign{\medskip}
            NGC4593$^{\dag}$ & $-$0.784$^{+0.166}_{-0.144}$ & $-$0.566$^{+0.302}_{-0.280}$ & $-$0.453$^{+0.131}_{-0.130}$ & $-$0.743$^{+0.000}_{-0.221}$ & $-$0.242$^{+0.051}_{-0.044}$ & $-$0.175$^{+0.093}_{-0.086}$ & $-$0.140$^{+0.040}_{-0.040}$ & $-$0.229$^{+0.000}_{-0.068}$ & 18.96 \\
            \noalign{\medskip}
            NGC4750$^{\dag}$ & $-$1.384$^{+0.309}_{-0.297}$ & \phantom{$-$}0.059$^{+0.129}_{-0.127}$ & $-$0.945$^{+0.047}_{-0.051}$ & $-$0.230$^{+0.014}_{-0.014}$ & $-$0.547$^{+0.122}_{-0.118}$ & \phantom{$-$}0.023$^{+0.051}_{-0.050}$ & $-$0.374$^{+0.019}_{-0.020}$ & $-$0.091$^{+0.006}_{-0.006}$ & 22.78 \\
            \noalign{\medskip}
            NGC5218$^{\dag}$ & $-$0.585$^{+0.127}_{-0.133}$ & $-$0.757$^{+0.000}_{-0.307}$ & $-$0.492$^{+0.111}_{-0.087}$ & $-$0.744$^{+0.017}_{-0.018}$ & $-$0.164$^{+0.035}_{-0.037}$ & $-$0.211$^{+0.000}_{-0.086}$ & $-$0.137$^{+0.031}_{-0.024}$ & $-$0.208$^{+0.005}_{-0.005}$ & 18.36 \\
            \noalign{\medskip}
            NGC5394 & $-$0.229$^{+0.068}_{-0.000}$ & $-$0.620$^{+0.000}_{-0.199}$ & $-$0.220$^{+0.063}_{-0.000}$ & $-$0.621$^{+0.008}_{-0.008}$ & $-$0.112$^{+0.033}_{-0.000}$ & $-$0.304$^{+0.000}_{-0.098}$ & $-$0.108$^{+0.031}_{-0.000}$ & $-$0.305$^{+0.004}_{-0.004}$ & 8.68 \\
            \noalign{\medskip}
            NGC5616 & \phantom{$-$}0.084$^{+0.033}_{-0.033}$ & $-$6.926$^{+0.139}_{-0.142}$ & $-$0.230$^{+0.022}_{-0.021}$ & $-$5.544$^{+0.079}_{-0.080}$ & \phantom{$-$}0.006$^{+0.002}_{-0.002}$ & $-$0.478$^{+0.010}_{-0.010}$ & $-$0.016$^{+0.002}_{-0.001}$ & $-$0.383$^{+0.005}_{-0.006}$ & 25.67 \\
            \noalign{\medskip}
            NGC5953$^{\dag}$ & \phantom{$-$}0.014$^{+0.030}_{-0.000}$ & $-$0.363$^{+0.003}_{-0.003}$ & \phantom{$-$}0.003$^{+0.028}_{-0.000}$ & $-$0.356$^{+0.000}_{-0.083}$ & \phantom{$-$}0.015$^{+0.034}_{-0.000}$ & $-$0.410$^{+0.003}_{-0.004}$ & \phantom{$-$}0.004$^{+0.032}_{-0.000}$ & $-$0.402$^{+0.000}_{-0.094}$ & 6.56 \\
            \noalign{\medskip}
            NGC5957$^{\dag}$ & $-$0.743$^{+0.135}_{-0.132}$ & $-$0.953$^{+0.043}_{-0.046}$ & $-$0.891$^{+0.114}_{-0.115}$ & $-$0.946$^{+0.030}_{-0.030}$ & $-$0.216$^{+0.039}_{-0.038}$ & $-$0.277$^{+0.013}_{-0.013}$ & $-$0.258$^{+0.033}_{-0.033}$ & $-$0.275$^{+0.009}_{-0.009}$ & 27.57 \\
            \noalign{\medskip}
            NGC5963 & \phantom{$-$}0.002$^{+0.021}_{-0.023}$ & $-$0.285$^{+0.000}_{-0.034}$ & \phantom{$-$}0.002$^{+0.021}_{-0.022}$ & $-$0.287$^{+0.000}_{-0.049}$ & \phantom{$-$}0.003$^{+0.032}_{-0.036}$ & $-$0.445$^{+0.000}_{-0.054}$ & \phantom{$-$}0.003$^{+0.033}_{-0.034}$ & $-$0.449$^{+0.000}_{-0.077}$ & 14.20 \\
            \noalign{\medskip}
            NGC6027 & $-$0.119$^{+0.112}_{-0.115}$ & $-$0.692$^{+0.000}_{-0.333}$ & $-$0.078$^{+0.064}_{-0.061}$ & $-$0.757$^{+0.000}_{-0.191}$ & $-$0.063$^{+0.059}_{-0.061}$ & $-$0.365$^{+0.000}_{-0.176}$ & $-$0.041$^{+0.034}_{-0.032}$ & $-$0.400$^{+0.000}_{-0.101}$ & 6.33 \\
            \noalign{\medskip}
            NGC6140 & \phantom{$-$}0.066$^{+0.036}_{-0.036}$ & $-$0.832$^{+0.008}_{-0.008}$ & \phantom{$-$}0.062$^{+0.035}_{-0.034}$ & $-$0.829$^{+0.008}_{-0.008}$ & \phantom{$-$}0.041$^{+0.023}_{-0.022}$ & $-$0.517$^{+0.005}_{-0.005}$ & \phantom{$-$}0.039$^{+0.022}_{-0.021}$ & $-$0.515$^{+0.005}_{-0.005}$ & 25.55 \\
            \noalign{\medskip}
            NGC6217$^{\dag}$ & \phantom{$-$}0.733$^{+0.123}_{-0.119}$ & $-$1.417$^{+0.028}_{-0.028}$ & \phantom{$-$}0.431$^{+0.106}_{-0.123}$ & $-$1.131$^{+0.015}_{-0.016}$ & \phantom{$-$}0.197$^{+0.033}_{-0.032}$ & $-$0.381$^{+0.008}_{-0.008}$ & \phantom{$-$}0.116$^{+0.028}_{-0.033}$ & $-$0.304$^{+0.004}_{-0.004}$ & 39.56 \\
            \noalign{\medskip}
            NGC6339 & \phantom{$-$}0.006$^{+0.027}_{-0.026}$ & $-$2.046$^{+0.020}_{-0.020}$ & \phantom{$-$}0.007$^{+0.026}_{-0.026}$ & $-$2.048$^{+0.020}_{-0.020}$ & \phantom{$-$}0.001$^{+0.006}_{-0.006}$ & $-$0.465$^{+0.004}_{-0.005}$ & \phantom{$-$}0.002$^{+0.006}_{-0.006}$ & $-$0.465$^{+0.004}_{-0.005}$ & 30.35 \\
            \noalign{\medskip}
            NGC6412 & $-$0.023$^{+0.097}_{-0.105}$ & $-$1.509$^{+0.020}_{-0.021}$ & $-$0.021$^{+0.091}_{-0.093}$ & $-$1.512$^{+0.020}_{-0.020}$ & $-$0.008$^{+0.032}_{-0.035}$ & $-$0.504$^{+0.007}_{-0.007}$ & $-$0.007$^{+0.030}_{-0.031}$ & $-$0.505$^{+0.007}_{-0.007}$ & 33.28 \\
            \noalign{\medskip}
            NGC7025 & \multicolumn{1}{c}{$-$} & \multicolumn{1}{c}{$-$} & \multicolumn{1}{c}{$-$} & \multicolumn{1}{c}{$-$} & \multicolumn{1}{c}{$-$} & \multicolumn{1}{c}{$-$} & \multicolumn{1}{c}{$-$} & \multicolumn{1}{c}{$-$} & 17.77 \\
            \noalign{\medskip}
            NGC7437 & \phantom{$-$}0.178$^{+0.633}_{-0.636}$ & $-$1.696$^{+0.510}_{-0.438}$ & \phantom{$-$}0.309$^{+0.317}_{-0.344}$ & $-$1.739$^{+0.274}_{-0.243}$ & \phantom{$-$}0.051$^{+0.183}_{-0.184}$ & $-$0.490$^{+0.147}_{-0.126}$ & \phantom{$-$}0.089$^{+0.092}_{-0.099}$ & $-$0.502$^{+0.079}_{-0.070}$ & 23.72 \\
            \noalign{\medskip}
            NGC7479$^{\dag}$ & $-$0.163$^{+0.269}_{-0.243}$ & $-$1.442$^{+0.000}_{-0.907}$ & $-$1.062$^{+0.197}_{-0.148}$ & $-$0.850$^{+0.000}_{-0.528}$ & $-$0.028$^{+0.046}_{-0.041}$ & $-$0.244$^{+0.000}_{-0.153}$ & $-$0.180$^{+0.033}_{-0.025}$ & $-$0.144$^{+0.000}_{-0.089}$ & 36.28 \\
            \noalign{\medskip}
            NGC7591$^{\dag}$ & $-$0.176$^{+0.128}_{-0.143}$ & $-$2.091$^{+0.349}_{-0.395}$ & $-$0.932$^{+0.024}_{-0.024}$ & $-$0.878$^{+0.024}_{-0.026}$ & $-$0.037$^{+0.027}_{-0.030}$ & $-$0.440$^{+0.073}_{-0.083}$ & $-$0.196$^{+0.005}_{-0.005}$ & $-$0.185$^{+0.005}_{-0.005}$ & 14.11 \\
            \noalign{\medskip}
            NGC7738 & \phantom{$-$}1.485$^{+0.969}_{-0.925}$ & $-$4.966$^{+2.236}_{-2.125}$ & $-$0.148$^{+0.210}_{-0.114}$ & $-$2.113$^{+0.048}_{-0.050}$ & \phantom{$-$}0.187$^{+0.122}_{-0.117}$ & $-$0.626$^{+0.282}_{-0.268}$ & $-$0.019$^{+0.027}_{-0.014}$ & $-$0.266$^{+0.006}_{-0.006}$ & 17.43 \\
            \noalign{\medskip}
            NGC7787 & \phantom{$-$}0.165$^{+0.061}_{-0.057}$ & $-$1.301$^{+0.050}_{-0.050}$ & \phantom{$-$}0.194$^{+0.036}_{-0.037}$ & $-$1.316$^{+0.028}_{-0.026}$ & \phantom{$-$}0.052$^{+0.019}_{-0.018}$ & $-$0.415$^{+0.016}_{-0.016}$ & \phantom{$-$}0.062$^{+0.012}_{-0.012}$ & $-$0.419$^{+0.009}_{-0.008}$ & 6.99 \\
            \noalign{\medskip}
            PGC066559 & \multicolumn{1}{c}{$-$} & \multicolumn{1}{c}{$-$} & \multicolumn{1}{c}{$-$} & \multicolumn{1}{c}{$-$} & $-$0.048$^{+0.036}_{-0.036}$ & $-$0.573$^{+0.042}_{-0.044}$ & $-$0.050$^{+0.036}_{-0.036}$ & $-$0.571$^{+0.043}_{-0.042}$ & $-$ \\
			\noalign{\medskip}
			\hline       
	\end{tabular}}
	\parbox{\textwidth}{\footnotesize \noindent  (1) galaxy name; (2) to (5) slope and y-intercept estimations corresponding to the interstellar gas metallicity gradient when only and not only (all) excited by photoionisation lines are used, computed as a function of the effective radius; (6) to (9) slope and y-intercept estimations corresponding to the interstellar gas metallicity gradient when only and not only (all) excited by photoionisation lines are used; (10) effective radius. $^{\dag}$ Galaxies classified as LINER or Seyfert in the literature.}
	
\end{table*}

\section{Discussion}
\label{section:Discussion}

In this section we analyse how the results obtained above provide some clues on the relative role of the different mechanisms involved in the evolution of galaxies and, in particular, in shaping their current day spectrophotometric, chemical and dynamical properties. 

Concerning the diagnosis of ionised gas, the different analyses performed by exploiting BPT diagrams of $[\ion{O}{III}]\lambda5007$/H$\beta$ versus $[\ion{N}{II}]\lambda6584$/H$\alpha$ and $[\ion{S}{II}]\lambda6717$+$[\ion{S}{II}]\lambda6731$ together with the study of their radial variations and their relationships with the velocity dispersion of the emission lines that populate them, have led us to attempt to differentiate the predominant excitation mechanisms in each spaxel on the basis of a purely dynamical criterion. This criterion, based on the values of the $[\ion{N}{II}]\lambda6584$ and H$\alpha$ velocity dispersion and the relation between them, has proved to be quite effective. This criterion, given the importance of properly distinguishing the dominant mechanisms responsible for the ionised gas emission, can save a lot of telescope time in cases where it is not possible to observe simultaneously the spectral range of $[\ion{O}{III}]\lambda5007$/H$\beta$ and $[\ion{N}{II}]\lambda6584$/H$\alpha$, as is the case of MEGARA.

Our diagnosis have shown that a large number of spaxels (at any galactocentric distances) have emission-line ratios that indicate the presence of shocks both based on their line ratios (typically high $[\ion{N}{II}]\lambda6584$/H$\alpha$ and intermediate $[\ion{O}{III}]\lambda5007$/H$\beta$) and widths (intermediate between photoionization by massive stars and AGN, either NLR or BLR). In this context, excitation with shocks comes naturally associated with the onset of galactic winds. Given the relatively high masses of these objects, these winds would not imply the loss of gas mass, probably not even the escape of chemically-enriched outflows to the intergalactic medium, but can certainly produce the redistribution of metals in scales of several kpc (see~\citealt{Christensen_2018}). Under this scenario, when a galactic wind is in place, some of the gas is blown into the circumgalactic environment but eventually it falls back into the plane of the galaxy, not necessarily at the same radii at which the wind was generated. In the case of a nuclear starburst that throws material into the circumgalactic medium that ends up falling into the outer parts of the disc, it enriches the outer parts and causes the metallicity gradients to flatten (dilute). Another piece of evidence for the presence of galactic winds (neutral in this case) in some of our galaxies is the finding of NaI D absorption with complex kinematics and large equivalent widths \citep{Chamorro_Cazorla_2023}. Another mechanism that can produce widespread shocked-like line ratios is the presence of large-scale shocks induced by bars (see e.g$.$ ~\citealt{Maciejewski_2002}). Note that also in cases like this, where the gravitational potential has a non-axisymmetric component, we would expect a flattening in the physical properties of the galaxies (age and metallicity) with galactocentric distance compared to those expected from the perspective of purely inside-out galaxy formation scenario, at least for a range of galaxy masses (see~\citealt{Zurita_2021} and references therein).

The availability of a dynamical criterion for selecting regions whose line emission arises from photoionization from massive stars, prevents the bias in the ionised-gas metallicity measurements that the selection purely on line ratios would impose on our results. Despite of that caveat, the impact of the spaxels dominated by other mechanisms is small, since there is little difference in the ionised-gas metallicity gradients derived with the two datasets (photoionised-only and all spaxels). Besides, given our limited range in galactocentric distance, we expect that such a potential bias would be even less of a problem in those works covering the (outer) disks, where emission due to AGN or shocks (or even from "retired" stellar populations) should be minimal. The correlation found between central abundances and gradients, especially in fast rotators, with a tendency to have higher negative gradients the higher the central abundance, supports the idea of inside-out formation and the presence of diffusion mechanisms in the sample galaxies (i.e$.$ redistribution of metals by galactic winds, gas radial transfer by bars, contribution of minor mergers). Under this scenario, a metallicity distribution with a certain gradient and central abundance would evolve through radial displacements of gas flattening the gradients and reducing the central abundances. The fact that we find a similar trend in the stellar metallicities indicates that this is not simply due to the gas having been diluted recently with externally accreted pristine gas. 

\begin{acknowledgements}

We/IPARCOS acknowledge/s the support from the “Tecnologías avanzadas para la exploración de universo y sus componentes" (PR47/21 TAU) project funded by Comunidad de Madrid, by the Recovery, Transformation and Resilience Plan from the Spanish State, and by NextGenerationEU from the European Union through the Recovery and Resilience Facility. The authors acknowledge the support of AC3, a project funded by the European Union's Horizon Europe Research and Innovation programme under grant agreement No 101093129. MCC, AGdP, ACM, JG, CCT, JZ, NC and SP acknowledge financial support from the Spanish Ministerio de Ciencia, Innovación y Universidades (MCIU) under the grant RTI2018-096188-B-I00, PID2021-123417OB-I00 and PID2022-138621NB-I00.

This research has made use of the NASA/IPAC Extragalactic Database, which is funded by the National Aeronautics and Space Administration and operated by the California Institute of Technology.

The Pan-STARRS1 Surveys (PS1) and the PS1 public science archive have been made possible through contributions by the Institute for Astronomy, the University of Hawaii, the Pan-STARRS Project Office, the Max-Planck Society and its participating institutes, the Max Planck Institute for Astronomy, Heidelberg and the Max Planck Institute for Extraterrestrial Physics, Garching, The Johns Hopkins University, Durham University, the University of Edinburgh, the Queen's University Belfast, the Harvard-Smithsonian Center for Astrophysics, the Las Cumbres Observatory Global Telescope Network Incorporated, the National Central University of Taiwan, the Space Telescope Science Institute, the National Aeronautics and Space Administration under Grant No. NNX08AR22G issued through the Planetary Science Division of the NASA Science Mission Directorate, the National Science Foundation Grant No. AST-1238877, the University of Maryland, Eotvos Lorand University (ELTE), the Los Alamos National Laboratory, and the Gordon and Betty Moore Foundation.
\end{acknowledgements}

\bibliographystyle{aa}
\bibliography{biblio}

\clearpage

\onecolumn
\begin{appendix}

\section{Gas metallicity gradients plots}
\label{section: gas_metallicity_gradients_plots}

\begin{figure}[h]
	\centering
	\includegraphics[trim={0cm 0mm 0cm 6.2mm},clip, width=0.44\linewidth]{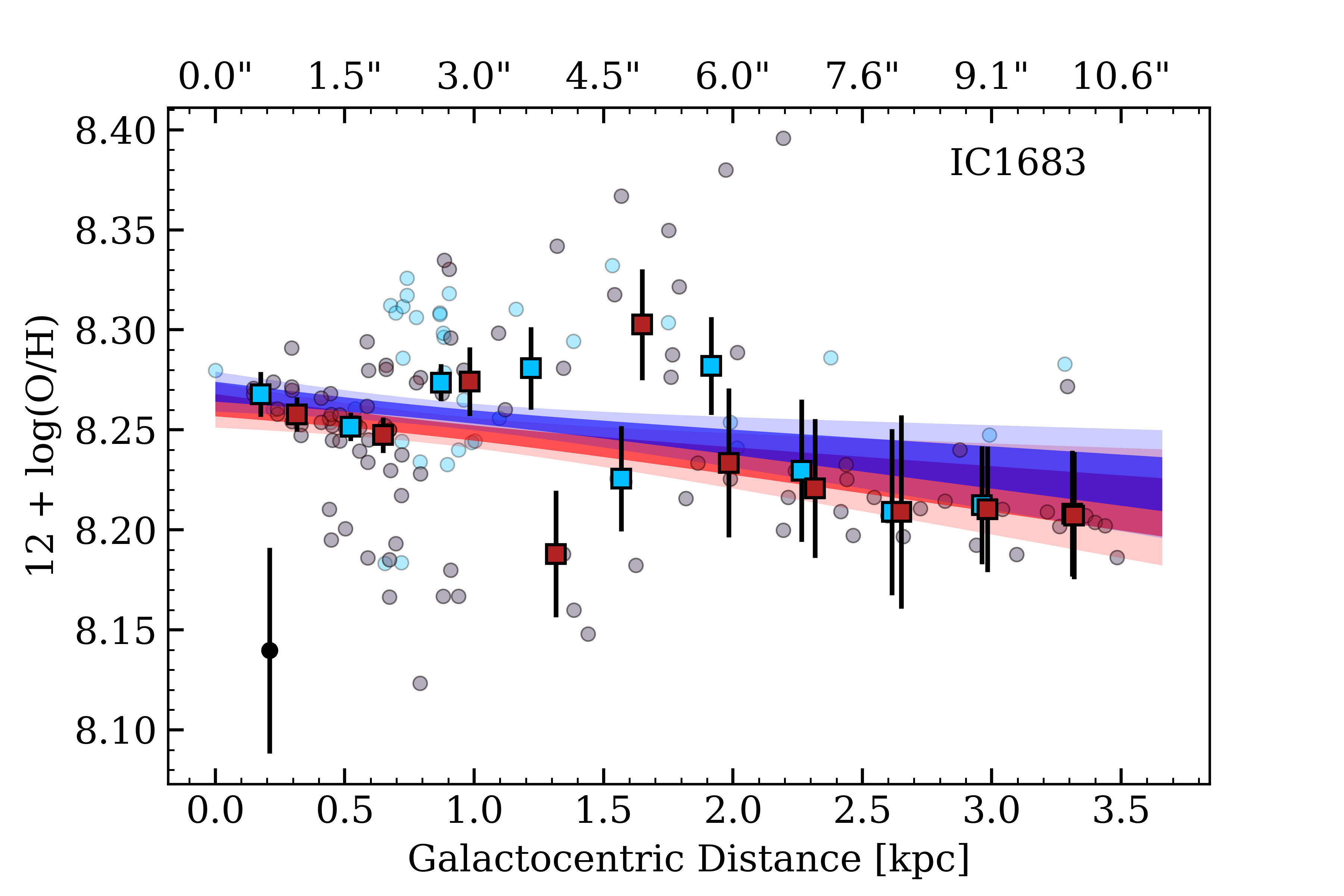}
	\includegraphics[trim={0cm 0mm 0cm 6.2mm},clip, width=0.44\linewidth]{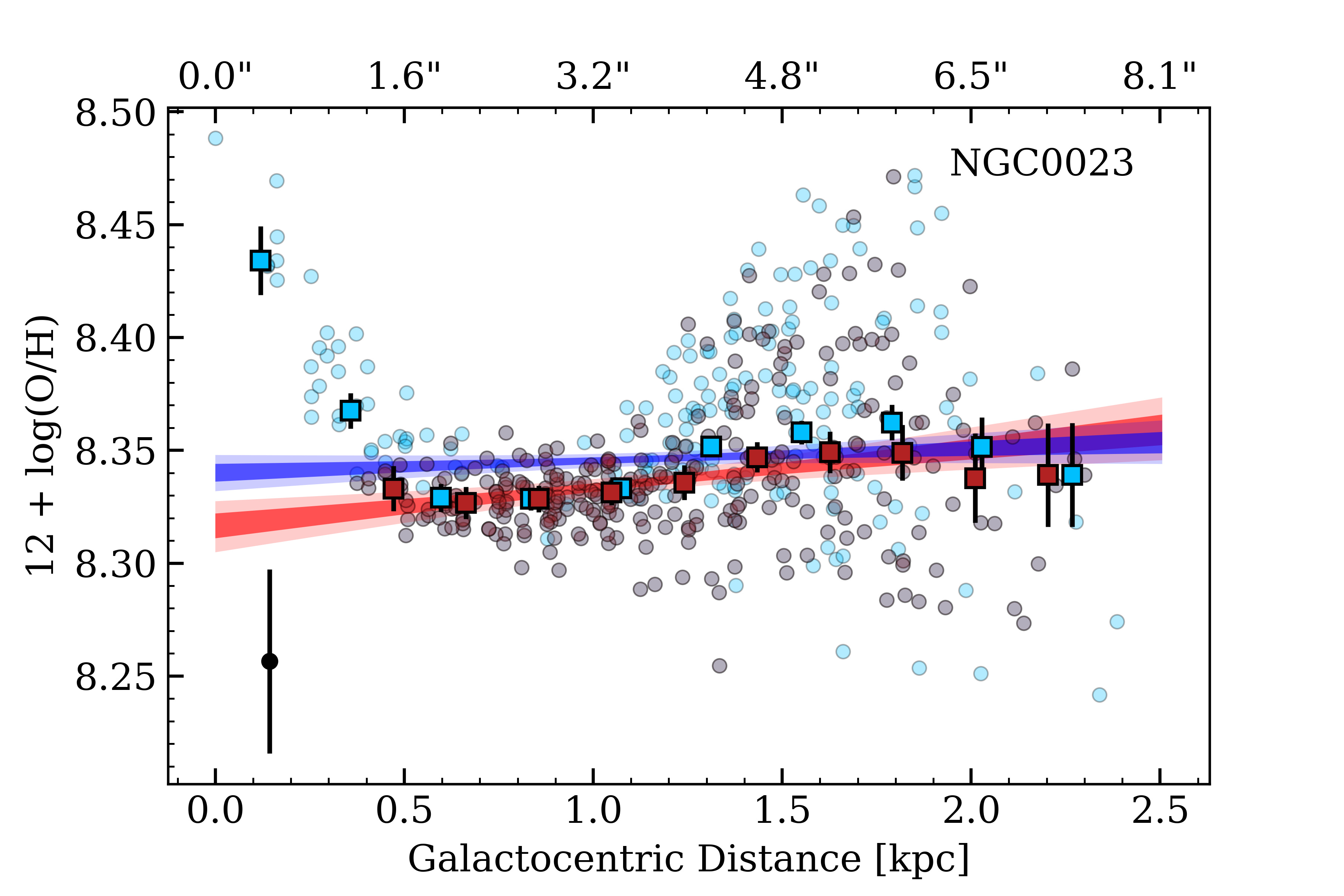}
    \includegraphics[trim={0cm 0mm 0cm 6.2mm},clip, width=0.44\linewidth]{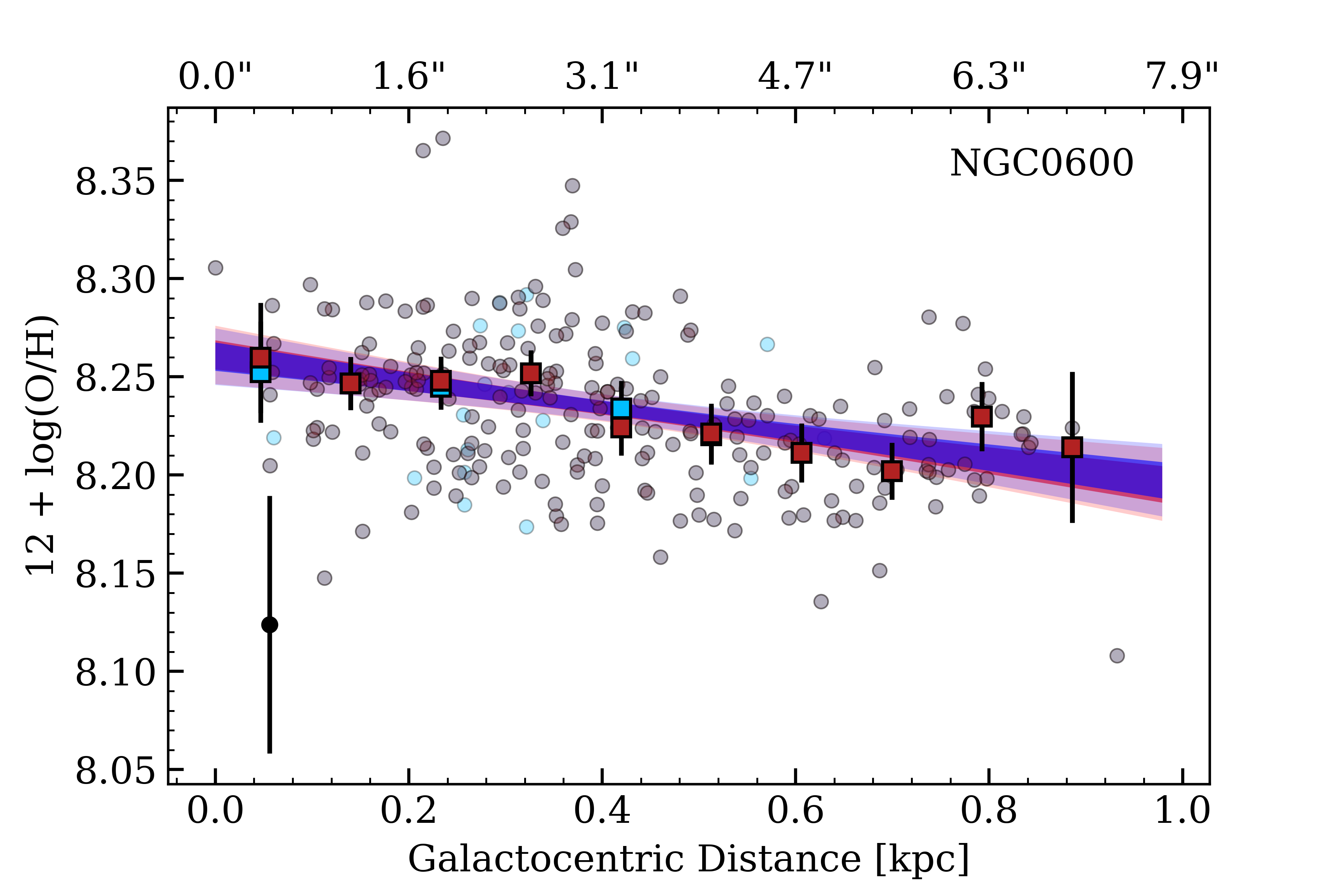}
    \includegraphics[trim={0cm 0mm 0cm 6.2mm},clip, width=0.44\linewidth]{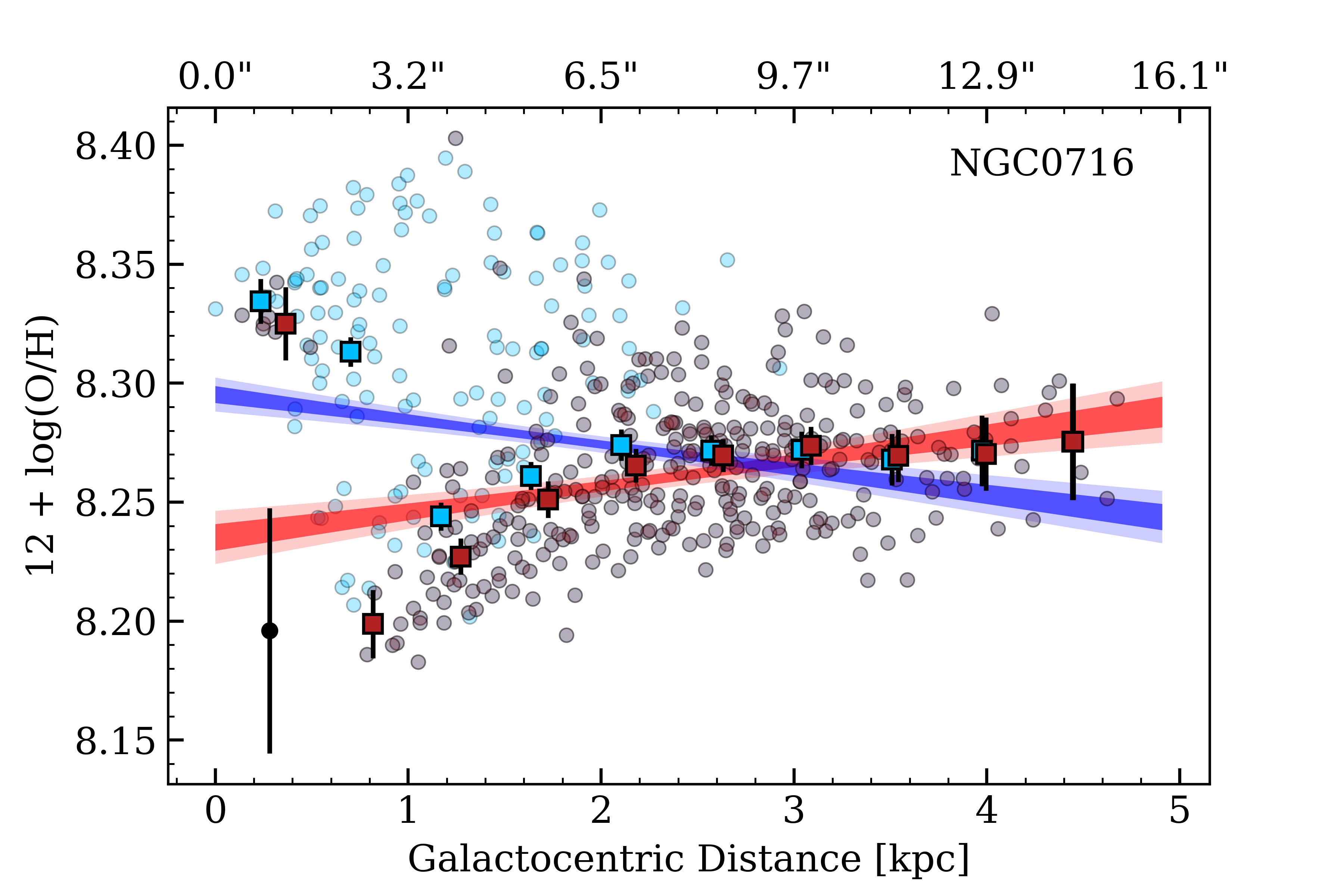}
    \includegraphics[trim={0cm 0mm 0cm 6.2mm},clip, width=0.44\linewidth]{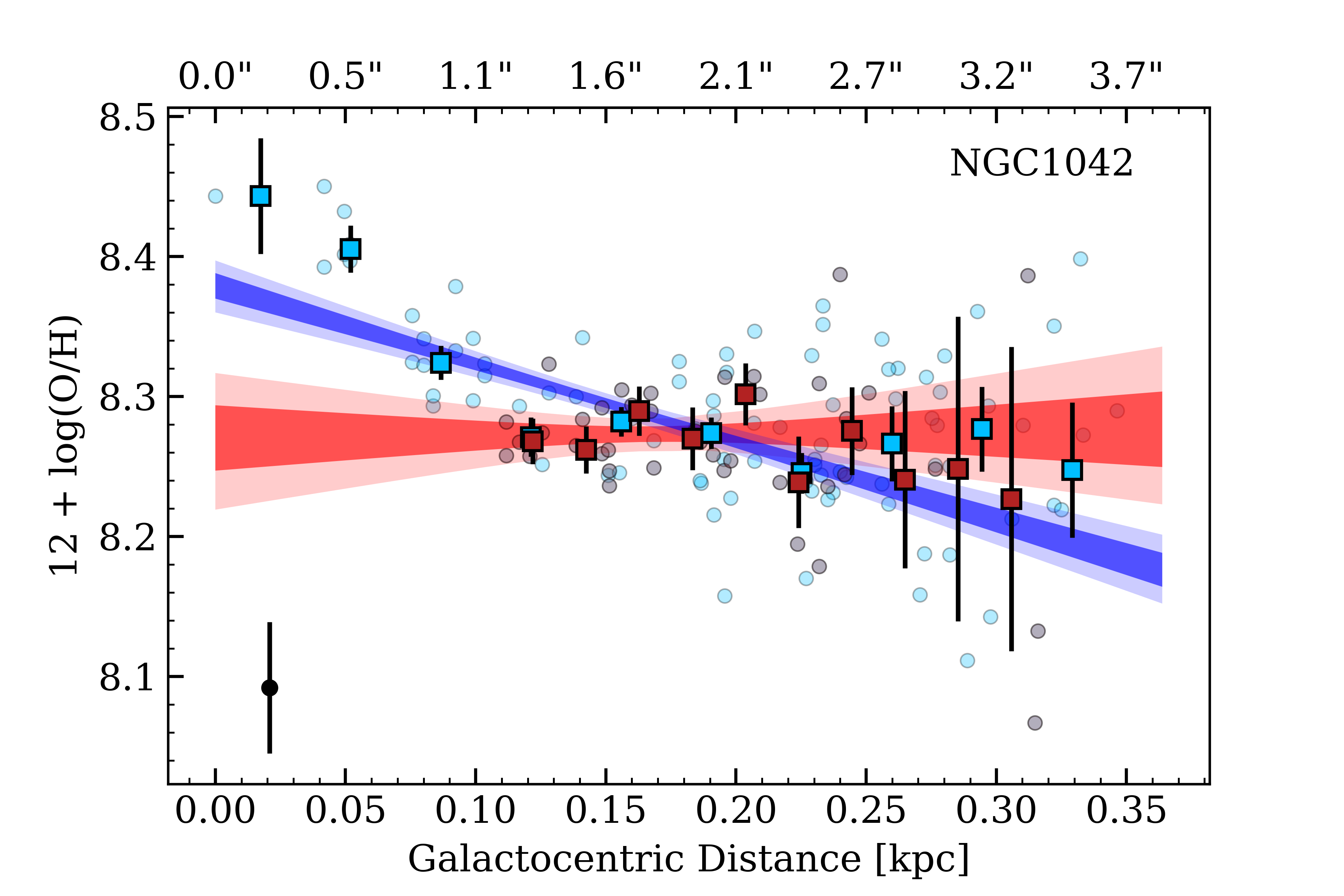}
    \includegraphics[trim={0cm 0mm 0cm 6.2mm},clip, width=0.44\linewidth]{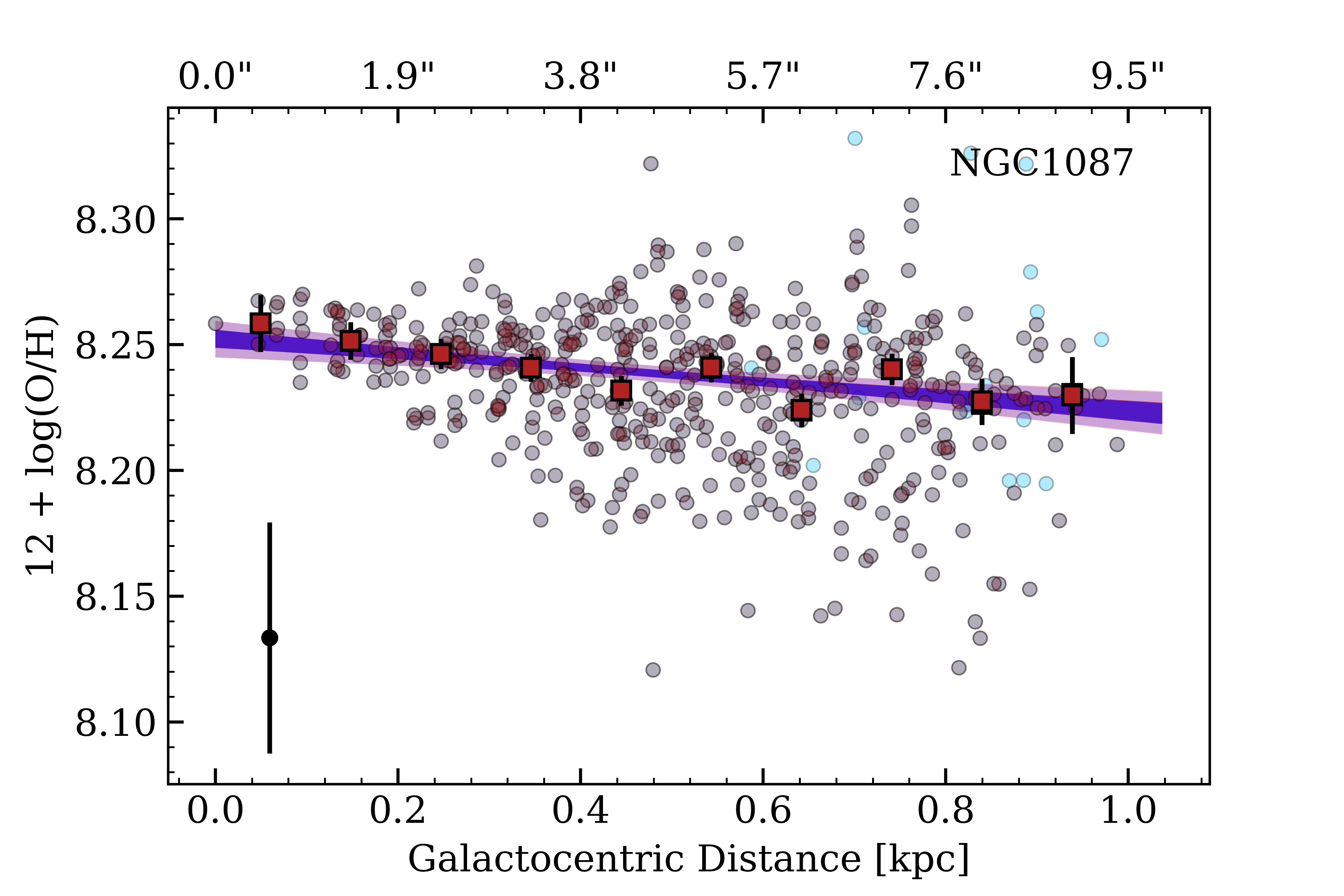}
    \includegraphics[trim={0cm 0mm 0cm 6.2mm},clip, width=0.44\linewidth]{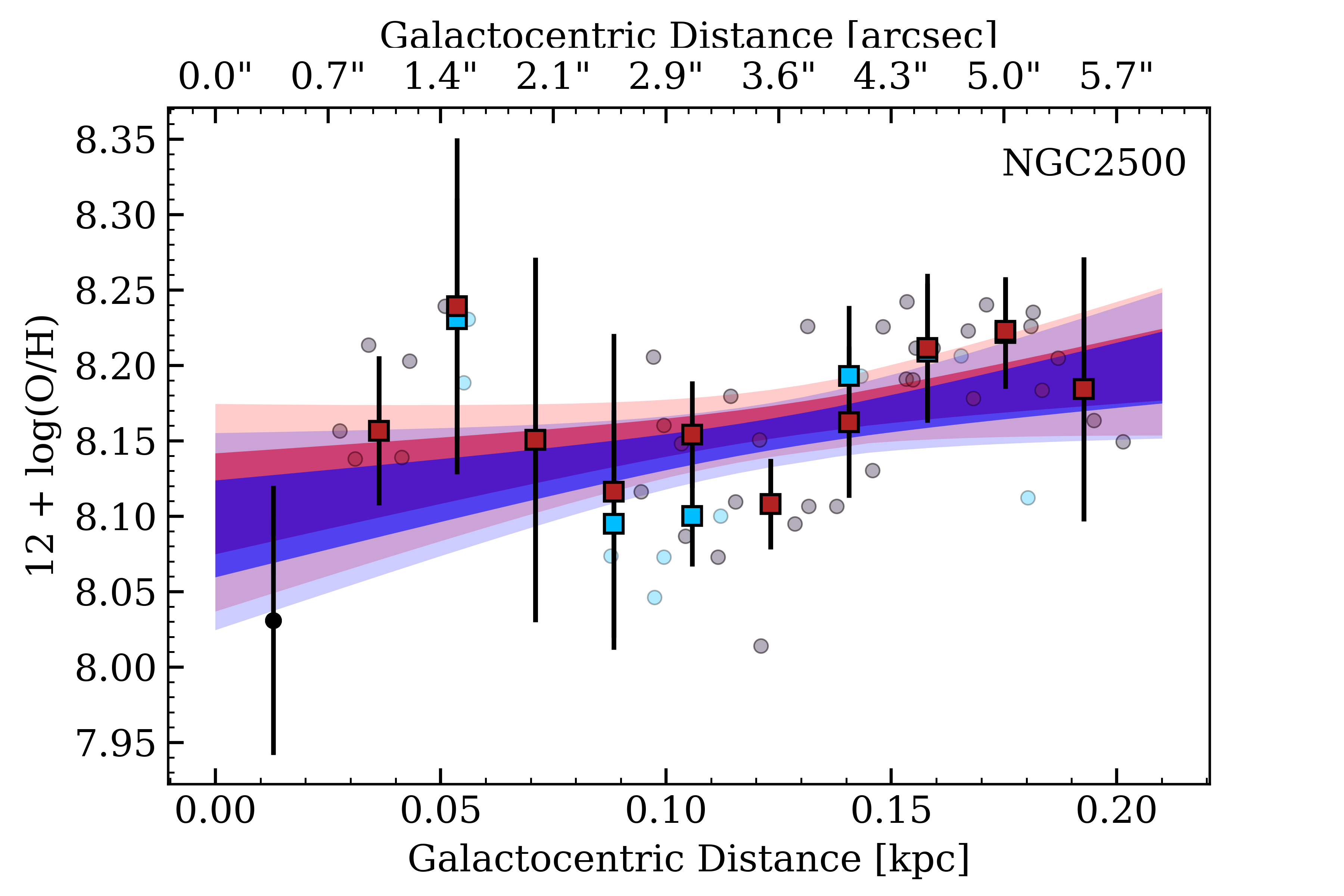}
    \includegraphics[trim={0cm 0mm 0cm 6.2mm},clip, width=0.44\linewidth]{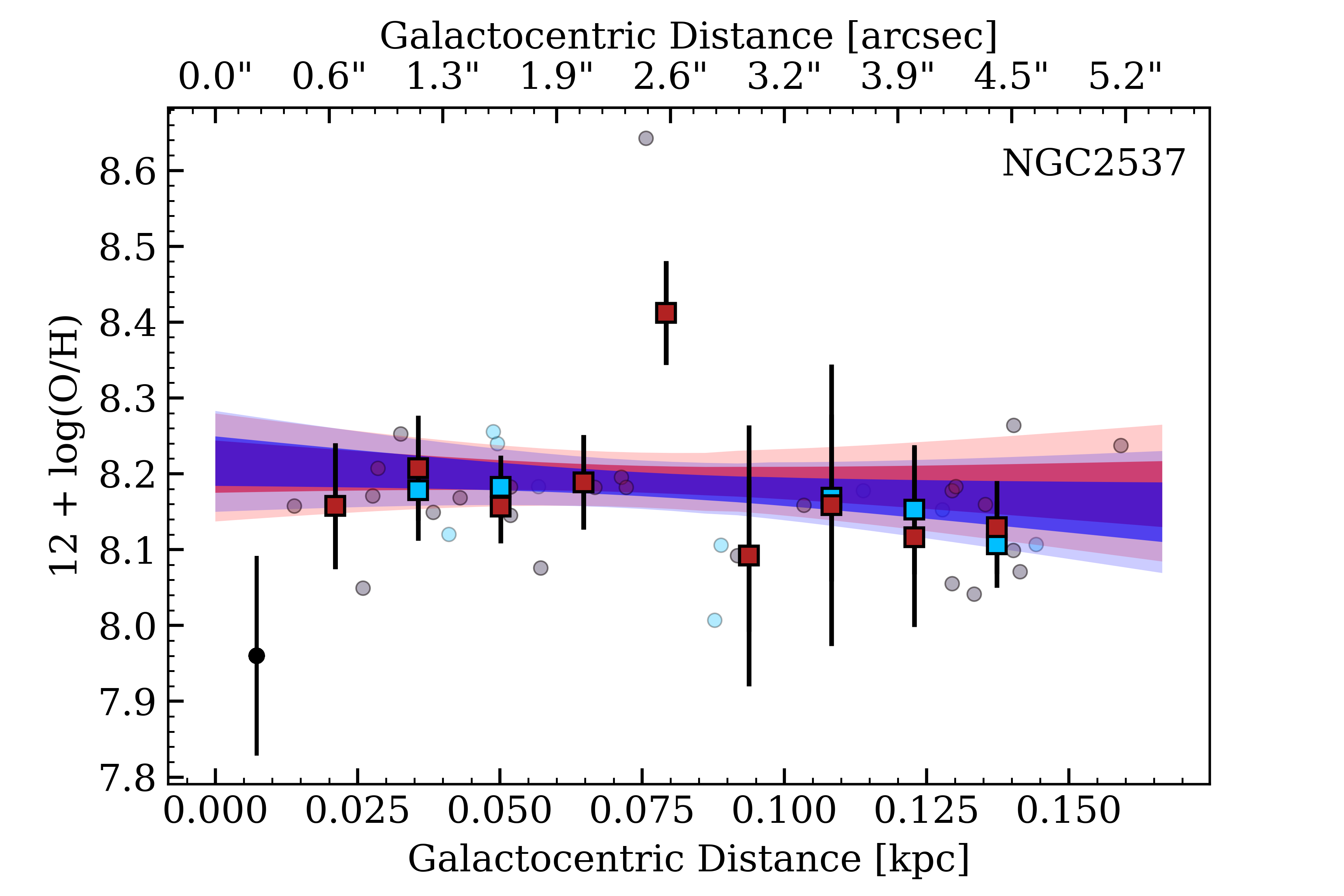}
	\caption{Oxygen abundance (i.e$.$ $\rm 12+log(O/H)$)  estimations based in the calibration by~\cite{Marino_2013} as a function of galactocentric distance. Gradients have been derived using the median values estimated for different galactocentric distance intervals. Within each galactocentric distance interval, the median values of the data are represented by blue (red) squares. The blue dots represent metallicity measurements obtained from all the spaxels while the red dots indicate those spaxels where the emission lines are originated in star-forming regions. The blue and red bands indicate the results of our Bayesian linear best fits to the data with the same colour coding as for the individual measurements. The dark and light shaded bands correspond to the confidence intervals at 1-$\sigma$ and 2-$\sigma$ levels, respectively. The black dot at the bottom left of the plot represents the median error of all (blue) points.}
	\label{fig:B1}
\end{figure}

\addtocounter{figure}{-1}
\begin{figure*}[h]
	\centering
    \includegraphics[trim={0cm 0mm 0cm 6.2mm},clip, width=0.44\linewidth]{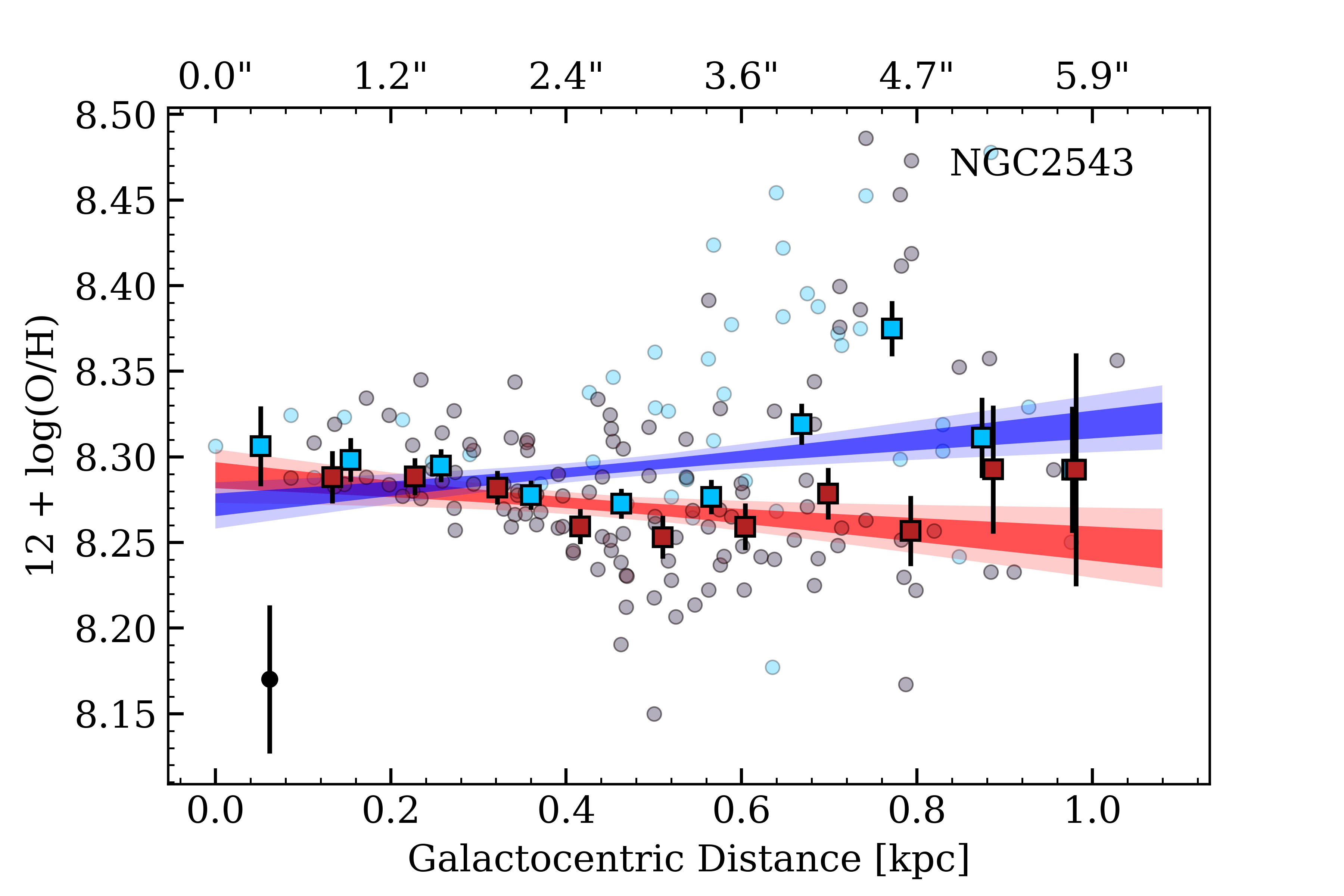}
    \includegraphics[trim={0cm 0mm 0cm 6.2mm},clip, width=0.44\linewidth]{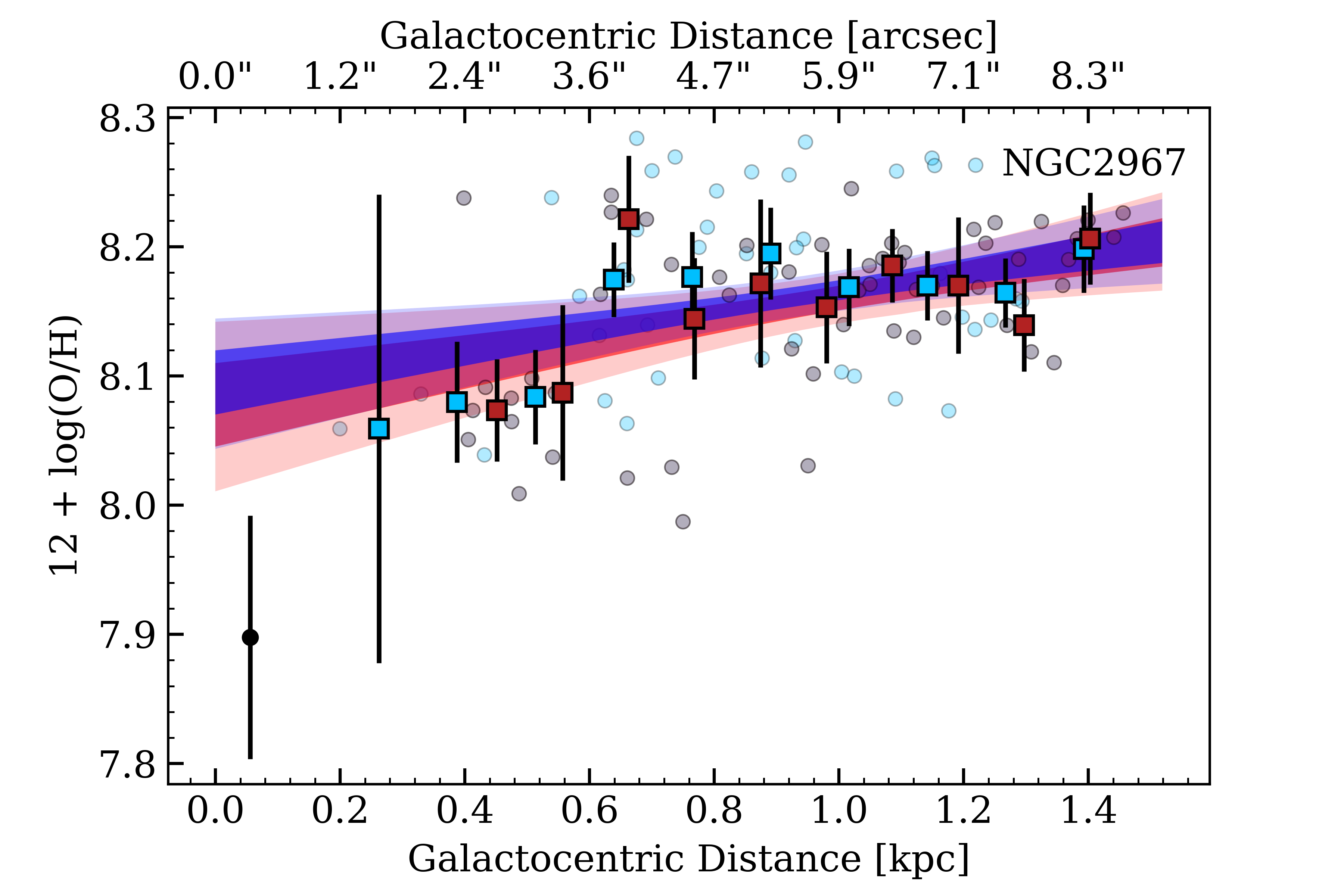}
    \includegraphics[trim={0cm 0mm 0cm 6.2mm},clip, width=0.44\linewidth]{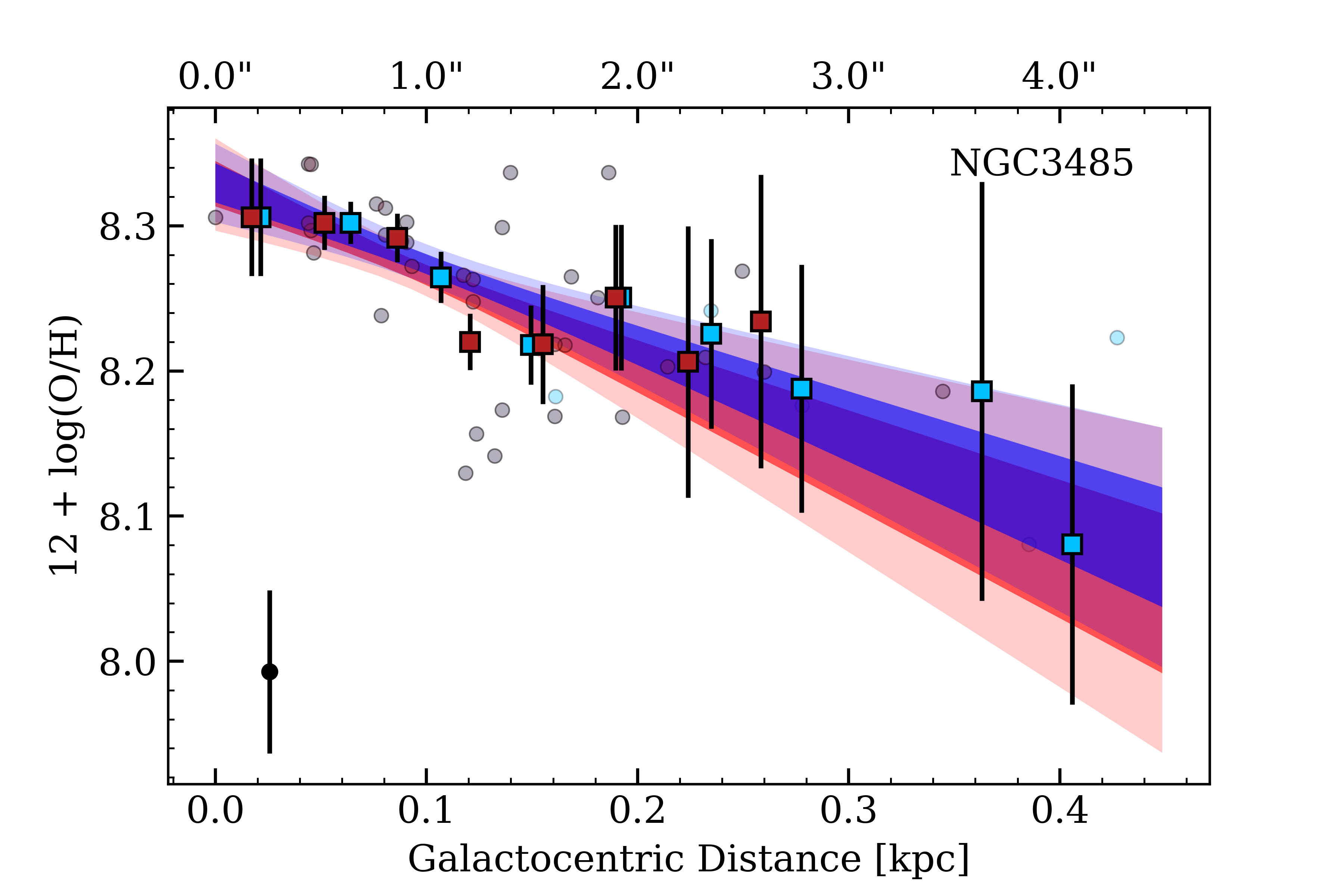}
    \includegraphics[trim={0cm 0mm 0cm 6.2mm},clip, width=0.44\linewidth]{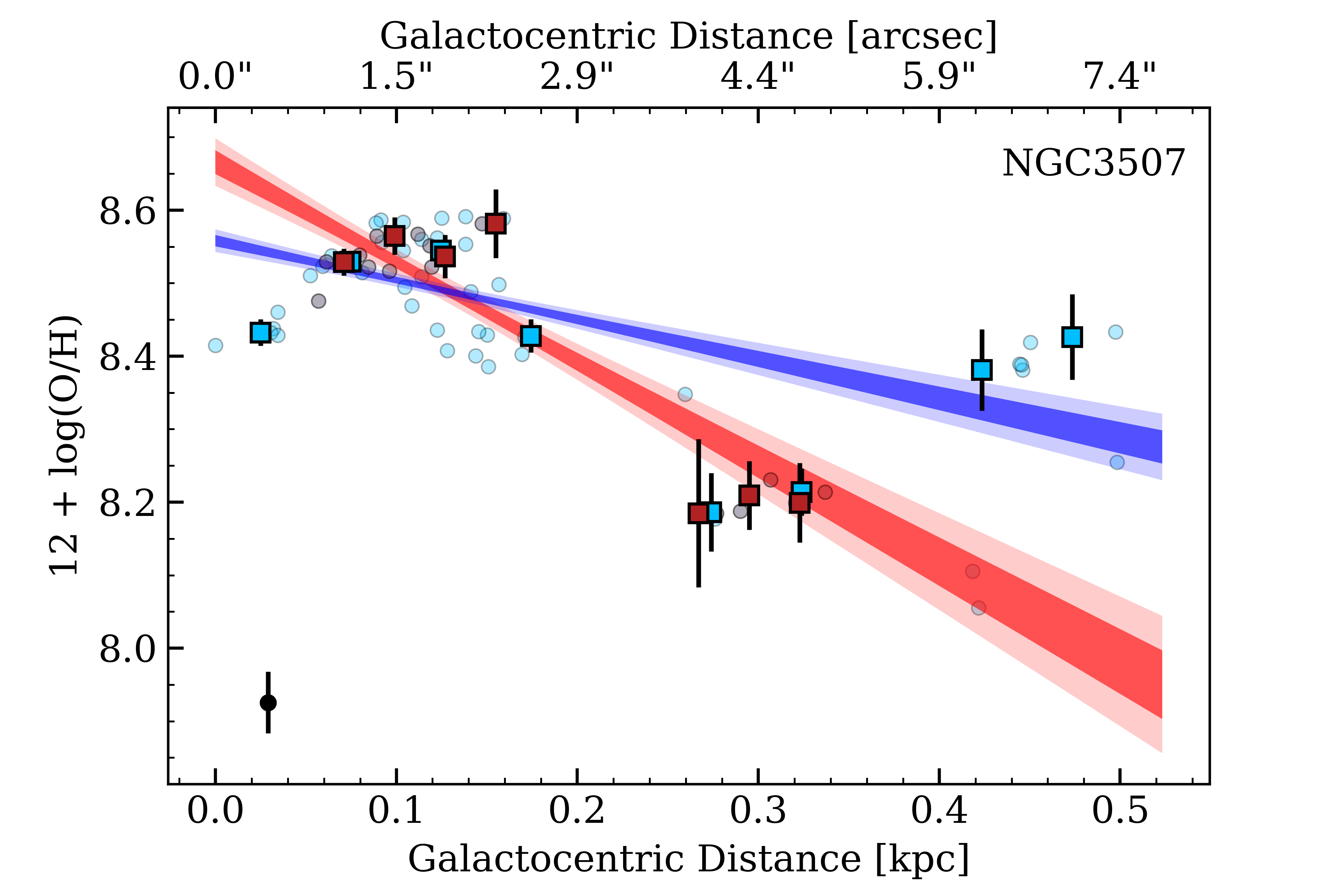}
    \includegraphics[trim={0cm 0mm 0cm 6.2mm},clip, width=0.44\linewidth]{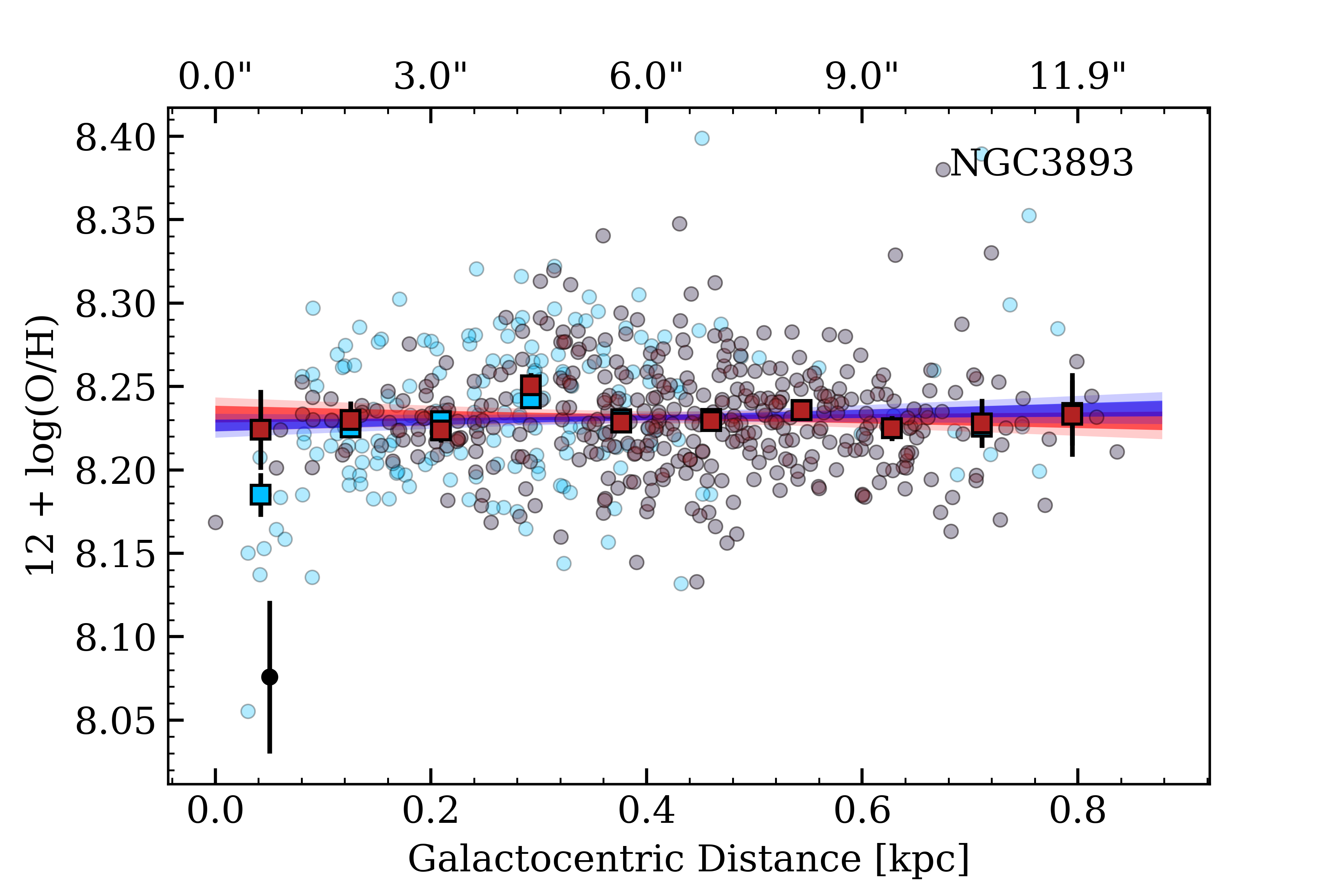}
    \includegraphics[trim={0cm 0mm 0cm 6.2mm},clip, width=0.44\linewidth]{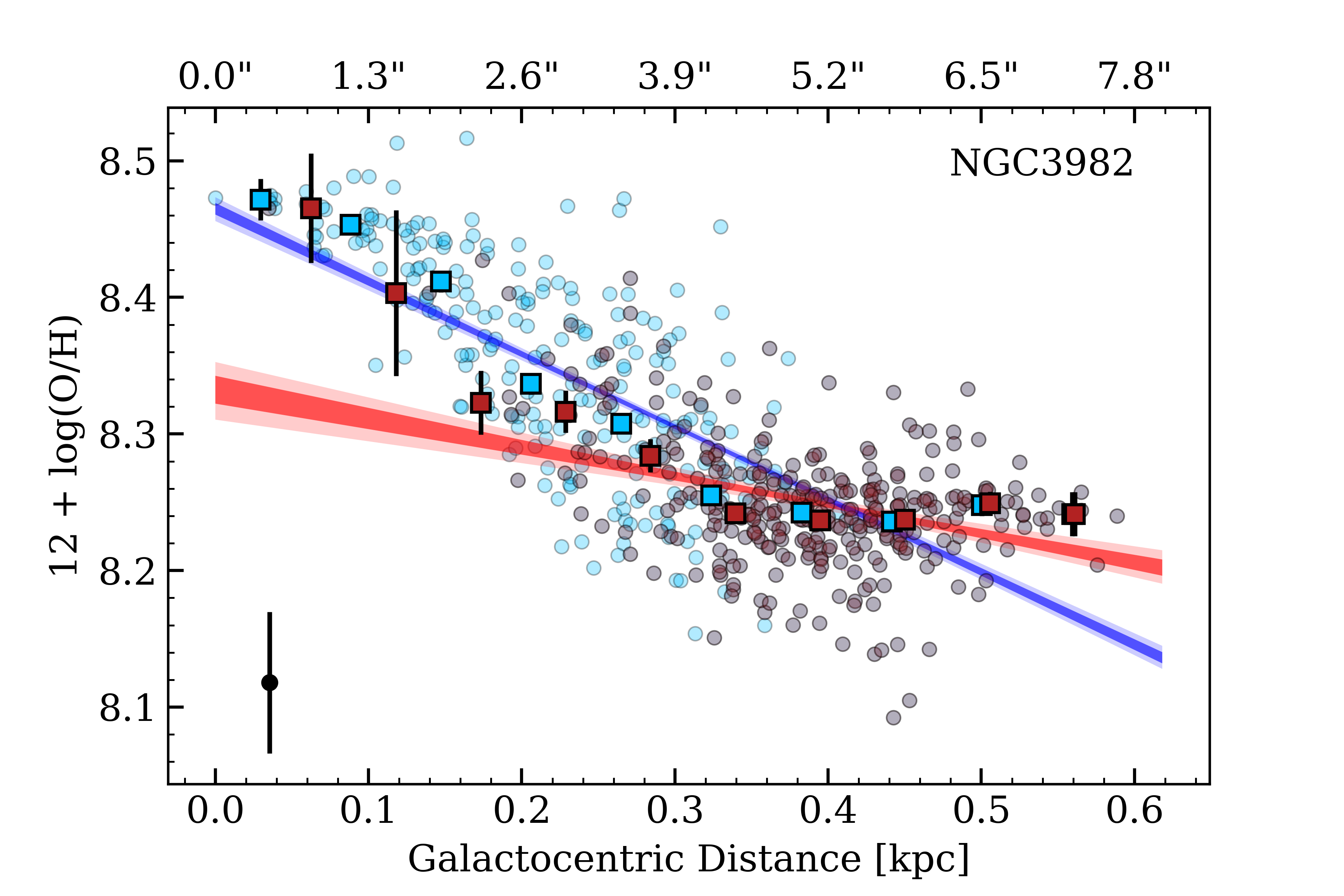}
    \includegraphics[trim={0cm 0mm 0cm 6.2mm},clip, width=0.44\linewidth]{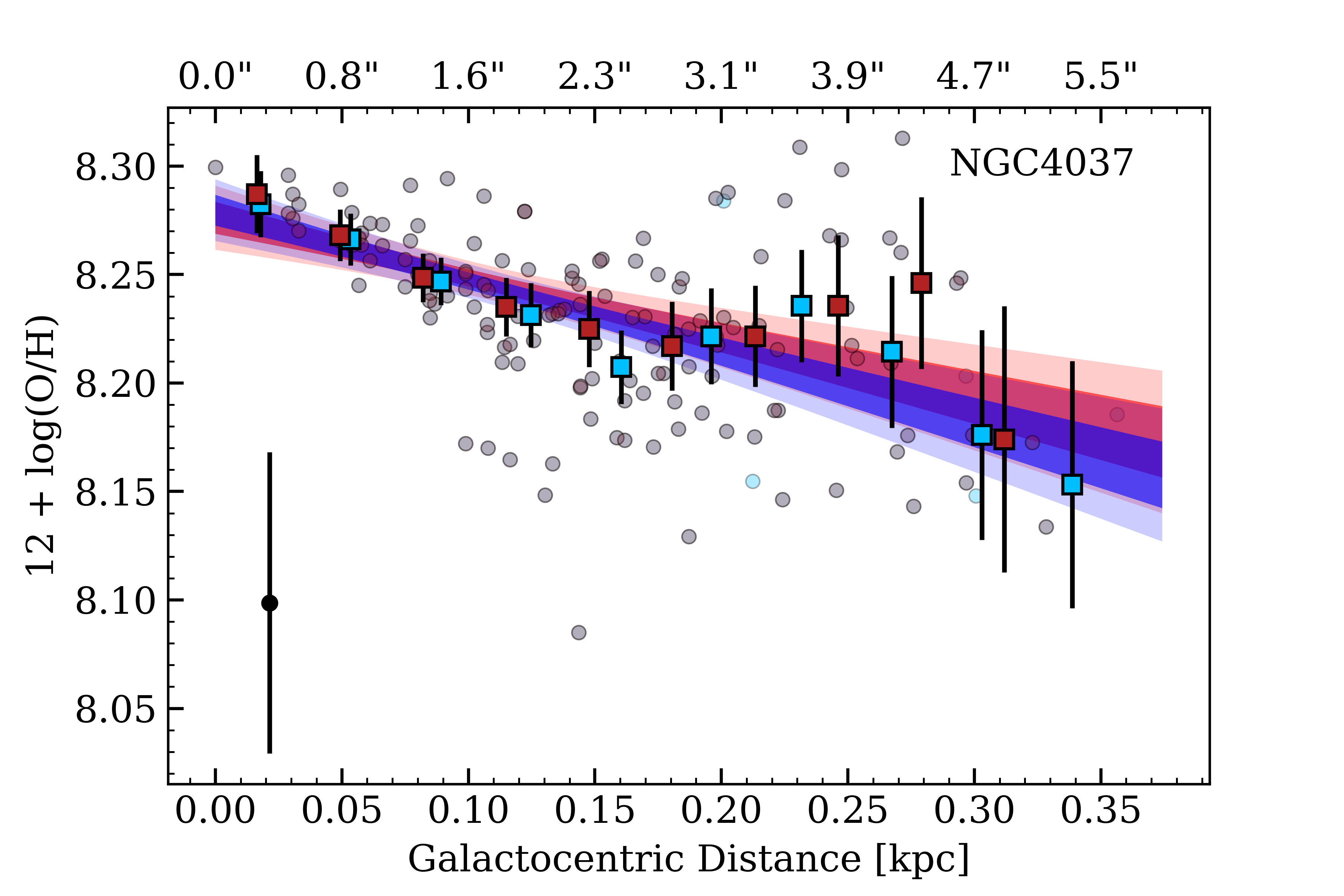}
    \includegraphics[trim={0cm 0mm 0cm 6.2mm},clip, width=0.44\linewidth]{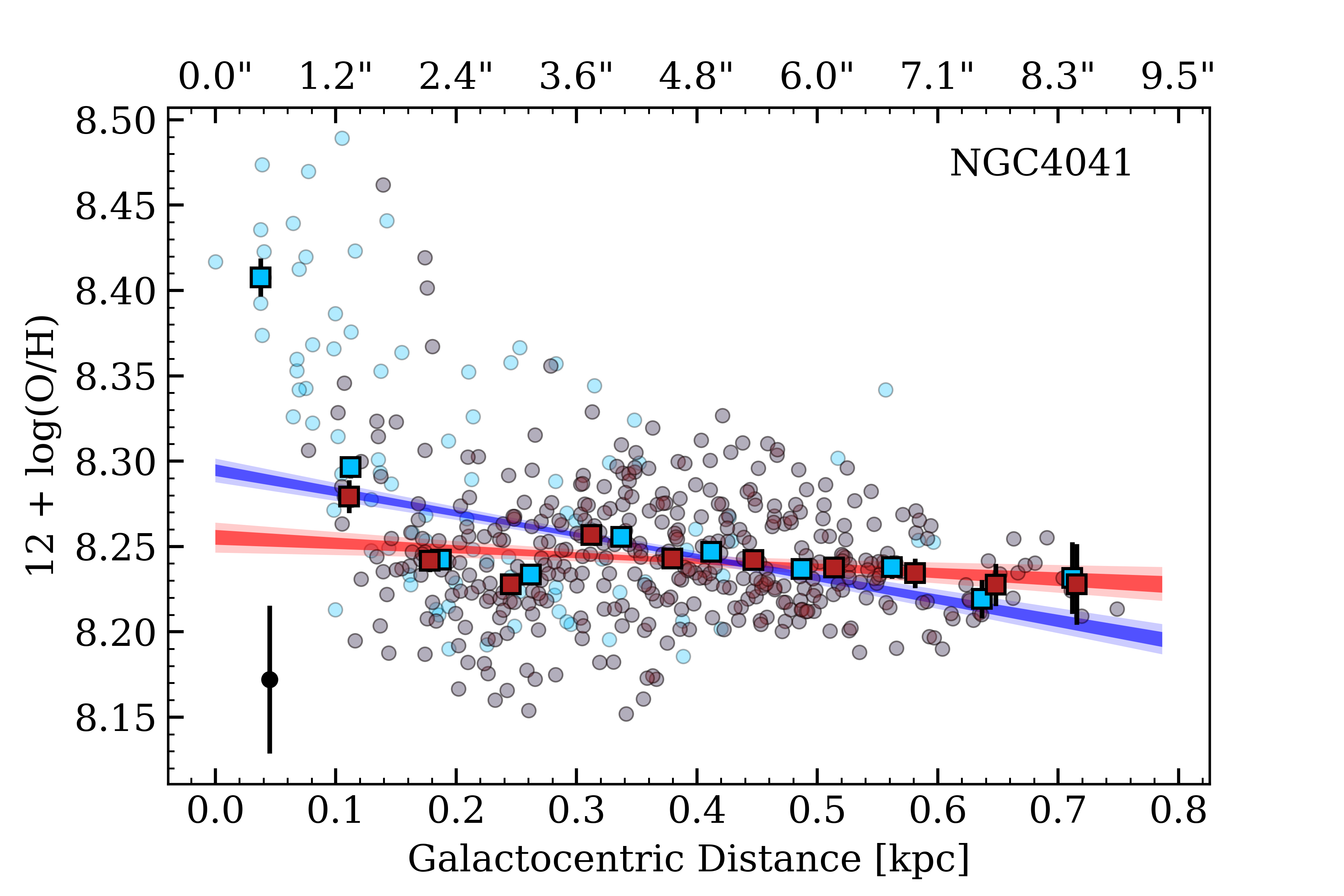}
	\caption{continued.}
	\label{fig:B1}
\end{figure*}
\addtocounter{figure}{-1}
\begin{figure*}[h]
	\centering
    \includegraphics[trim={0cm 0mm 0cm 6.2mm},clip, width=0.44\linewidth]{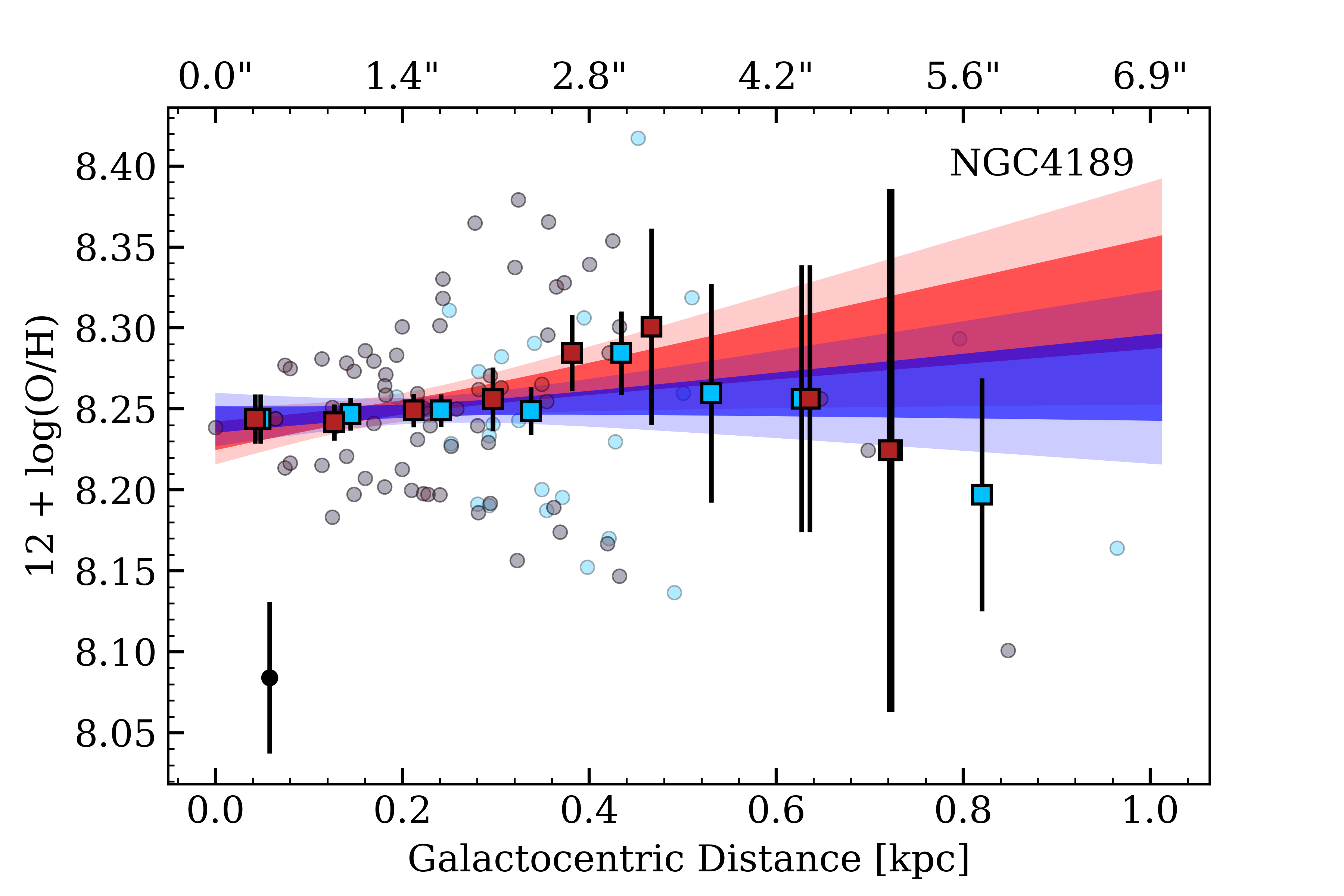}
    \includegraphics[trim={0cm 0mm 0cm 6.2mm},clip, width=0.44\linewidth]{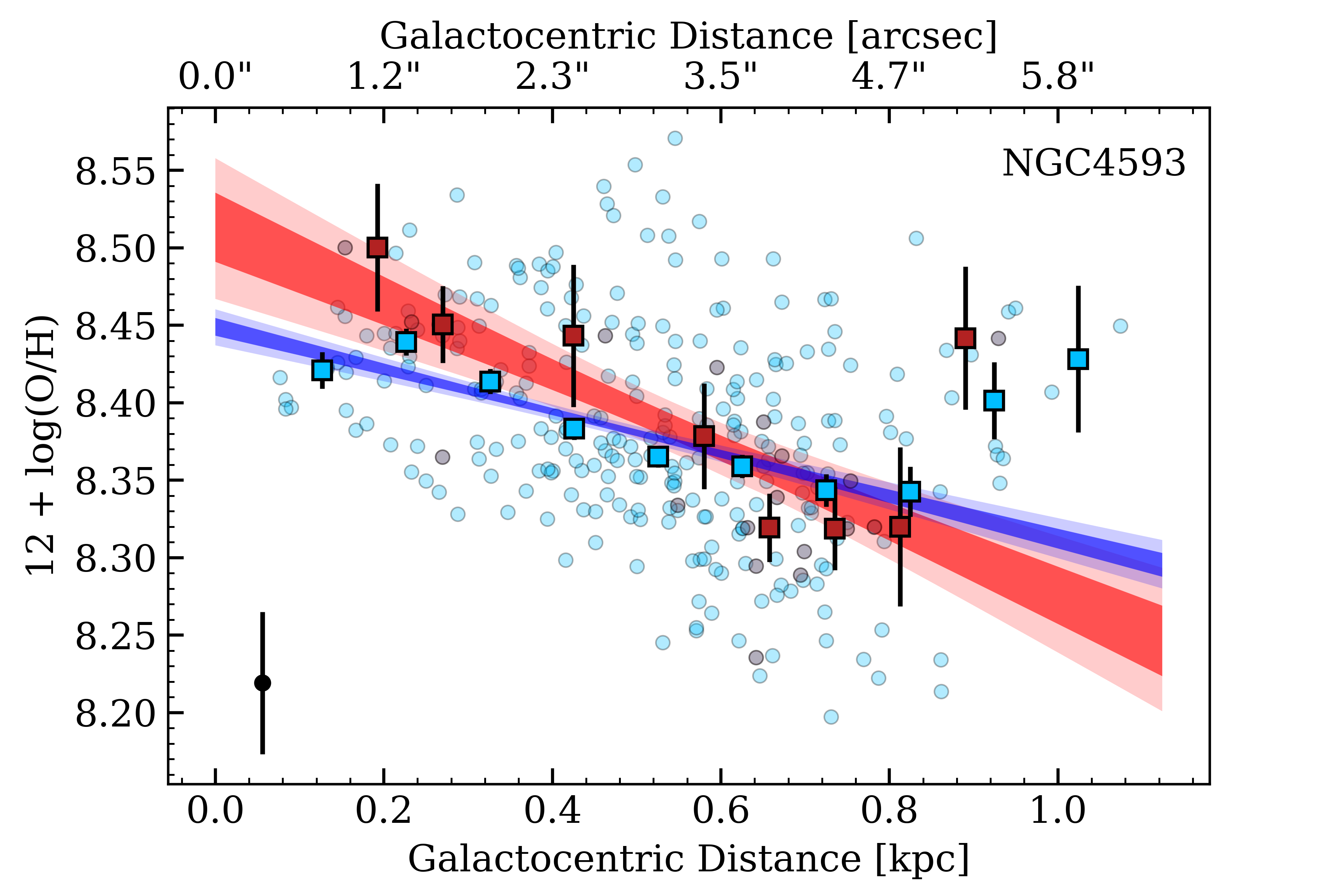}
    \includegraphics[trim={0cm 0mm 0cm 6.2mm},clip, width=0.44\linewidth]{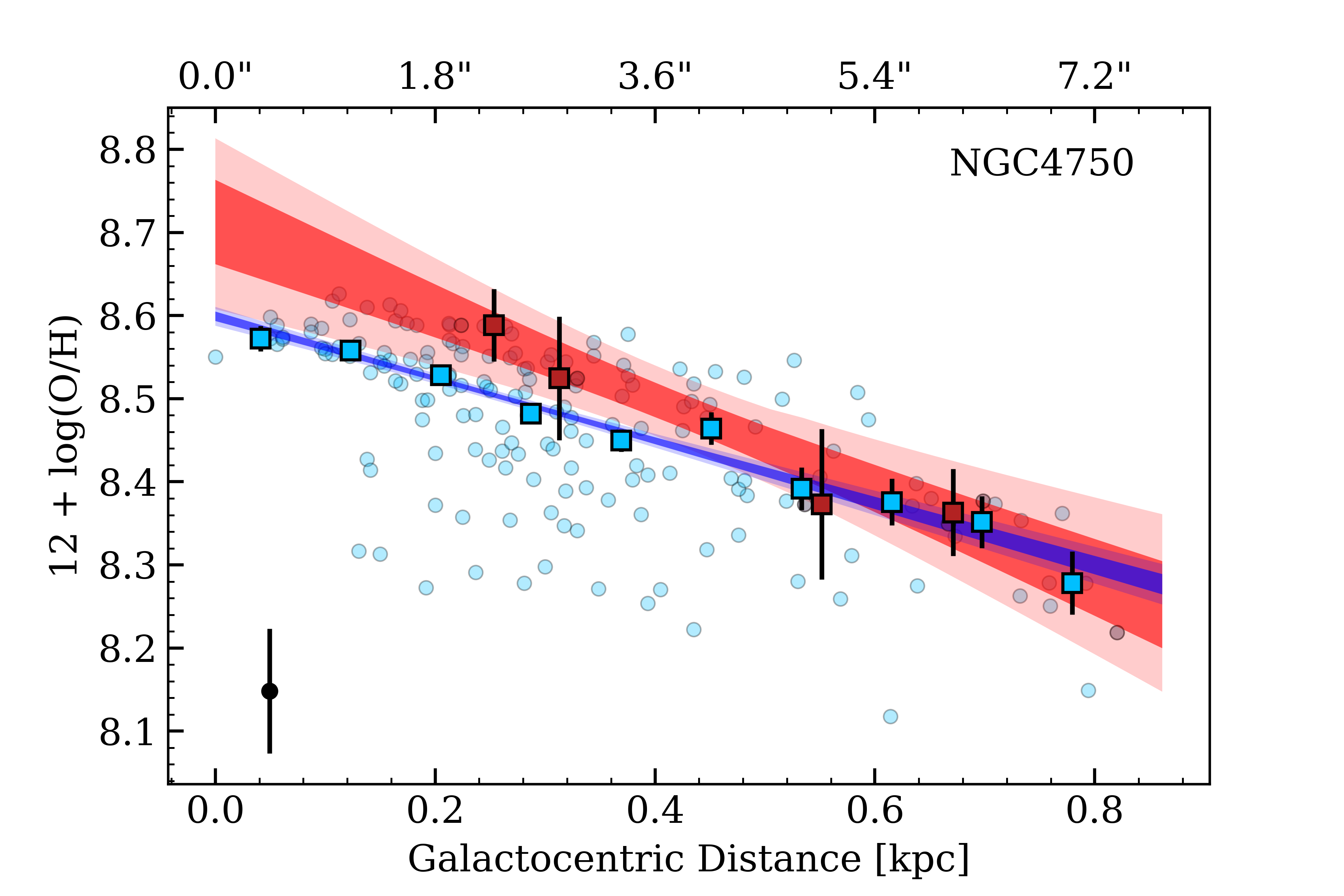}
    \includegraphics[trim={0cm 0mm 0cm 6.2mm},clip, width=0.44\linewidth]{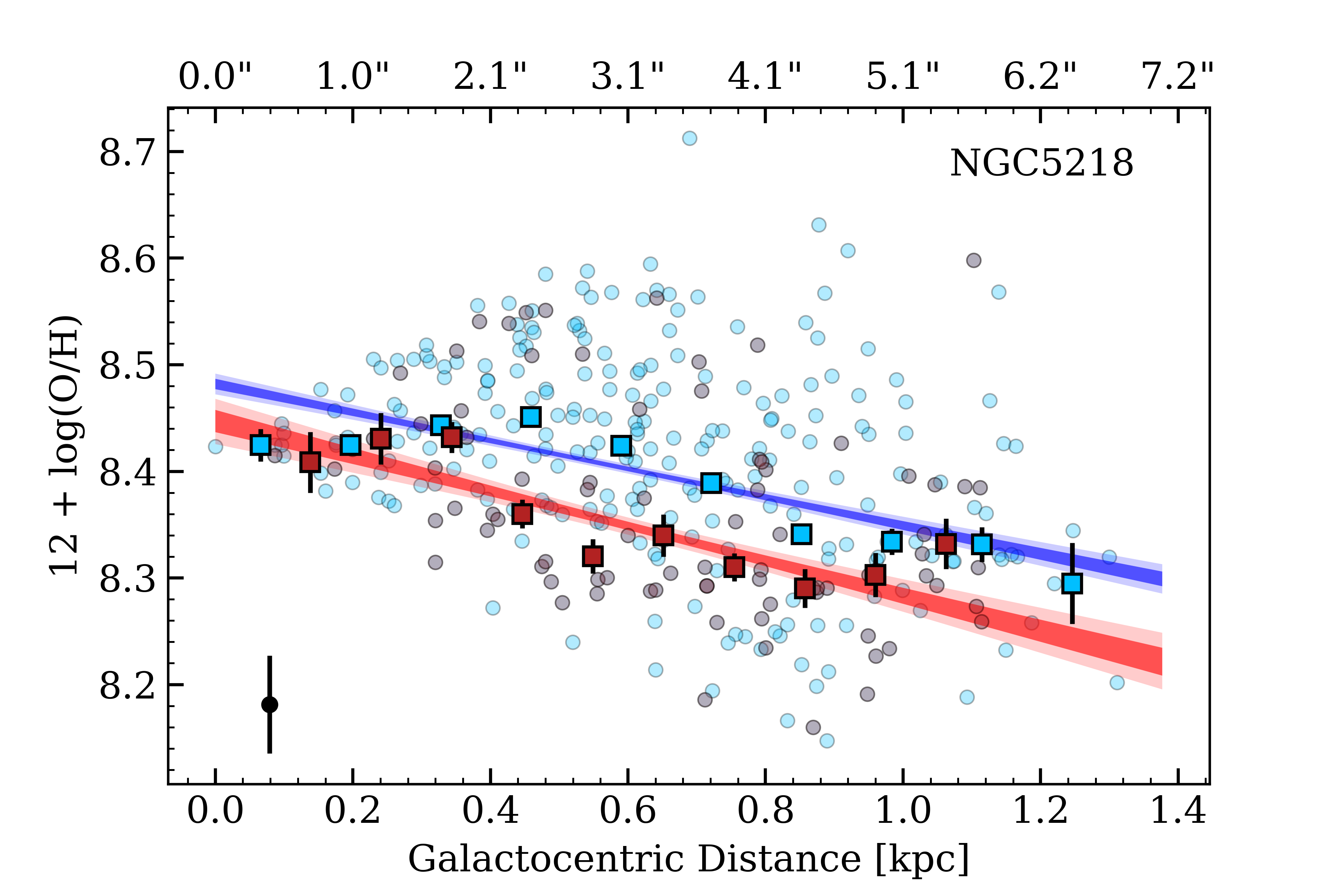}
    \includegraphics[trim={0cm 0mm 0cm 6.2mm},clip, width=0.44\linewidth]{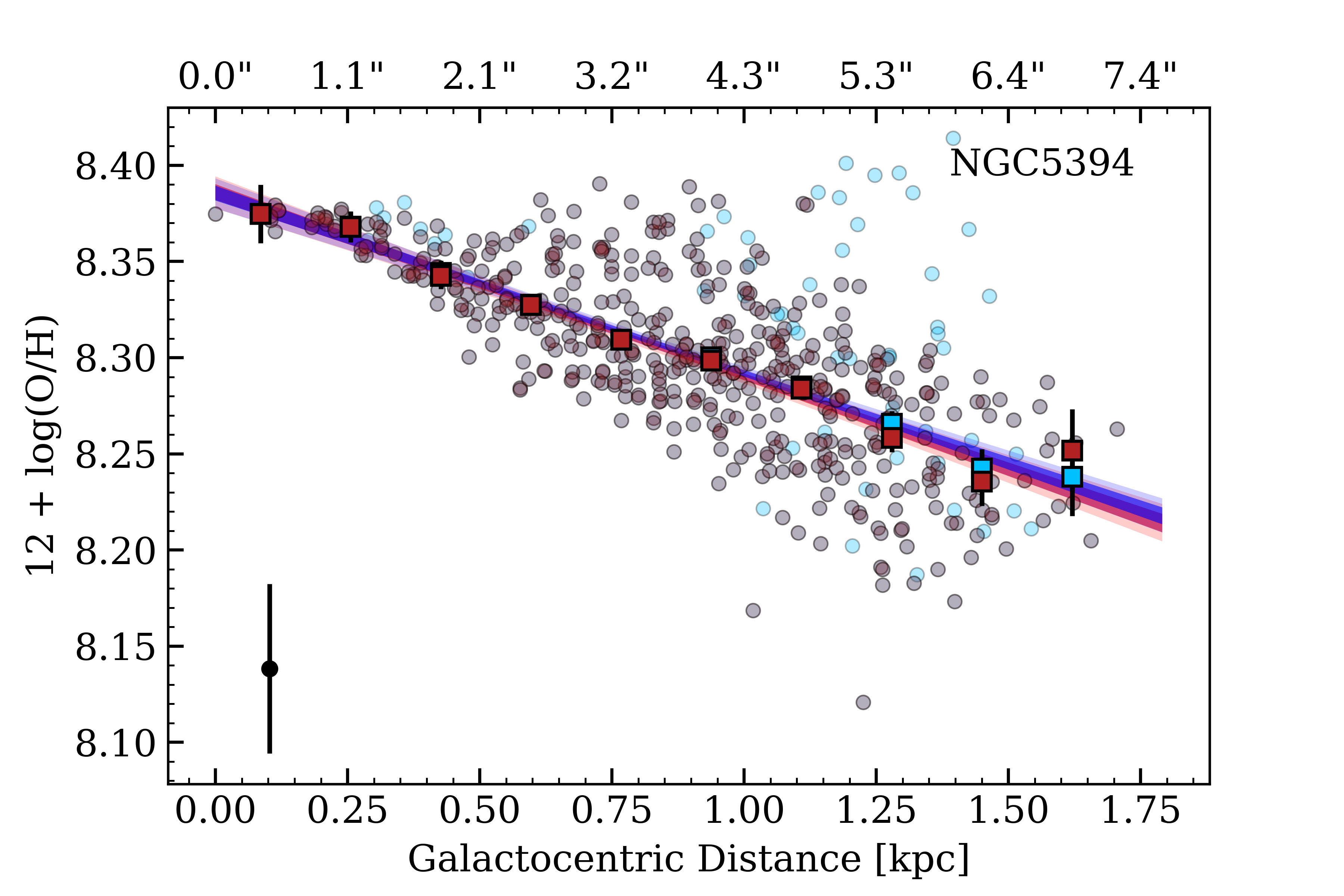}
    \includegraphics[trim={0cm 0mm 0cm 6.2mm},clip, width=0.44\linewidth]{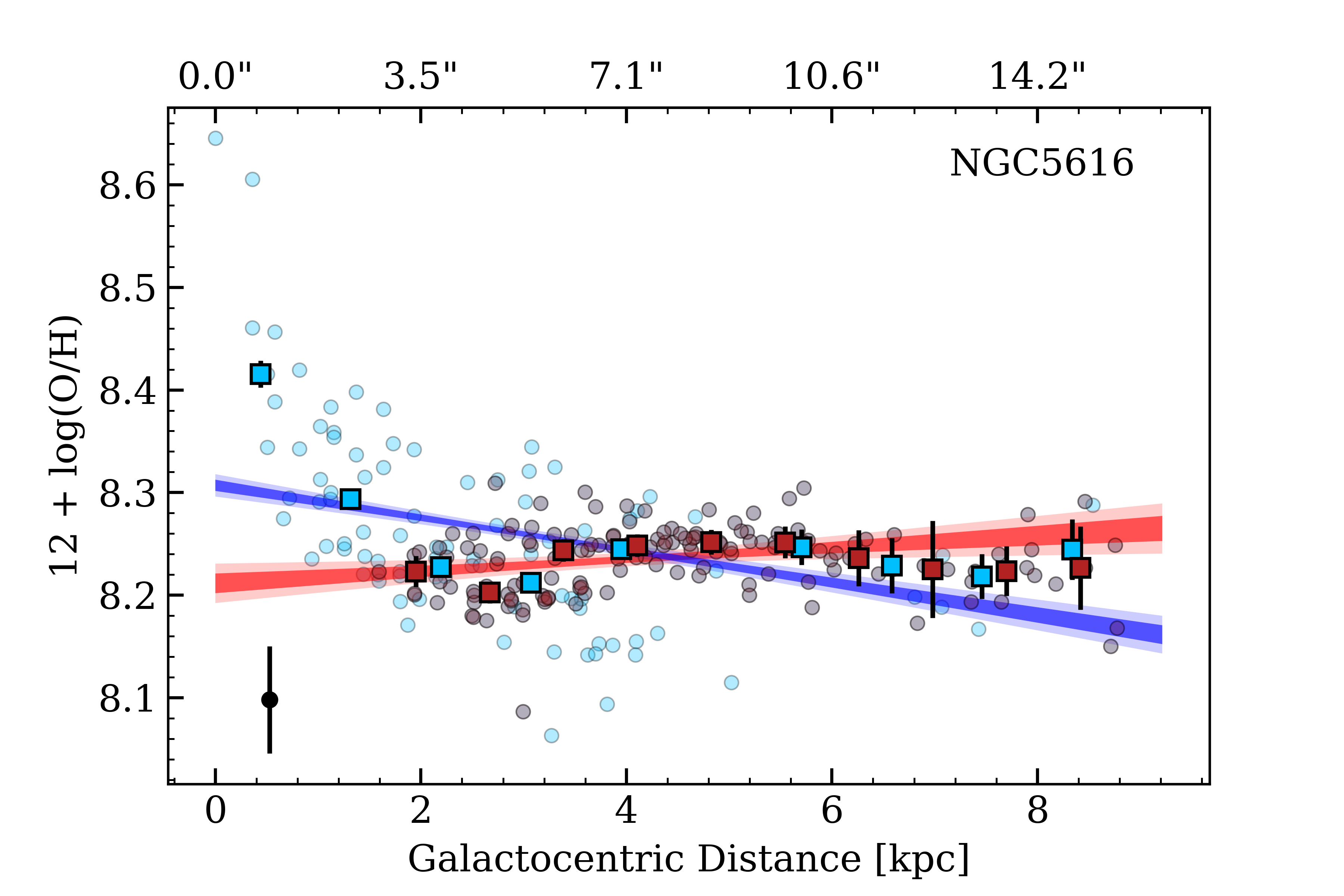}
    \includegraphics[trim={0cm 0mm 0cm 6.1mm},clip, width=0.43\linewidth]{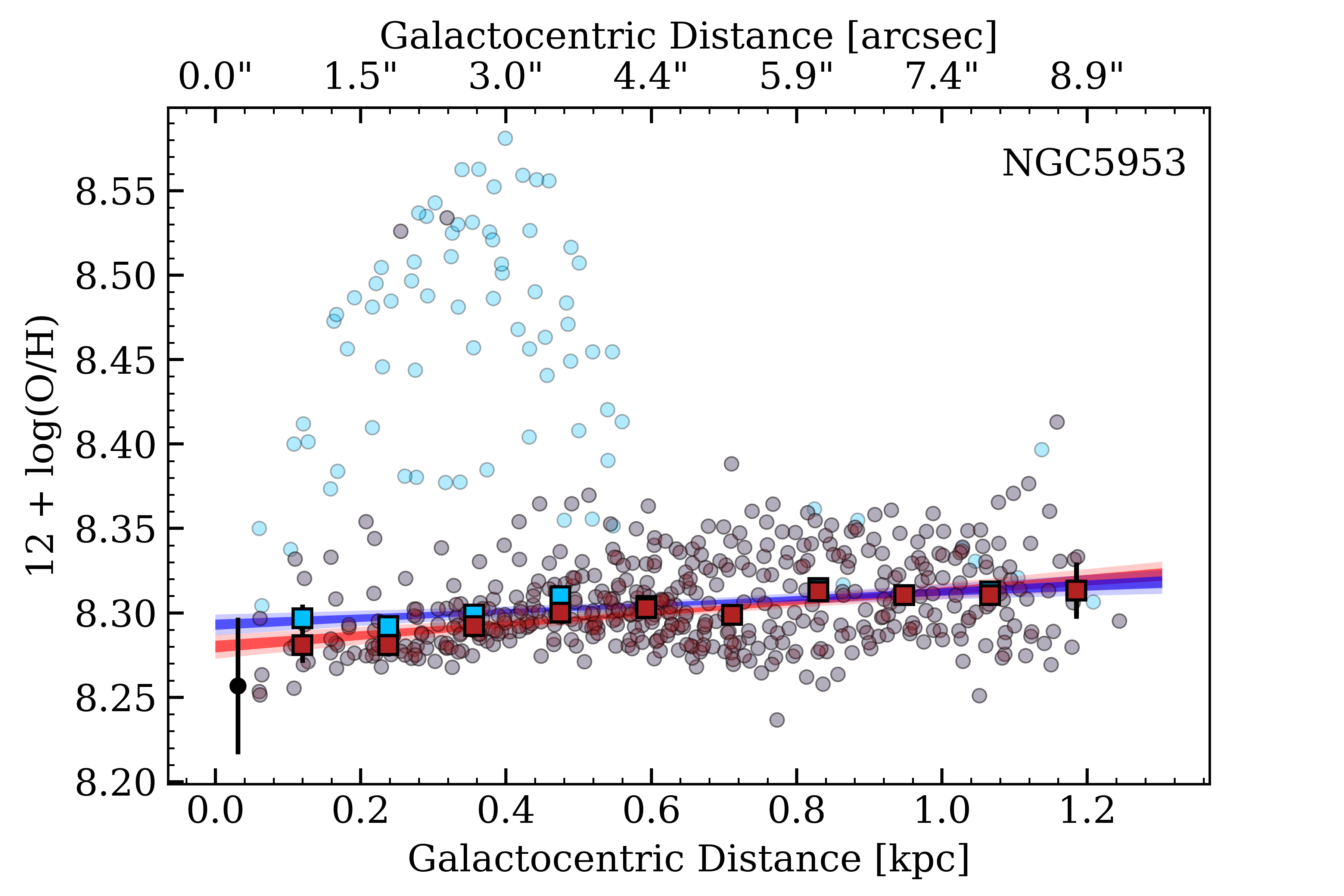}
    \includegraphics[trim={0cm 0mm 0cm 6.2mm},clip, width=0.44\linewidth]{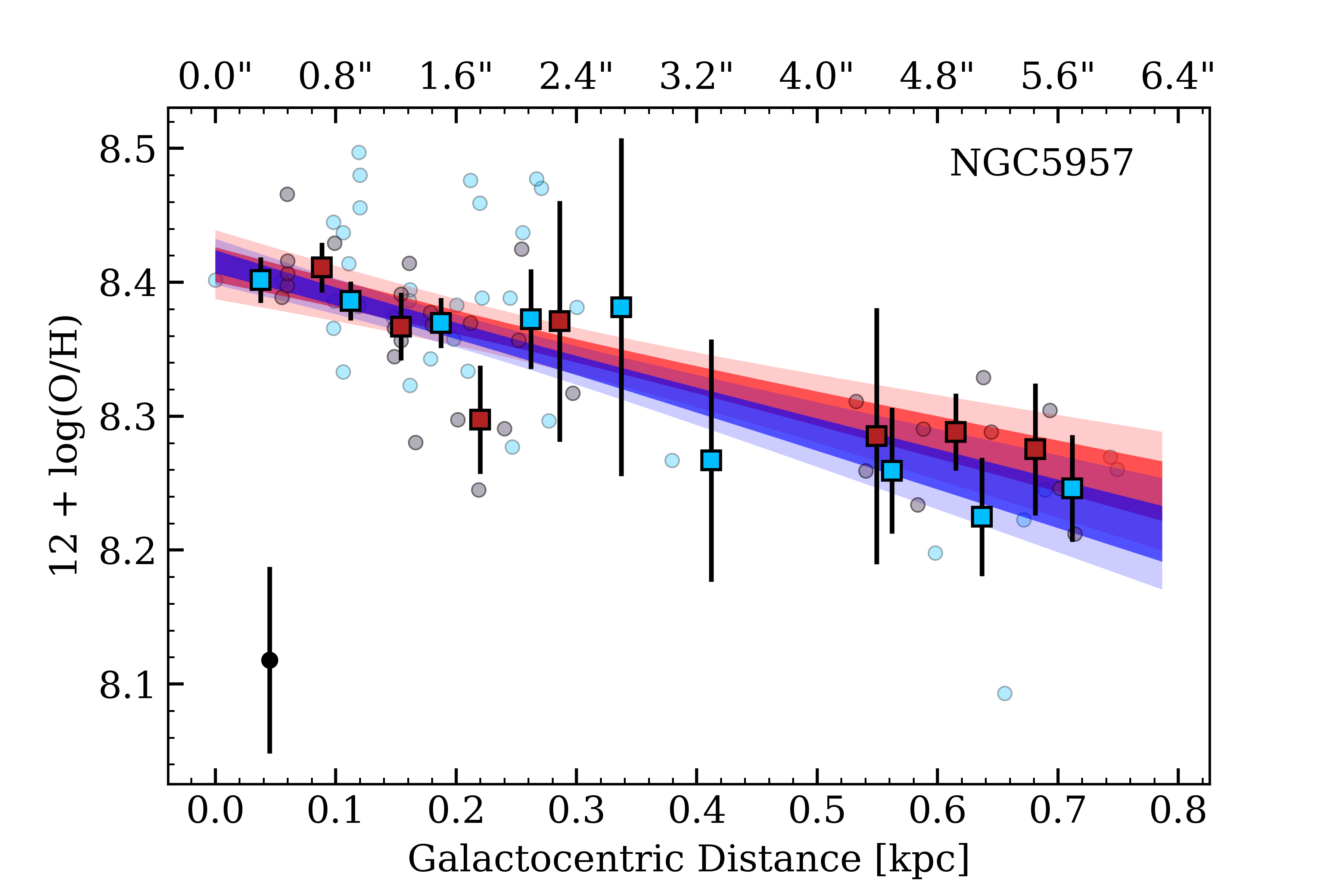}
	\caption{continued.}
	\label{fig:B1}
\end{figure*}
\addtocounter{figure}{-1}
\begin{figure*}[h]
	\centering
    \includegraphics[trim={0cm 0mm 0cm 6.2mm},clip, width=0.44\linewidth]{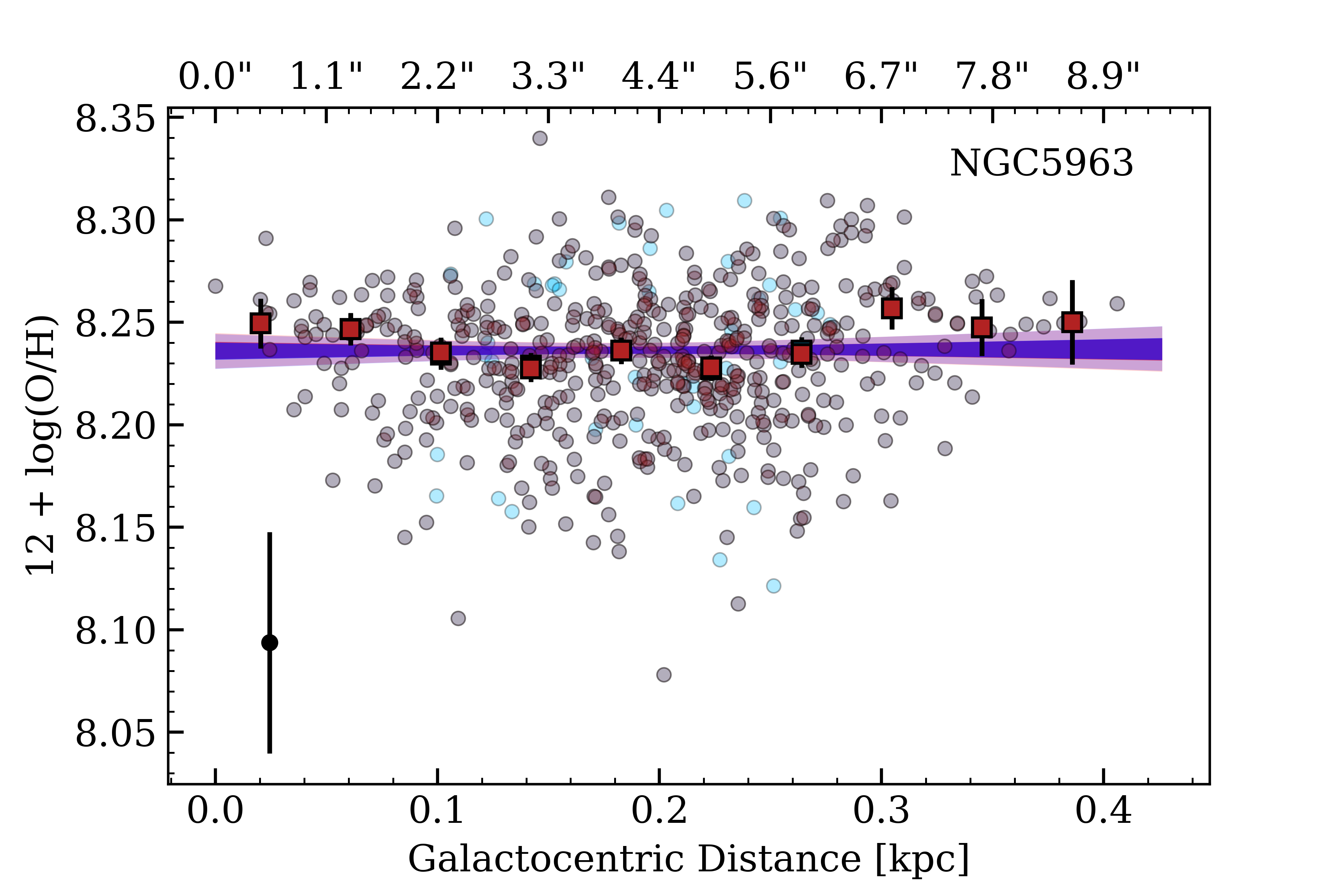}
    \includegraphics[trim={0cm 0mm 0cm 6.2mm},clip, width=0.44\linewidth]{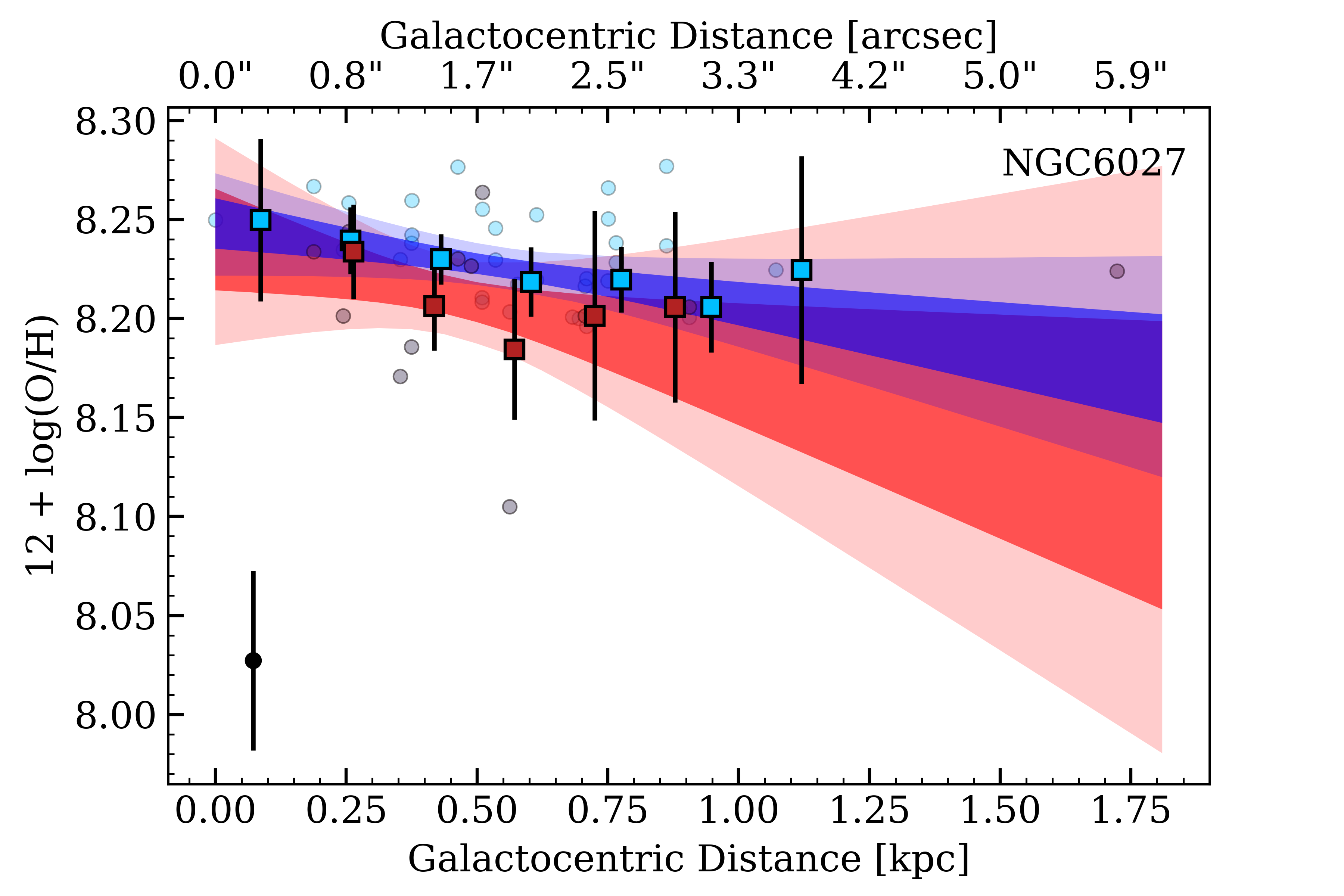}
    \includegraphics[trim={0cm 0mm 0cm 6.2mm},clip, width=0.44\linewidth]{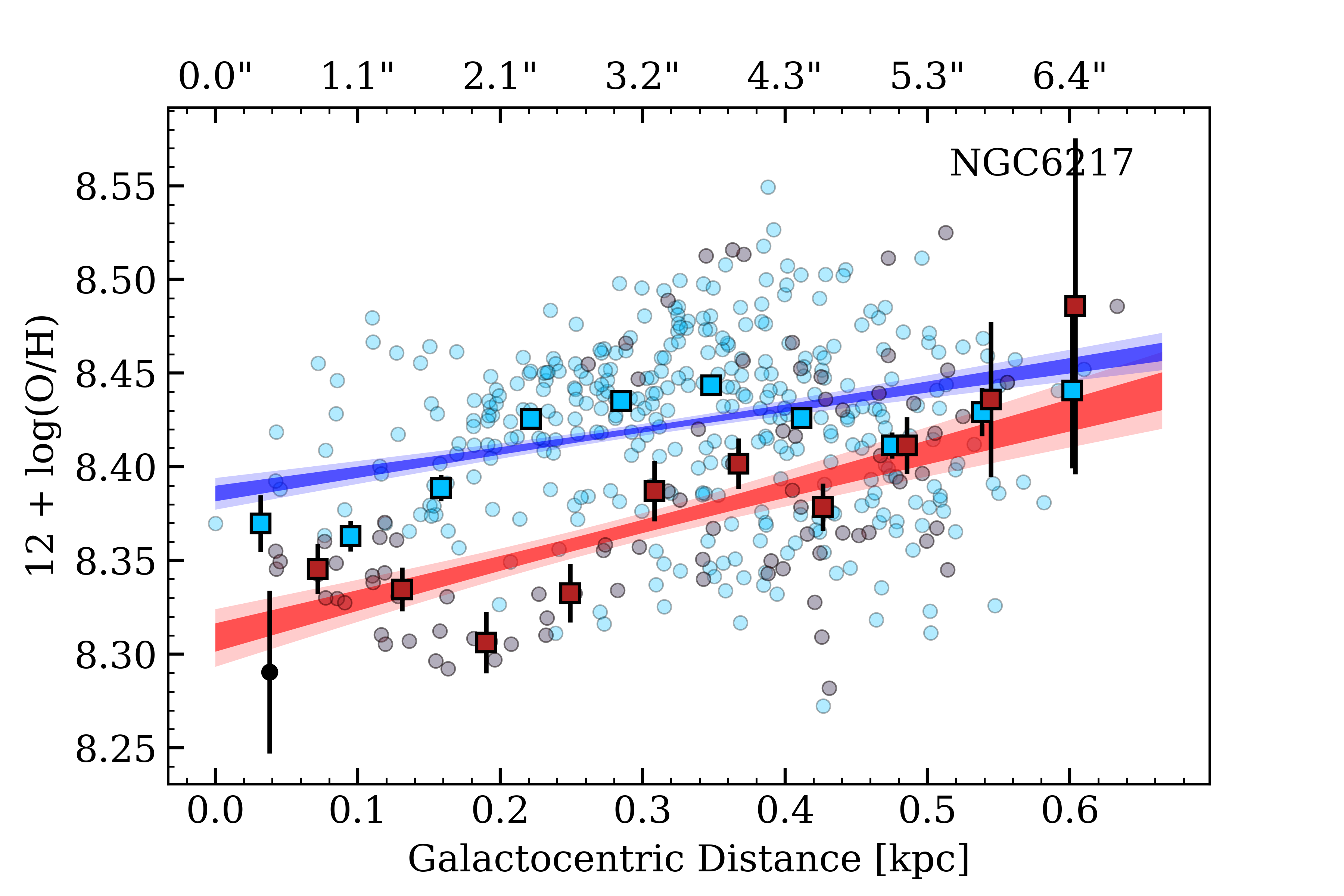}
    \includegraphics[trim={0cm 0mm 0cm 6.2mm},clip, width=0.44\linewidth]{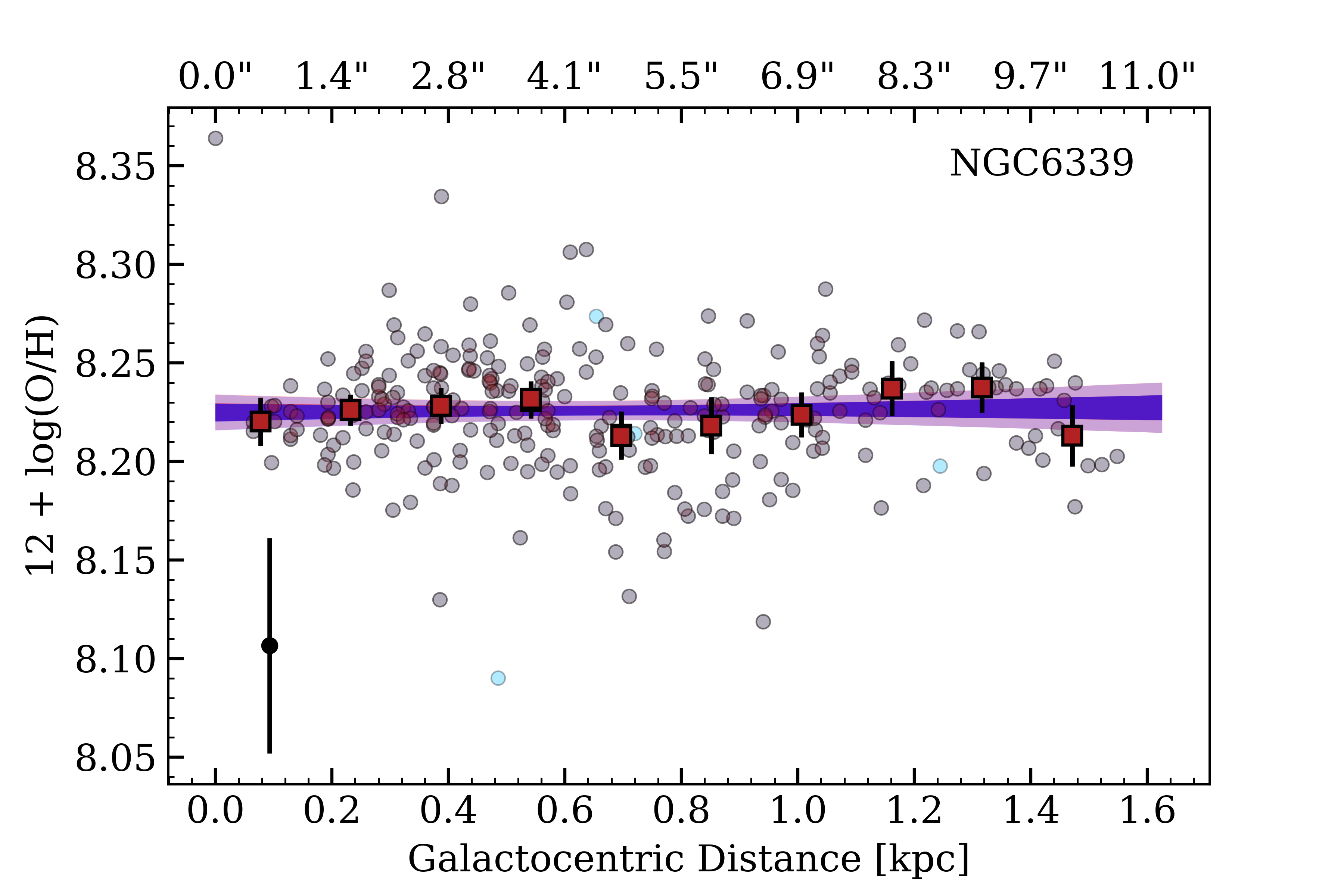}
    \includegraphics[trim={0cm 0mm 0cm 6.2mm},clip, width=0.44\linewidth]{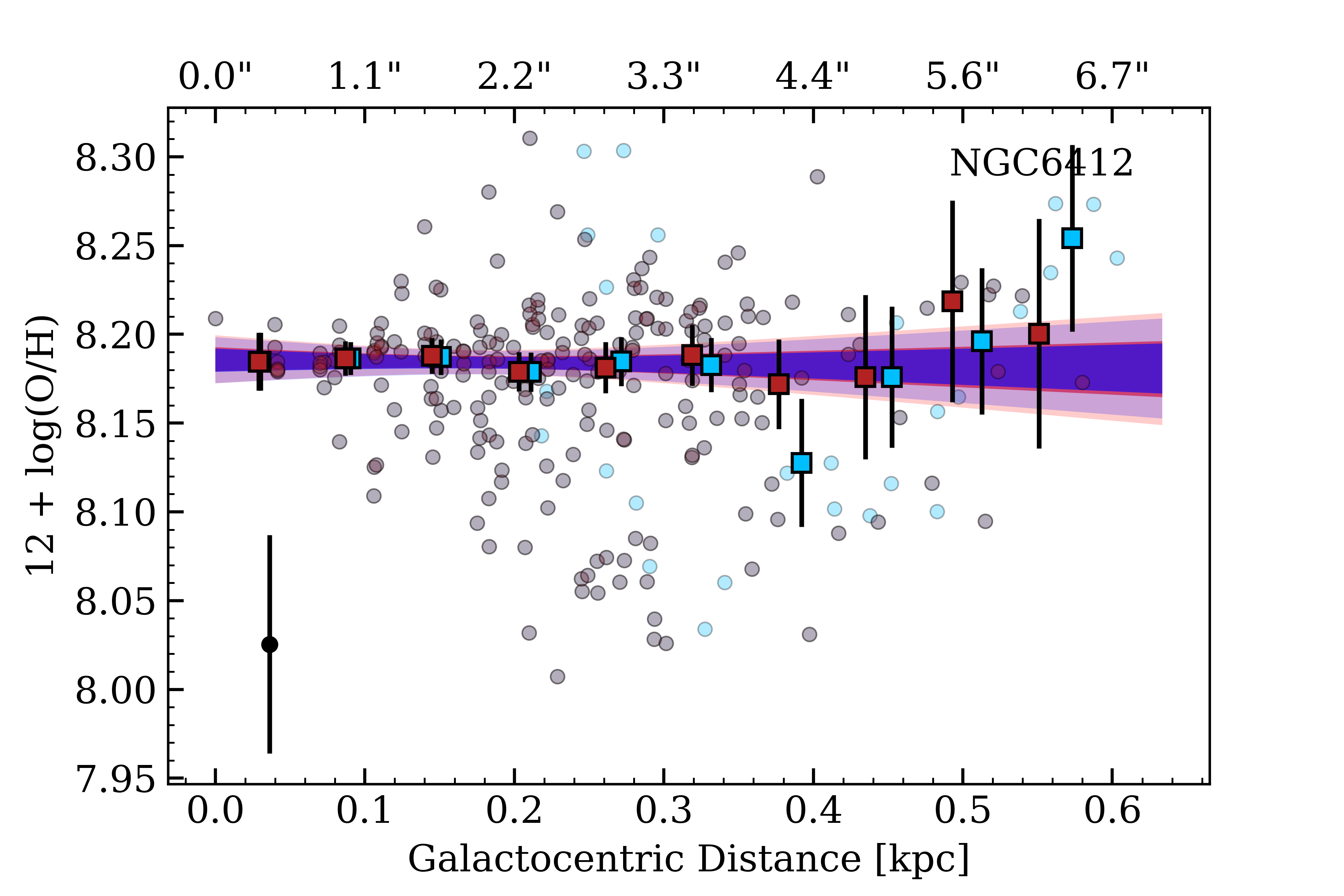}
    \includegraphics[trim={0cm 0mm 0cm 6.2mm},clip, width=0.44\linewidth]{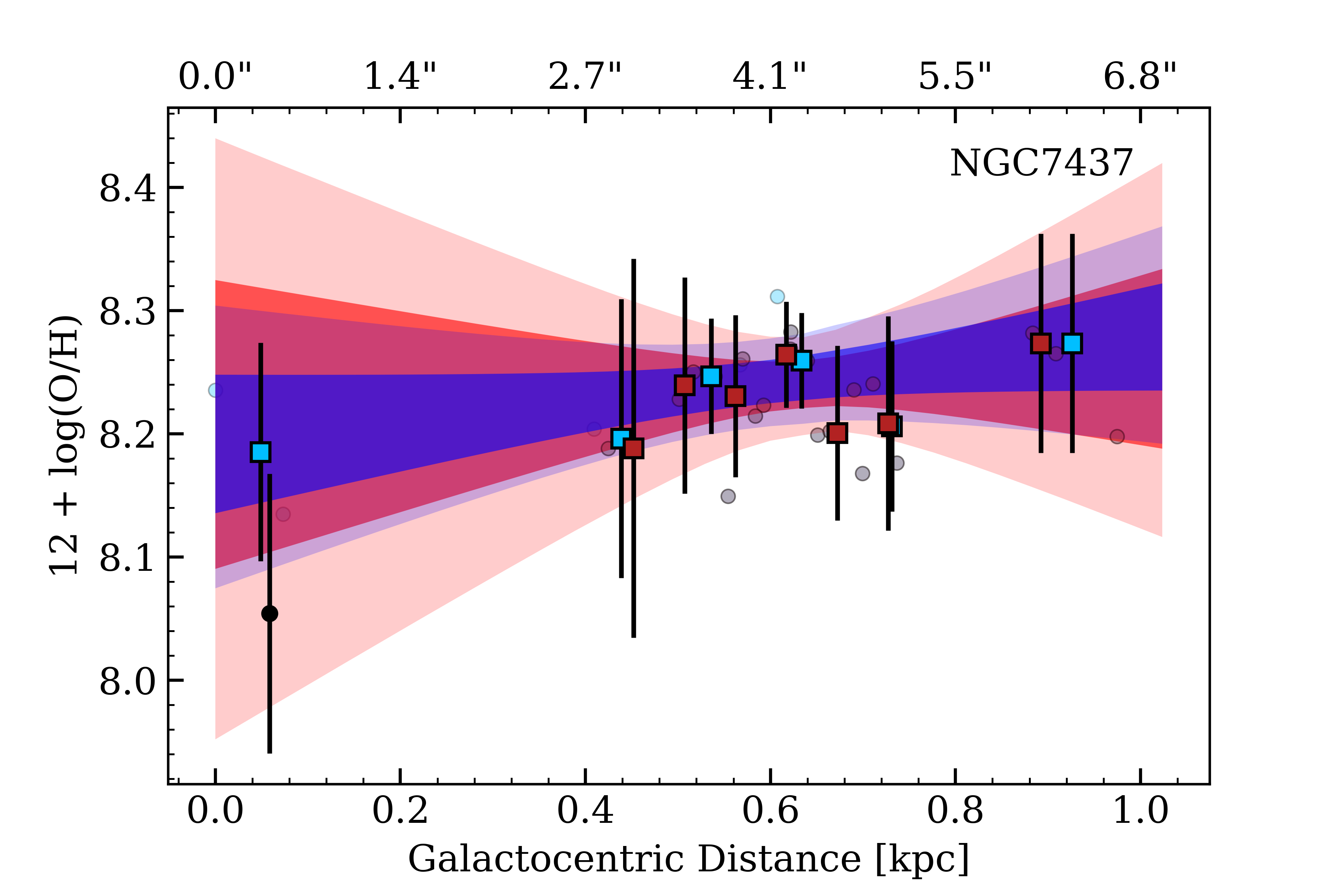}
    \includegraphics[trim={0cm 0mm 0cm 6.2mm},clip, width=0.44\linewidth]{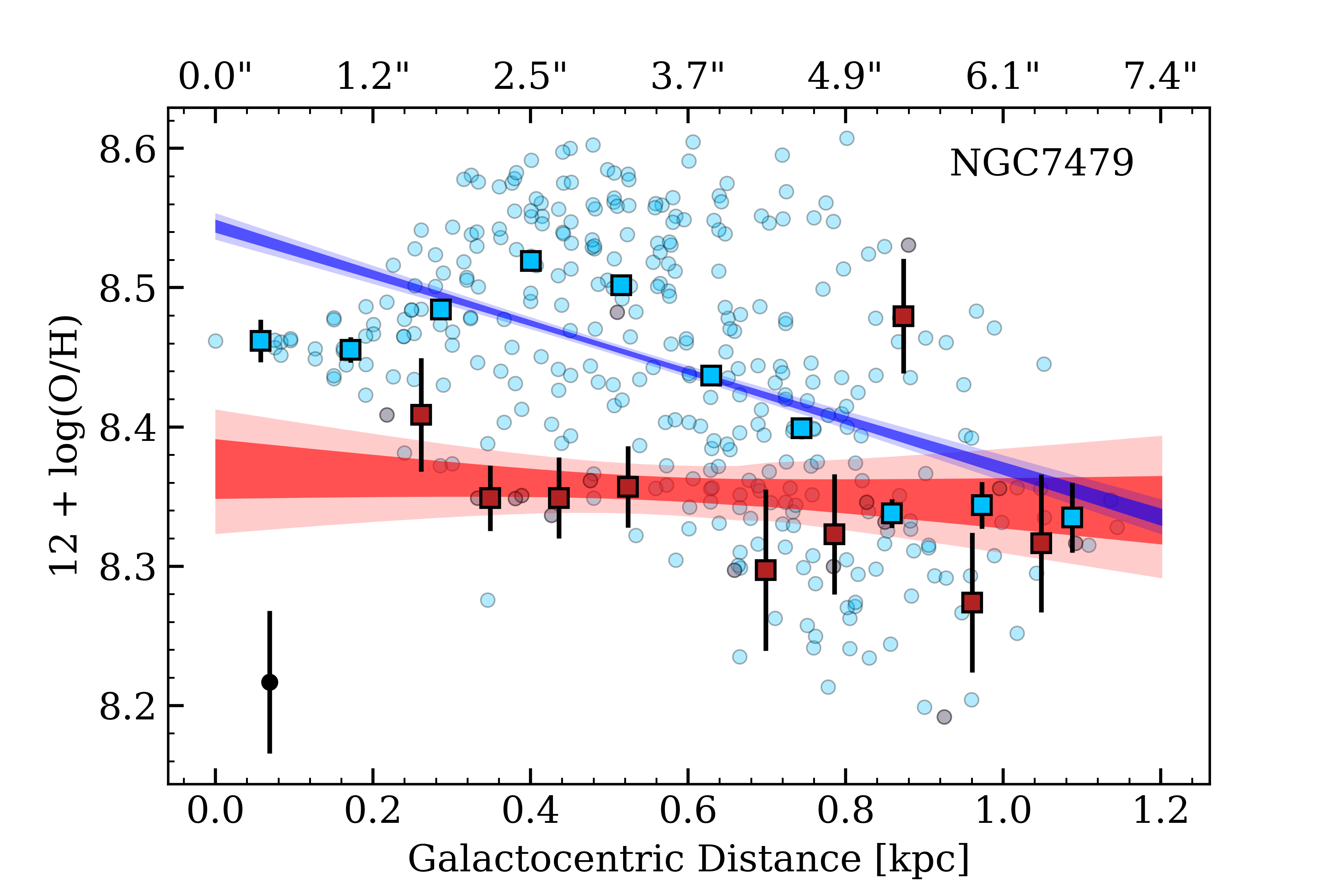}
    \includegraphics[trim={0cm 0mm 0cm 6.2mm},clip, width=0.44\linewidth]{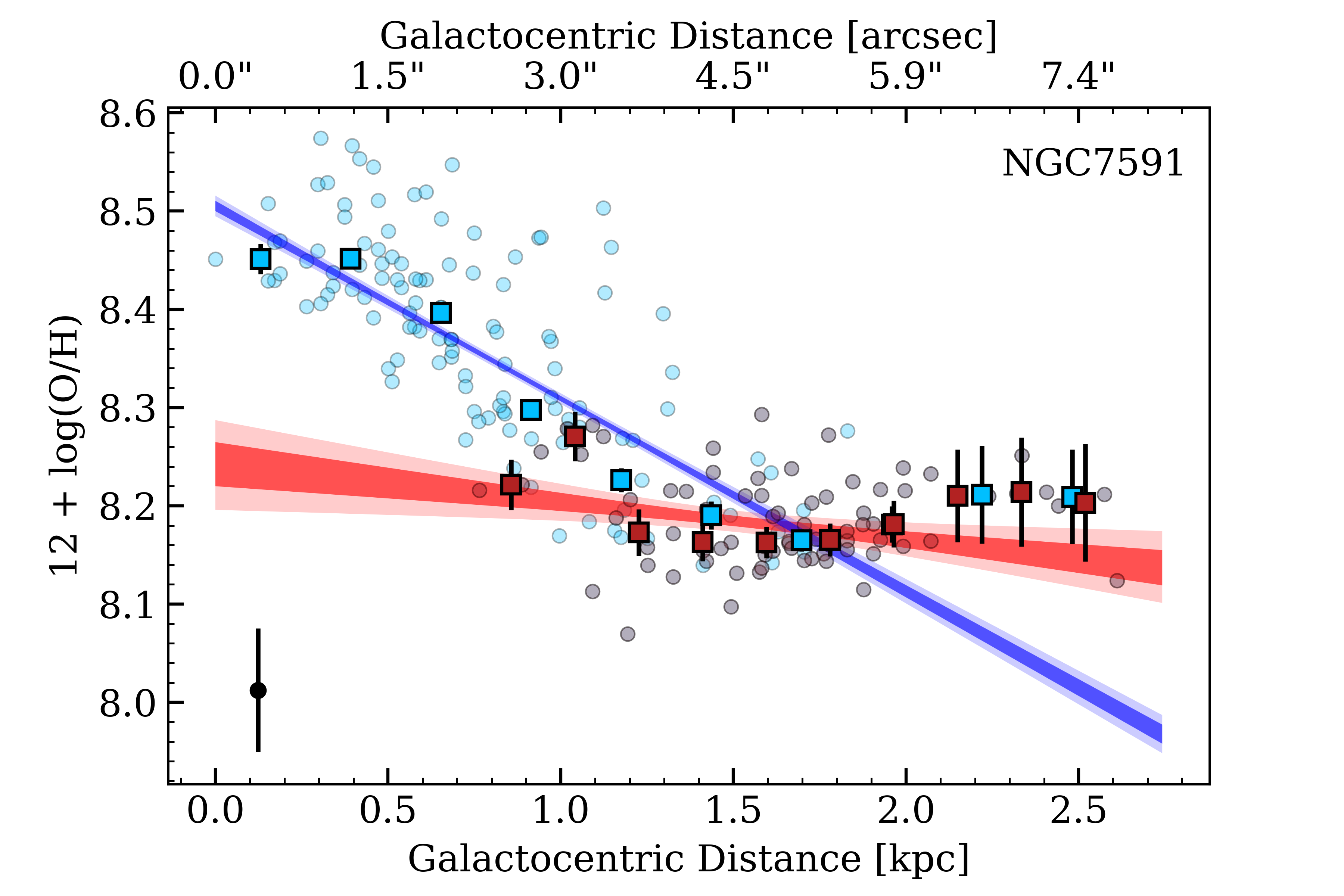}
	\caption{continued.}
	\label{fig:B1}
\end{figure*}
\addtocounter{figure}{-1}
\begin{figure*}[h]
	\centering
    \includegraphics[trim={0cm 0mm 0cm 6.2mm},clip, width=0.44\linewidth]{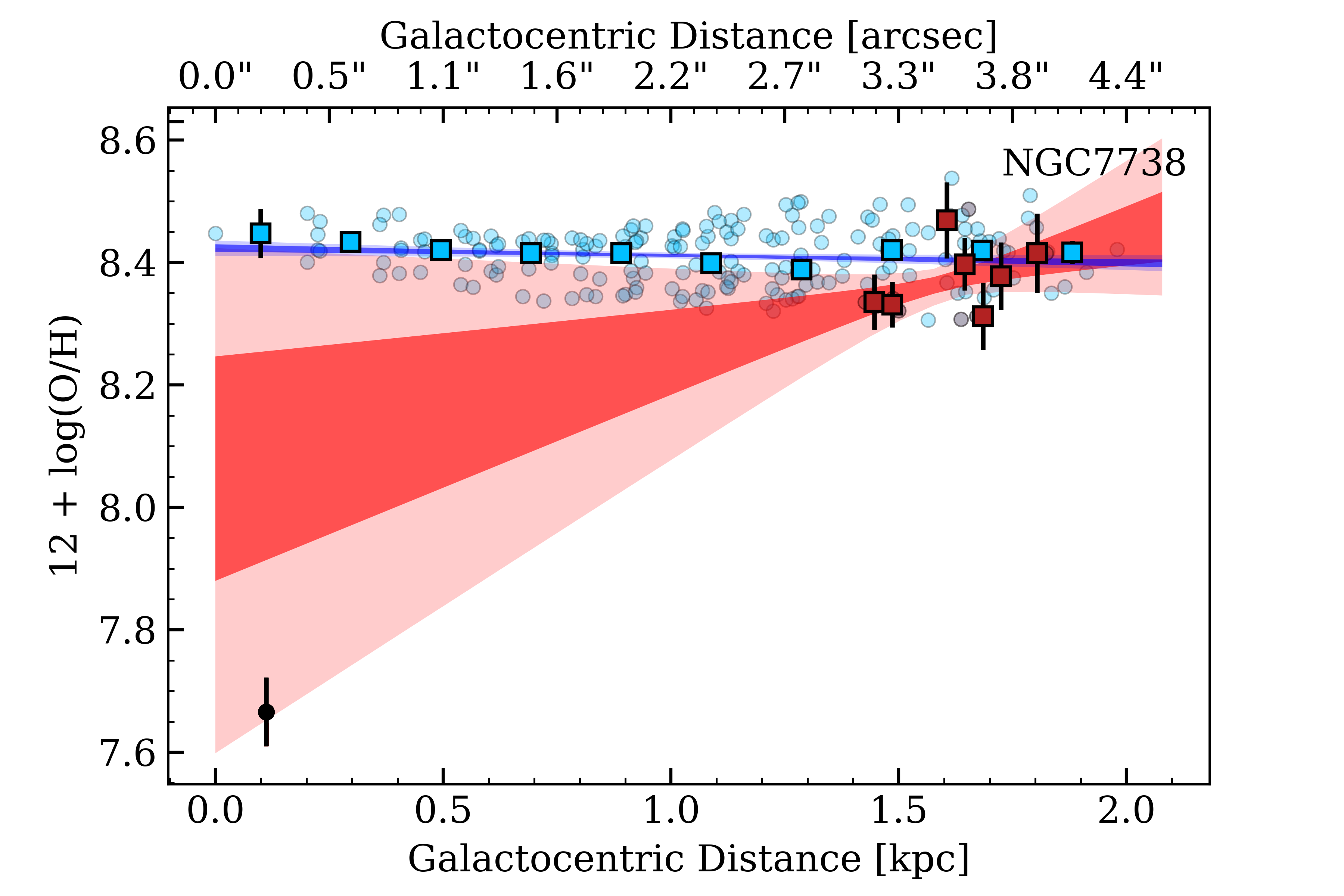}
    \includegraphics[trim={0cm 0mm 0cm 6.2mm},clip, width=0.44\linewidth]{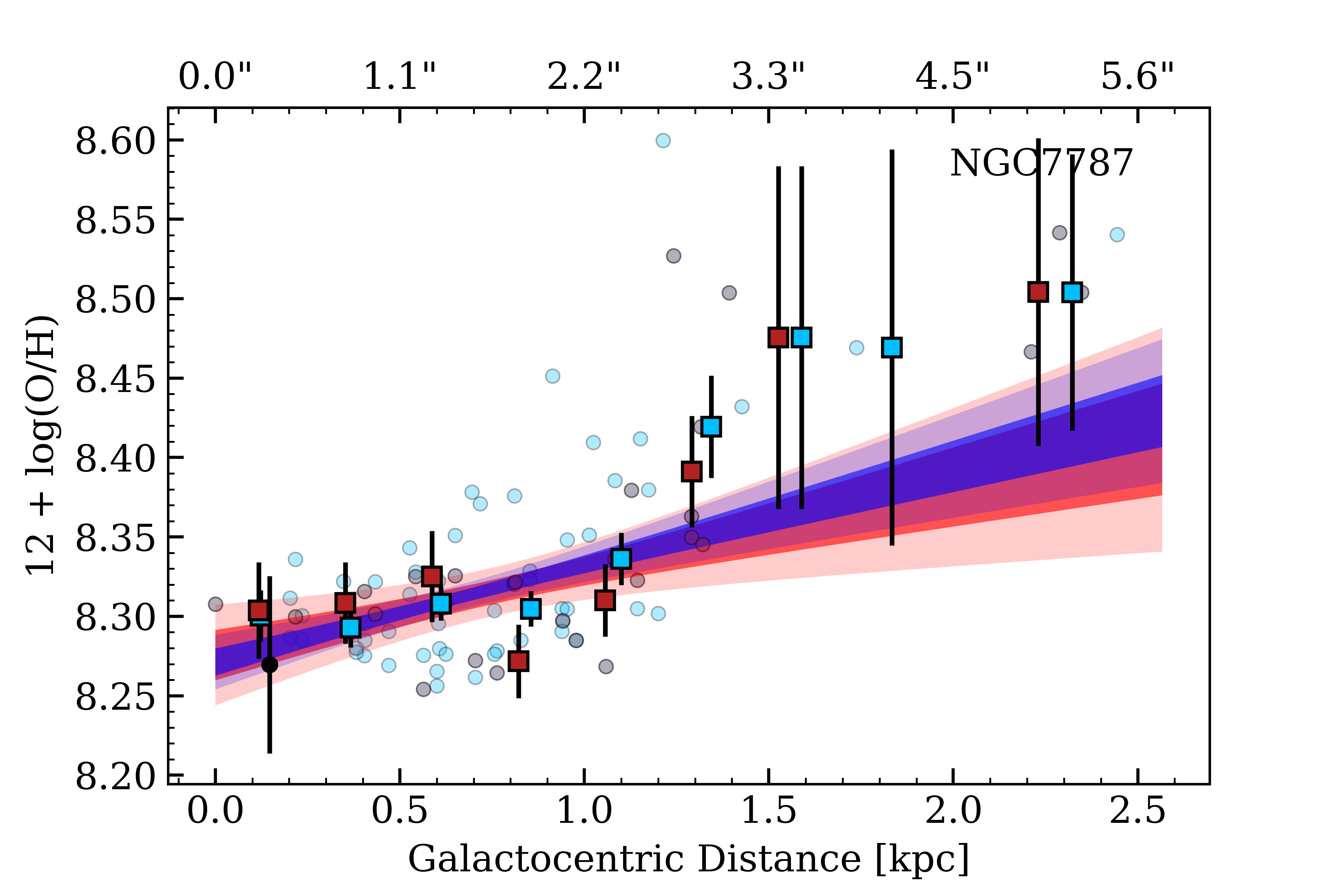}
    \includegraphics[trim={0cm 0mm 0cm 6.2mm},clip, width=0.44\linewidth]{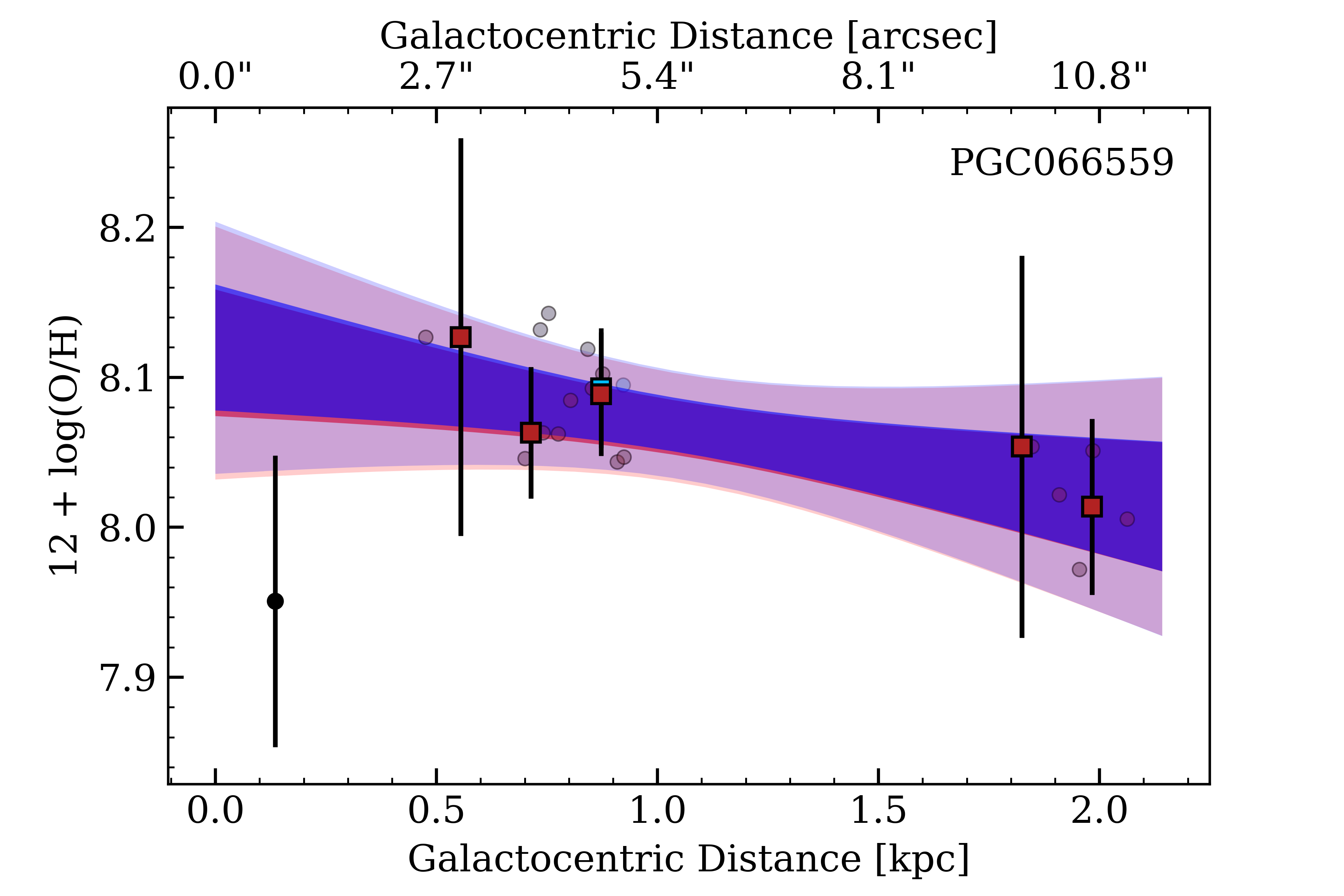}
	\caption{continued.}
	\label{fig:B1}
\end{figure*}

\end{appendix}

\end{document}